\documentclass[apj, letterpaper]{emulateapj}  
\pdfoutput=1

\usepackage{apjfonts}  
\usepackage{subfigure}

\usepackage{amsmath}  

\newcommand{\m}{\rm \,m}
\newcommand{\K}{\rm \,K}
\newcommand{\J}{\rm \,J}
\newcommand{\kg}{\rm \,kg}

\newcommand{\km}{\rm \, km}
\def\R{{\cal P}}

\begin{document}

\slugcomment{\bf}
\slugcomment{Submitted to the Astrophysical Journal}

\title{Equatorial Superrotation on tidally locked exoplanets}

\shorttitle{Equatorial superrotation on hot exoplanets}

\shortauthors{Showman and Polvani}

\author{Adam P.\ Showman\altaffilmark{1,2},
Lorenzo M.\ Polvani\altaffilmark{3}}

\altaffiltext{1}{Department of Planetary Sciences and Lunar and Planetary
Laboratory, The University of Arizona, 1629 University Blvd., Tucson, AZ 85721 USA; showman@lpl.arizona.edu}
\altaffiltext{2}{Part of the work was carried out while APS was on sabbatical at the Department of Applied Physics and Applied Mathematics, Columbia University, New York, NY}
\altaffiltext{3}{Department of Applied Physics and Applied Mathematics, Columbia University, New York, NY and Lamont-Doherty Earth Observatory, Palisades, NY}

\begin{abstract}
\label{abstract}
The increasing richness of exoplanet observations has motivated a
variety of three-dimensional atmospheric circulation models of these
planets. Under strongly irradiated conditions, models of tidally
locked, short-period planets (both hot Jupiters and terrestrial planets)
tend to exhibit a circulation
dominated by a fast eastward, or ``superrotating,'' jet stream at the
equator.  When the radiative and advection time scales are comparable,
this phenomenon can cause the hottest regions to be displaced eastward
from the substellar point by tens of degrees longitude. Such an offset
has been subsequently observed on HD 189733b, supporting the
possibility of equatorial jets on short-period exoplanets. Despite its
relevance, however, the dynamical mechanisms responsible for
generating the equatorial superrotation in such models have not
been identified. Here, we show that the equatorial jet results from the
interaction of the mean flow with standing Rossby waves induced by the
day-night thermal  forcing. The strong longitudinal variations in
radiative heating---namely intense dayside heating and nightside
cooling---trigger the formation of standing, planetary-scale
equatorial Rossby and Kelvin waves. The Rossby waves develop phase
tilts that pump eastward momentum from high latitudes to the equator,
thereby inducing equatorial superrotation. We present an analytic theory
demonstrating this mechanism and explore its properties in a 
hierarchy of one-layer (shallow-water) calculations and
fully 3D models. The wave-mean-flow interaction produces an
equatorial jet whose latitudinal width is comparable to that of the
Rossby waves, namely the equatorial Rossby deformation radius modified
by radiative and frictional effects. For conditions typical of
synchronously rotating hot Jupiters, this length is comparable to a
planetary radius, explaining the broad scale of the equatorial jet
obtained in most hot Jupiter models. Our theory illuminates
the dependence of the equatorial jet speed
 on forcing amplitude, strength of friction, and other parameters, 
as well as the conditions under which jets can form at all.
\end{abstract}

\keywords{hydrodynamics -- methods: analytical -- methods: numerical --
planets and satellites: atmospheres -- planets and satellites:
individual (HD 189733b) -- waves}


\section{Introduction}
\label{Introduction}

The past few years have witnessed major strides in our efforts to
understand the atmospheric circulation of short-period
exoplanets---both gas giants (hot Jupiters) and smaller terrestrial
planets.  Infrared photometry, spectra, and
light curves from the {\it Spitzer} and {\it Hubble} Space Telescopes now 
provide constraints on the three-dimensional temperature structure of
several hot Jupiters, which hint at a vigorous atmospheric circulation on
these bodies \citep[e.g.,][]{knutson-etal-2007b, knutson-etal-2009a,
charbonneau-etal-2008, harrington-etal-2006,
cowan-etal-2007, swain-etal-2009, crossfield-etal-2010}.   These
observations have motivated a growing effort to model the atmospheric 
circulation on these objects: to date, many three-dimensional atmospheric circulation models 
of hot Jupiters have been published \citep{showman-guillot-2002, 
cooper-showman-2005, cooper-showman-2006, showman-etal-2008a, 
showman-etal-2009, dobbs-dixon-lin-2008, menou-rauscher-2009, 
rauscher-menou-2010, dobbs-dixon-etal-2010, thrastarson-cho-2010,
lewis-etal-2010, perna-etal-2010, heng-etal-2010}.  These models have 
emphasized synchronously rotating hot Jupiters in circular, 
approximately 2--5-day orbits.

Just as the last decade witnessed the first characterization of hot Jupiters,
the next decade will see a shift toward characterizing 
``super Earths'' (planets of 1--10 Earth
masses) and terrestrial planets.  To date, roughly 30 super Earths have been
discovered, including several that transit their host stars
\citep{charbonneau-etal-2009, leger-etal-2009, batalha-etal-2011}
with hundreds of additional candidates recently announced from
the NASA {\it Kepler} mission
\citep{borucki-etal-2011}.  Attempts to observationally characterize
their atmospheres have already begun \citep{bean-etal-2010}.  In
anticipation of this vanguard, several three-dimensional circulation
models of tidally locked, short-period terrestrial exoplanets have
been published \citep{joshi-etal-1997, joshi-2003, merlis-schneider-2011,
heng-vogt-2011}.

Intriguingly, the flows in most of these three-dimensional
models---both hot Jupiters and terrestrial planets---develop 
a fast eastward, or {\it superrotating}, jet stream at the equator, with 
 westward flow typically occurring at deeper levels and/or higher latitudes.
In hot-Jupiter models, 
the superrotating jet extends from the equator to latitudes of typically
20--$60^{\circ}$ and is perhaps the dominant dynamical feature
of the modeled flows.  In some cases (depending on the strength of
the imposed stellar heating and other factors), this jet causes an eastward 
displacement of the hottest regions 
from the substellar point by $\sim$$10^{\circ}$ to $60^{\circ}$ longitude.
\citet{showman-guillot-2002} first predicted this feature and suggested
that, if it existed on hot Jupiters, it would have important implications for 
infrared spectra and light curves.  This prediction has been confirmed
in Spitzer infrared observations of HD 189733b
\citep{knutson-etal-2007b, knutson-etal-2009a}, suggesting that this
planet may indeed exhibit a superrotating jet.

However, despite the ubiquity of equatorial superrotation in three-dimensional
models of synchronously rotating exoplanets---and its 
relevance for observations---the mechanisms that produce
this superrotation have yet to be identified.  As demonstrated in
a theorem due to \citet{hide-1969}, such superrotation cannot result
from atmospheric circulations that are longitudinally
symmetric or that conserve angular momentum per mass about the planetary 
rotation axis.   The equatorial atmosphere is the region of the planet 
farthest from the planetary rotation axis, and a superrotating equatorial
jet therefore corresponds to a local maximum in the angular momentum per 
mass about the planetary rotation axis.  Thus, any angular-momentum-conserving 
circulation that moves air to the equatorial atmosphere from higher 
latitudes or deeper levels tends to produce {\it westward} equatorial flow.  
Equivalently, Coriolis forces 
always induce westward acceleration for air moving equatorward or upward,
so an eastward equatorial jet cannot result from Coriolis forces
acting on air that moves into the equatorial atmosphere from higher latitudes
or deeper levels.  In Earth's equatorial troposphere, for example,
the flow is westward, which results from the tropospheric Hadley
cell on Earth (a regime where a mean overturning circulation and its
Coriolis accelerations plays the defining role \citep{held-hou-1980}).  
To maintain equatorial superrotation, a mechanism is needed that
pumps angular momentum per mass from regions where it
is {\it small} (outside the jet) to regions where it is {\it large} (within
the jet)---a so-called ``up-gradient''
momentum transport.  According to Hide's theorem, this transport can
only be accomplished by waves or eddies.

Equatorial superrotation exists in several atmospheres of the 
Solar System---the equatorial atmospheres of Venus, Titan, Jupiter, and 
Saturn all superrotate.  Even localized layers within Earth's equatorial 
stratosphere exhibit superrotation, part of a phenomenon called
the ``Quasi-Biennial Oscillation'' or QBO \citep{andrews-etal-1987}.
The mechanisms for driving the equatorial superrotation on
these planets are diverse.  Possible mechanisms include eddy transport 
associated with baroclinic instabilities, barotropic instabilities, turbulence,
and the absorption/radiation of various types of atmospheric waves 
\citep[e.g.,][]{williams-2003a, williams-2003b, 
lian-showman-2008, lian-showman-2010, schneider-liu-2009, delgenio-etal-1993,
delgenio-zhou-1996, andrews-etal-1987, mitchell-vallis-2010}.

Here, we demonstrate how the equatorial superrotation in three-dimensional
models of synchronously rotating, short-period exoplanets can result from the
existence of standing, planetary-scale Rossby waves;
such waves are naturally excited by the longitudinally dependent
heating patterns---dayside heating and nightside cooling---that
accompany the photospheric regions of short-period, synchronously rotating 
planets.  Section 2 provides background.  In Section 3, we present an analytic
theory demonstrating the mechanism, and we systematically explore its behavior 
in idealized, nonlinear one-layer models.  In Section 4 we extend the
analysis to three-dimensional circulation models.  In Section 5, we 
summarize and discuss implications.

\section{Background Theory}
\label{background}

The ability of Rossby waves to accelerate jets can be schematically
illustrated using the
two-dimensional non-divergent barotropic vorticity equation, 
which is the simplest model for the global-scale flow of a 
planetary atmosphere \citep[see discussion in][]{vallis-2006, showman-etal-2010}.  
The equation reads
\begin{equation}
{d(\zeta + f)\over dt} = F
\label{barotropic-vorticity-eq}
\end{equation}
where $\zeta\equiv {\bf k}\cdot\nabla\times {\bf v}$ is the relative
vorticity, ${\bf v}$ is the horizontal wind velocity, ${\bf k}$ is
the local upward unit vector, $f\equiv 2\Omega\sin\phi$ is the
Coriolis parameter, $\Omega$ is the planetary rotation rate ($2\pi$
over the rotation period), $\phi$ is latitude, and $d/dt = \partial/\partial t +
{\bf v}\cdot\nabla$ is the material derivative (i.e., the derivative
following the flow).  The equation states that individual fluid parcels
conserve the absolute vorticity, $\zeta + f$, save for vorticity
sources/sinks, which are represented by the term $F$.  The equation
can equivalently be written
\begin{equation}
{\partial\zeta\over\partial t} + {\bf v}\cdot \nabla\zeta + v\beta = F
\label{barotropic-vorticity-eq2}
\end{equation}
where $v$ is the meridional (northward) wind speed and $\beta=df/dy$
is the gradient of the Coriolis parameter with northward distance
$y$.  Because the flow in this simple model is horizontally non-divergent,
we can define a streamfunction $\psi$ such that $u=-\partial\psi/\partial y$
and $v=\partial \psi/\partial x$, where $x$ is the eastward
coordinate and $u$ is the zonal (eastward) wind speed. This allows
the equation to be written as a function of one variable, $\psi$. 

For purposes of illustration, consider the linearized version of
Eq.~(\ref{barotropic-vorticity-eq2}) with no sources and sinks.
The solutions to this linearized, unforced equation are Rossby waves.
For simplicity, we 
consider Cartesian geometry with constant $\beta$, representing a local
region on the sphere.  Decomposing variables into zonal means (denoted
with overbars) and deviations therefrom (denoted with primes), and
assuming that the mean flow is zero, leads to a solution
given by $\psi'=\hat\psi \exp[i(kx+ ly)]$, 
where $i$ is the imaginary number and $k$ and $l$ are the zonal and meridional 
wavenumbers.  The dispersion relation is
\begin{equation}
\omega = -{\beta k\over k^2 + l^2}.
\end{equation}
These waves propagate meridionally with a group velocity given by
\begin{equation}
{\partial\omega\over\partial l}={2\beta kl\over (k^2+l^2)^2}.
\label{group-velocity}
\end{equation}

A simple argument, first clearly presented by \citet{thompson-1971}
and reviewed in \citet{held-2000} and
\citet{vallis-2006}, shows how these waves can produce an east-west
acceleration of the zonal-mean flow.  The latitudinal transport of eastward 
eddy momentum per mass is $\overline{u'v'}$, where $u'$ and $v'$ are
the deviations of the zonal and meridional winds from their zonal means,
respectively,
and the overbar denotes a zonal average.  Using the solutions for the 
wave-induced zonal and meridional wind, 
$u'=-il\hat\psi\exp[i(kx+ly)]$ and $v'=ik\hat\psi\exp[i(kx +ly)]$,
yields a momentum flux
\begin{equation}
\overline{u'v'}=-{1\over 2}\hat\psi^2 kl.
\label{uprime-vprime}
\end{equation}
Since the group velocity must point away from the region of wave
generation (which we call the ``wave source''), and
since $\beta$ is positive, we
must have $kl>0$ north of the source and $kl<0$ south of the source. 
Thus, {\it north} of the source,
$\overline{u'v'}$ is negative, implying southward transport of
eastward momentum.  But {\it south} of the source, $\overline{u'v'}$
is positive, implying northward transport of eastward momentum.
Rossby waves therefore transport eastward momentum into the wave source
region.   An eastward acceleration must therefore occur in the wave
source region and a westward acceleration must occur in the region
of wave breaking or dissipation.  This can lead to the formation
of zonal (east-west) jet streams.\footnote{The dynamical picture outlined 
above is not limited to small-amplitude
disturbances, as can be shown with a simple argument described for 
example in \citet{held-2000} and \citet{vallis-2006}.  Imagine
an initially undisturbed latitude, where the absolute-vorticity contour
initially aligns with the latitude circle, and suppose a disturbance---of any
amplitude---propagates into that latitude from elsewhere.  The disturbance 
will perturb the absolute vorticity contours, causing northward transport 
of air in some regions and southward transport in others.  Because 
absolute vorticity generally increases northward,
the northward advection carries with it air of low absolute vorticity,
whereas the southward advection carries with it air of high absolute
vorticity.  Thus, this process will generally cause a southward flux of 
absolute vorticity, thereby decreasing the areal integral of the absolute
vorticity over the polar cap bounded by the latitude circle in question.
By Stokes' theorem,
this implies that the zonal-mean zonal wind
{\it decelerates} (i.e., accelerates westward) because of this vorticity 
flux.  In absence of dissipative processes, this deceleration would reverse
if the disturbance exited the region.  However, when mixing occurs
(e.g., if the wave breaks), or if the disturbance is damped before
air parcels can return to their original latitudes, then the areal
integral of the vorticity inside the latitude circle has been irreversibly
decreased, and the westward impulse cannot be undone.  Thus, we
again recover the result that westward acceleration occurs in the
region of wave dissipation; if momentum is conserved, eastward acceleration
would then occur in the wave-source region.}

\begin{figure}
\vskip 10pt
\includegraphics[scale=0.5, angle=0]{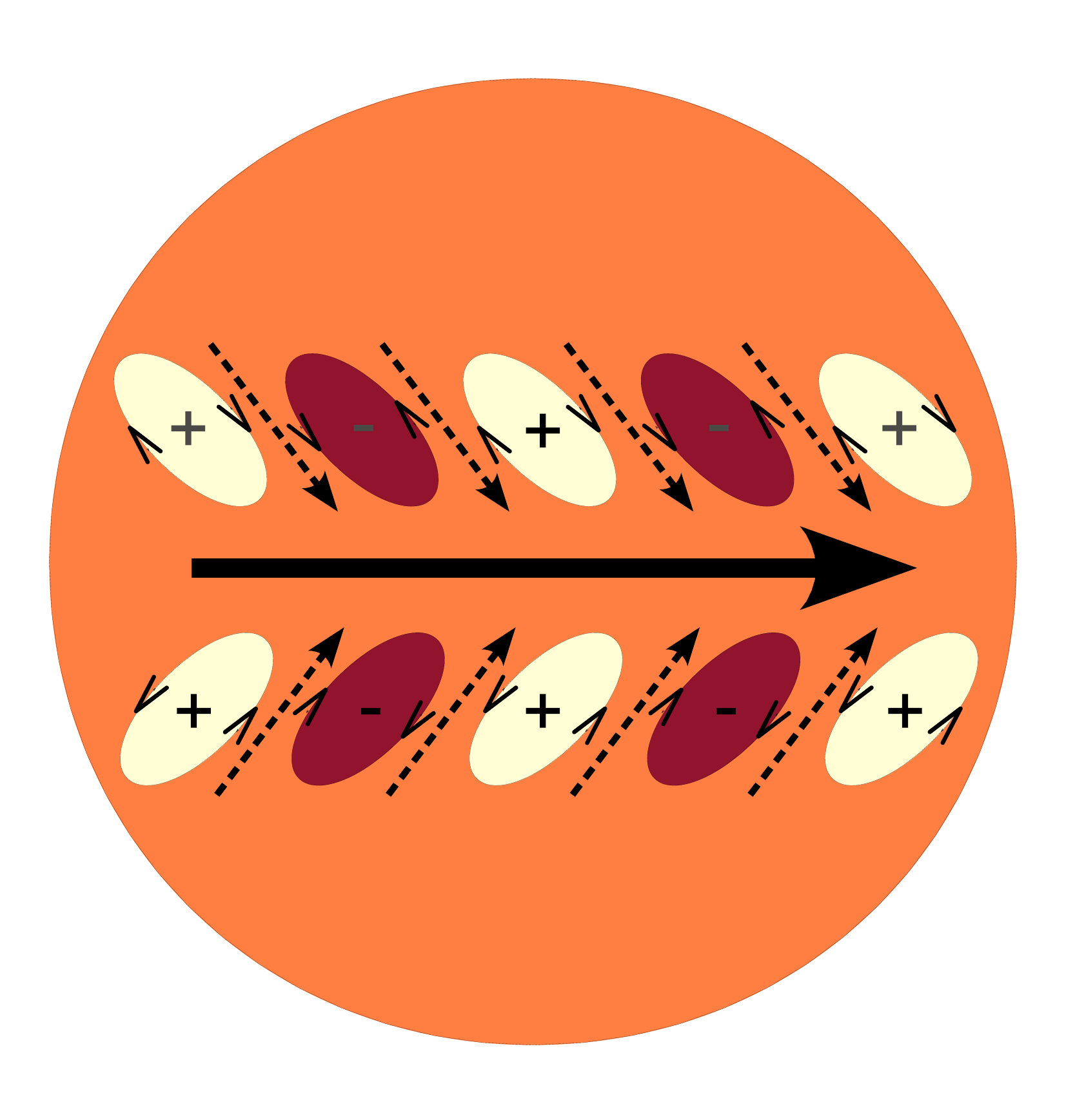}
\caption{Illustration of the dynamical mechanism for generating equatorial
superrotation on tidally locked short-period exoplanets, including hot
Jupiters and super Earths.  The intense day-night heating gradient
generates standing, planetary-scale Rossby and Kelvin waves.  These
waves develop a structure with velocities tilting northwest-to-southeast 
in the northern hemisphere and southwest-to-northeast in the southern
hemisphere (yellow and red ovals).  In turn, these patterns 
transport eddy momentum from high latitudes to the equator (dashed arrows). 
Equatorial superrotation therefore emerges (thick, right-pointing arrow).  }
\label{qualitative-scenario}
\end{figure}

Rossby waves correspond to latitudinal oscillations in
surfaces of constant potential vorticity\footnote{Potential vorticity
is a quantity related to vorticity
that is conserved in adiabatic, frictionless, stratified
flow.  For the barotropic system it is simply the absolute
vorticity $\zeta+f$; for the shallow-water system it is absolute
vorticity over layer thickness $(\zeta+f)/h$, and for a 
three-dimensional stratified atmosphere is it given by 
$\rho^{-1}(\nabla\times{\bf v}+2{\bf \Omega})\cdot\nabla\theta$,
where $\rho$ is density, $\Omega$ is the planetary rotation vector,
and $\theta$ is potential temperature.  For discussion of the 
conservation of potential vorticity and its uses in dynamics, see
\citet{pedlosky-1987} or \citet{vallis-2006}.}; thus, any process
that triggers such oscillations at large scales will tend to excite 
Rossby waves.  In Earth's atmosphere, one of the predominant sources 
is baroclinic instability, which occurs 
in the mid-latitude troposphere where latitudinal temperature gradients 
are large.  Spatially varying tropospheric heating and cooling 
(e.g., due to land-sea contrasts)
or flow over topography also perturb the potential vorticity contours
and can therefore trigger Rossby waves.  In the atmospheres of tidally
locked, hot exoplanets,
on the other hand, the day-night heating pattern constitutes the
overriding dynamical forcing.  For such planets, we expect this
heating/cooling pattern to trigger Rossby waves at low latitudes
(Fig.~\ref{qualitative-scenario}).

The above theory is for free waves.  Consider now the extension
to an atmosphere forced by vorticity sources/sinks and damped by frictional
drag.  The zonal-mean zonal momentum equation of the barotropic system
reads
\begin{equation}
{\partial \overline{u}\over\partial t}=-{\partial (\overline{u'v'})\over
\partial y} - {\overline{u}\over\tau_{\rm drag}}
\label{barotropic-zonal-mean-momentum}
\end{equation}
where overbars denote zonal means and primes denote deviations
therefrom.  The equation states that accelerations of the zonal-mean zonal flow result
from convergences of the latitudinal eddy momentum flux and from drag,
which we have parameterized as a term that relaxes the
zonal-mean zonal wind toward zero over a drag time constant $\tau_{\rm drag}$.
The relationship between the eddy acceleration in
Eq.~(\ref{barotropic-zonal-mean-momentum}) and the vorticity
sources/sinks can be made in two steps.  First, we note that
the definition of vorticity implies that $\overline{v'\zeta'}
= - \partial(\overline{u'v'})/\partial y$.   Second, we multiply
the linearized version of Eq.~(\ref{barotropic-vorticity-eq2})
by $\zeta'$ and zonally average.  This leads to an equation
for the budget of the so-called ``pseudomomentum''
\citep[][p.~493]{vallis-2006}:
\begin{equation}
{\partial{\cal A}\over\partial t} + \overline{v'\zeta'} = 
    {\overline{\zeta'F'}\over
2(\beta-{\partial^2 \overline{u}\over\partial y^2})}.
\label{pseudomomentum}
\end{equation}
 For the two-dimensional non-divergent model, 
${\cal A}=(\beta - \partial^2 \overline{u}/\partial y^2)^{-1}
\overline{\zeta'^2}/2$ is the pseudomomentum, which is a measure
of wave activity.  By combining 
Eqs.~(\ref{barotropic-zonal-mean-momentum})--(\ref{pseudomomentum})
and supposing that the wave amplitudes and zonal-mean zonal wind
are statistically steady, i.e., $\partial {\cal A}/\partial t \approx 0$ and
$\partial\overline{u}/\partial t \approx 0$, we obtain
\begin{equation}
 {\overline{u}\over\tau_{\rm drag}}= {\overline{\zeta'F'}\over
2(\beta-{\partial^2 \overline{u}\over\partial y^2})}.
\label{barotropic-steady}
\end{equation}
This equation relates the vorticity sources/sinks and drag to
the zonal-mean zonal wind, $\overline{u}$. 
When eddy sources/sinks of relative vorticity on average
exhibit the same sign as the vorticity itself (i.e. $\overline{\zeta'F'}>0$), 
the eddy acceleration is eastward, and in steady state results in an
eastward zonal-mean zonal wind.  When sources/sink of relative vorticity tend to exhibit
the opposite sign as the vorticity ($\overline{\zeta'F'}<0$), the eddy
acceleration is westward, and in steady state results in westward
zonal-mean zonal wind.\footnote{These arguments assume that 
$\beta-\partial^2\overline{u}/\partial y^2 > 0$, which is generally the case.}
In analogy with the free solutions, this behavior is typically interpreted in
terms of the generation, latitudinal propagation, and dissipation
of Rossby waves.

This mechanism is thought to be responsible for the eddy-driven
jet streams (and the associated eastward surface winds) in 
Earth's midlatitudes: baroclinic instability
generates Rossby waves that radiate away from the midlatitudes,
causing eastward eddy acceleration there and leading to eastward
surface flow in steady state \citep{held-2000, vallis-2006}.  At
the equator, Earth's troposphere does not superrotate;
nevertheless, idealized Earth general circulation models (GCMs) have
shown that the presence of strong zonally varying heating and cooling in the tropics 
can cause equatorial superrotation to emerge \citep{suarez-duffy-1992, 
saravanan-1993, kraucunas-hartmann-2005, norton-2006}.  In analogy
with the theory described above, \citet{held-1999b}  suggested heuristically that the 
superrotation in these models results from the generation and poleward 
propagation of Rossby waves by the tropical heating sources.  

Still, application of this barotropic theory to tidally locked
exoplanets is problematic.  
Most models of atmospheric circulation on synchronously
rotating, zero-eccentricity hot Jupiters exhibit relatively steady 
circulation patterns whose velocity and temperature patterns are
approximately symmetric about the equator \citep{showman-guillot-2002,
cooper-showman-2005, cooper-showman-2006, showman-etal-2008a,
showman-etal-2009, dobbs-dixon-lin-2008, rauscher-menou-2010}.
In such models, the relative vorticity is approximately antisymmetric
about, and zero at, the equator.
Eq.~(\ref{barotropic-steady}) predicts that $\overline{u}=0$ at the
equator for this situation.  Thus, while attractive, this theory fails
to explain the equatorial superrotation in these hot Jupiter models.

There are other difficulties. First, the theory presented here assumes 
the flow is barotropic---i.e. that it can
be described by Eq.~(\ref{barotropic-vorticity-eq})---and thereby
ignores the role of finite Rossby deformation radius in shaping
the wave properties.  Second, vertical motions normally accompany
regions of heating and cooling, and these vertical motions lead
to nonzero horizontal divergence.  The 
low-latitude flow is thus inherently divergent in the presence
of heating/cooling, in conflict with the stated assumptions.  
Moreover, the theory cannot account in any rigorous way 
for the generation of the Rossby waves by thermal forcing.  The above 
equations include
no thermodynamics and only the effect of such forcing in producing
relative vorticity is represented.  Away from the equator, one expects
that heating produces rising motion and horizontal divergence aloft, which
generates anticylonic relative vorticity by the action of Coriolis
forces.  At the equator, however, this vorticity source is small,
leaving unclear the applicability of this barotropic theory in
producing eastward flow {\it at the equator}.\footnote{Within
the context of barotropic theory, one can relax the nondivergent assumption
by resolving the horizontal velocity into rotational and divergent
components, specifying the horizontal divergence field, and then 
solving Eq.~(\ref{barotropic-vorticity-eq2}) for the rotational component
of the flow \citep[see][]{sardeshmukh-hoskins-1985, sardeshmukh-hoskins-1988}.
In this context, the specified divergence field represents the spatial field
of heating and cooling.  The vorticity source in 
Eq.~(\ref{barotropic-vorticity-eq2}) would then be $F=-(\zeta + f)
\nabla\cdot{\bf v_{\chi}} - {\bf v_{\chi}}\cdot\nabla(\zeta +f)$, where
${\bf v_{\chi}}$ is the specified divergent component of the flow field
and the velocity and vorticity on the lefthand side of 
Eq.~(\ref{barotropic-vorticity-eq2}) represent only the rotational
component.  One can then rework 
Eqs.~(\ref{barotropic-zonal-mean-momentum})--(\ref{barotropic-steady})
under the usual barotropic assumptions that vertical momentum/vorticity 
transport and vortex tilting
are negligible. However, such a 
divergent barotropic model suffers from the same 
failing as the nondivergent version:  when the pattern of heating/cooling 
and flowfield are mirror symmetric about the equator---as in most hot-Jupiter 
models---the vorticity source $F$ is zero at the equator, 
leading to a Rossby-wave source
$\overline{\zeta'F'}$ that is likewise zero at the equator.  In
the presence of frictional drag, one again obtains  
the result that $\overline{u}=0$
at the equator.  Thus, even such an extended barotropic treatment is 
insufficient to explain the equatorial
superrotation in hot-Jupiter models.}
When the geostrophic assumption is relaxed and finite deformation 
radius is included,
analytic solutions of freely propagating equatorial waves show
that such waves tend to be trapped within an equatorial waveguide
and cannot propagate away from the equator \citep{holton-2004,
andrews-etal-1987}; such wave solutions
involve no net meridional (north-south) momentum transports, thus
raising the question of whether the above mechanism is viable at the
equator in the presence of finite deformation radius.   Finally, 
for hot Jupiters, the Rossby waves are expected to be global in scale, 
and it is not clear {\it a priori} whether there is {\it room} for them
to propagate poleward from the equatorial regions.

We present a sequence of models in the following sections that
overcome these obstacles and provide a theoretical foundation
for understanding equatorial superrotation on tidally locked
exoplanets.

\section{Shallow-water model of equatorial superrotation}
\label{sw}

Full GCM solutions, although useful, involve so many interacting
processes that it is often difficult to cleanly identify specific
dynamical mechanisms from such solutions \citep[see, e.g.,][]
{showman-etal-2010}.   Simplified models therefore play an important
role in atmospheric dynamics.  Here, we present a highly idealized
model intended to capture the mechanism in the simplest possible context.

We adopt a two-layer model, with constant densities in each
layer, where the upper layer represents the meteorologically active 
atmosphere and the lower layer represents the quiescent deep atmosphere and
interior.  In the limit where the lower layer becomes infinitely
deep and the lower-layer winds and pressure gradients remain steady
in time (which requires the upper layer to be isostatically balanced),
this two-layer system reduced to the shallow-water equations
for the flow in the upper layer \citep[e.g.,][p.~129-130]{vallis-2006}:
\begin{equation}
{d{\bf v}\over dt}+g\nabla h + f{\bf k}\times {\bf v} = {\bf R} 
- {{\bf v}\over{\tau_{\rm drag}}}
\label{momentum}
\end{equation}
\begin{equation}
{\partial h\over\partial t} + \nabla\cdot ({\bf v}h) = 
{h_{\rm eq}(\lambda,\phi) - h\over\tau_{\rm rad}}
\equiv Q
\label{continuity}
\end{equation}
where ${\bf v}(\lambda,\phi,t)$ is horizontal velocity, $h(\lambda,\phi,t)$ 
is the upper layer thickness, 
$t$ time, $g$ is the (reduced) gravity\footnote{$g$ in
  Eq.~(\ref{momentum}) is the actual gravity times the fractional
  density difference between the layers, and is therefore called the
  ``reduced gravity.''  When interpreting this shallow-water model 
in the context of a three-dimensional atmosphere, this density
difference should be interpreted as (for example) the fractional
  change in potential temperature across a scale height.  This is of
  order unity for a hot Jupiter.}, $f=2\Omega\sin\phi$ 
is Coriolis parameter, ${\bf k}$ is the upward unit vector,
$\Omega$ is planetary rotation rate, and $\phi$ is latitude.
Again, $d/dt=\partial/\partial t + {\bf v}\cdot\nabla$ is the
material derivative.
The boundary between the layers represents an atmospheric isentrope,
across which mass is exchanged in the presence of heating or cooling.
Heating and cooling are therefore represented in the shallow-water 
system using mass sources and sinks, represented here as a Newtonian
relaxation of the height toward a specified radiative-equilibrium height
$h_{\rm eq}$---thick on the dayside and thin on the nightside---over 
a radiative time scale $\tau_{\rm rad}$.
The momentum equations (\ref{momentum}) include drag with a timescale 
$\tau_{\rm drag}$, which could represent the potentially
important effects of magnetohydrodynamic friction \citep{perna-etal-2010},
vertical turbulent mixing \citep[e.g.][]{li-goodman-2010}, or 
momentum transport by breaking gravity waves \citep{lindzen-1981, watkins-cho-2010}.

The term ${\bf R}$ in Eq.~(\ref{momentum}) represents the effect on the 
upper layer of momentum advection from the lower layer, and takes the 
same form as in \citet{shell-held-2004} and \citet{showman-polvani-2010}:
\begin{equation}
{\bf R}(\lambda,\phi,t)=
\begin{cases}
   -{Q{\bf v}\over h},  &Q>0;\\
   0, &Q<0
\end{cases}
\label{R}
\end{equation}
where $\lambda$ and $\phi$ are longitude and latitude.
Air moving out of the upper layer $(Q<0)$ does not locally affect the upper 
layer's specific angular momentum or wind speed, hence ${\bf R}=0$ 
for that case.   But air transported into the upper layer carries 
lower-layer momentum with it and thus alters the local specific 
angular momentum and zonal wind in the upper layer.  For the 
simplest case where the lower layer winds are assumed to be zero, this 
process preserves the column-integrated ${\bf v}h$ of the upper layer, 
leading to the expression in (\ref{R}) for $Q>0$.
Importantly, the expression for ${\bf R}$ follows directly from
the momentum budget and contains no free parameters.\footnote{The
condition that air moving out of the upper layer induces no change in the
upper layer's specific momentum requires that such momentum advection 
cause no accelerations, implying that ${\bf R}=0$ for $Q<0$.  
To derive the expression for the case $Q>0$, add ${\bf v}$ times
the continuity equation to $h$ times the momentum equation,
thus yielding an equation for the time rate of change of ${\bf v}h$.
Terms involving heating/cooling are $h{\bf R} + {\bf v}Q$.  In
the special case where the lower-layer winds are zero, mass transport
into the upper layer does not change the column-integrated horizontal
wind, ${\bf v}h$, of the upper layer.  This 
implies that $h{\bf R} + {\bf v}Q=0$, thus yielding Eq.~(\ref{R}).}

The relative roles of the two terms on the right side of 
the momentum equation can be clarified by rewriting it
as follows:
\begin{equation}
{d{\bf v}\over dt}+g\nabla h + f{\bf k}\times {\bf v} = 
-{\bf v}\left[{1\over{\tau_{\rm drag}}} +
{1\over \tau_{\rm rad}}\left({h_{\rm eq}-h\over h}\right)
{\cal H}(h_{\rm eq}-h)\right]
\label{momentum2}
\end{equation}
where ${\cal H}(h_{\rm eq}-h)$ is the Heaviside
step function, defined as 1 when $h_{\rm eq}-h>0$ and 0 otherwise. 
Dynamically, it should be clear that both terms on the right side
play a role analogous
to drag; one can define the entire quantity in square braces as one
over an {\it effective} drag time constant.  Still, the second term 
(${\bf R}$) is spatially heterogeneous and only exists in regions of heating,
and we will show that its effect on the zonal-mean flow is qualitatively
different than that of the first term (frictional drag).  
For a strongly irradiated
hot Jupiter, we might expect $(h_{\rm eq}-h)/h \sim 0.01$--1, and if so
then the first term would dominate if $\tau_{\rm drag} \ll \tau_{\rm rad}$
whereas the second term would dominate if $\tau_{\rm drag} \gg \tau_{\rm rad}$.

We present linear, analytic solutions and fully nonlinear, numerically
determined solutions of Eqs.~(\ref{momentum})--(\ref{R}) in the next two
subsections.

\subsection{Linear solutions}
\label{linear-sw}

To enable analytic solutions, we solve Eqs.~(\ref{momentum})--(\ref{R}) in
Cartesian geometry assuming that the Coriolis parameter can be
approximated as $f=\beta y$, where $y$ is northward distance from the
equator and $\beta$ (the gradient of Coriolis parameter with northward 
distance) is assumed constant.  This approximation, called the
``equatorial $\beta$-plane,'' is strictly valid only at low latitudes,
but we will see in \S\ref{nonlinear-sw} that the qualitative features 
of these solutions are recovered by the full solutions in spherical
geometry.  

We now linearize Eqs.~(\ref{momentum})--(\ref{R}) about a state of
rest.  By definition, all the terms in
the linearized equations have magnitudes that scale with the forcing
amplitude.  Note that the term ${\bf R}$  
involves the product of the velocity and forcing amplitude,
and therefore is quadratic in the forcing amplitude and does
not appear in the linearized equations.  
(We will come back to it when evaluating the implications of the 
solutions for the zonal-mean zonal wind).  The linearized
equations read
\begin{eqnarray}
\label{u-mom-linear}
{\partial u\over\partial t} + g {\partial \eta\over\partial x} -\beta y v
= -{u\over\tau_{\rm drag}} \\
\label{v-mom-linear}
{\partial v\over\partial t} + g{\partial\eta\over\partial y} + \beta y u
= -{v \over\tau_{\rm drag}} \\
\label{h-linear}
{\partial \eta\over\partial t} + 
H\left({\partial u\over\partial x}+{\partial v\over\partial y}\right)
= S(x,y) - {\eta\over \tau_{\rm rad}}
\end{eqnarray}
where $x$ is eastward distance and $\eta$ is the deviation of the 
thickness from its constant reference value $H$, such that $h=H + \eta$.
The quantity $S\equiv (h_{\rm eq}-H)/\tau_{\rm rad}$ is the forcing,
which can also be expressed as $S\equiv \eta_{\rm eq}/\tau_{\rm rad}$,
where $\eta_{\rm eq}\equiv h_{\rm eq}-H$ is the deviation of the
radiative-equilibrium height from $H$.

The free solutions to these equations (i.e., when
the right-hand sides are set to zero) are the well-known equatorially
trapped wave modes, described for example in \citet{holton-2004}
and \citet{andrews-etal-1987}.  Given the intense heating and cooling
experienced by hot, tidally locked exoplanets, however, 
we seek solutions to the forced problem.
Most three-dimensional dynamical models of hot Jupiters exhibit
relatively steady circulation patterns \citep{showman-guillot-2002,
cooper-showman-2005, cooper-showman-2006, dobbs-dixon-lin-2008,
dobbs-dixon-etal-2010, showman-etal-2008a, showman-etal-2009,
rauscher-menou-2010}, and so we seek steady solutions in the
presence of forcing and damping. 

We nondimensionalize Eqs.~(\ref{u-mom-linear})--(\ref{h-linear}) with
a length scale $L=(\sqrt{gH}/\beta)^{1/2}$, a velocity scale
$U=\sqrt{gH}$, and a time scale ${\cal T}=(\sqrt{gH}\beta)^{-1/2}$,
which correspond respectively to the equatorial Rossby deformation
radius, the gravity wave speed, and the time for a gravity wave
to cross a deformation radius in the shallow-water system.  The
thickness is nondimensionalized with $H$, the drag and thermal
time constants with ${\cal T}$, and the forcing with
$H/{\cal T}$.  This yields, for steady flows,
\begin{eqnarray}
\label{u-mom-nd}
{\partial \eta\over\partial x} -yv = - {u\over\tau_{\rm drag}}\\
\label{v-mom-nd}
{\partial\eta\over\partial y} + yu = - {v\over\tau_{\rm drag}}\\
\label{h-nd}
\left({\partial u\over\partial x} + {\partial v\over\partial y}\right)
= S(x,y) - {\eta\over\tau_{\rm rad}}
\end{eqnarray}
where all quantities, including $\tau_{\rm rad}$ and $\tau_{\rm drag}$, 
are now nondimensional.

In pioneering investigations, \citet{matsuno-1966} and \citet{gill-1980}
obtained analytic solutions to Eqs.~(\ref{u-mom-nd})--(\ref{h-nd})
for the special case where the drag and
radiative time constants are equal and drag is
neglected from the meridional momentum equation (\ref{v-mom-nd});
the latter assumption is called the ``longwave approximation'' because it
is valid in the limit where zonal length scales greatly exceed
meridional ones.  On tidally locked exoplanets, however, the drag and radiative
time scales can differ greatly, and the longwave approximation may
not apply, because the flow exhibits comparable zonal and meridional scales.
We therefore retain the full form of Eqs.~(\ref{u-mom-nd})--(\ref{h-nd}).

Equations~(\ref{u-mom-nd})--(\ref{h-nd}) can be combined to
yield a single differential equation for the meridional velocity
$v$ \citep[e.g.,][]{wu-etal-2001}:
\begin{eqnarray}
\nonumber
{1\over\tau_{\rm drag}}
\left({\partial^2v \over\partial y^2} + {\partial^2 v\over\partial x^2}\right) + 
{\partial v\over\partial x} - {1\over\tau_{\rm rad}}\left(y^2 + {1\over\tau_{\rm drag}^2}\right)v=\\
\left(-y{\partial S\over\partial x}
+{1\over\tau_{\rm drag}}{\partial S\over\partial y}\right).
\label{v-equation}
\end{eqnarray}

If one seeks separable solutions, then, as described by 
\citet{gill-1980} and \citet{wu-etal-2001},
the meridional structure of the solutions to this equation with finite 
$\tau_{\rm drag}$ are
the parabolic cylinder functions $\psi_n(y)$, which are simply Gaussians 
times Hermite polynomials\footnote{The first few Hermite polynomials are 
$H_0(\xi)=1$, $H_1(\xi)=2\xi$, $H_2(\xi)=4\xi^2 -2$, and
$H_3(\xi)=8\xi^3-12\xi$.}:

\begin{equation}
\label{parabolic-cylinder-functions}
\psi_n(y)=\exp\left(-{y^2\over 2 \R^2}\right) H_n\left({y\over \R}\right)
\end{equation}
where $\R\equiv (\tau_{\rm rad}/\tau_{\rm drag})^{1/4}$ is the fourth root of 
a Prandtl number.

Our goal is to specify the thermal forcing, $S(x,y)$, and solve for the 
unknowns $u$, $v$, and $\eta$.  
In general, any desired pattern of thermal forcing can be represented
as 
\begin{equation}
S(x,y) = \sum_{n=0}^{\infty} S_n(x) \psi_n(y).
\end{equation}
For tidally locked exoplanets, we expect  this pattern to consist
of a day-night variation in heating/cooling whose amplitude peaks
at low latitudes and diminishes near the poles.  We take the
forcing to be symmetric about the equator (appropriate for 
a planet with zero obliquity) and, to keep the mathematics tractable,
retain solely the term $S_0$, corresponding to pattern of heating 
and cooling that is a Gaussian, centered about the equator,
with a latitudinal half-width of the equatorial
Rossby radius of deformation modified by frictional and radiative
effects.  While the full solution would require consideration
of $S_n$ for all $n\ge0$, the first term, $S_0$, will be the dominant
term for cases where the deformation radius is similar to a planetary
radius, as is the case on typical hot Jupiters. 
Consideration of this term alone will therefore suffice to illustrate the 
qualitative features relevant for inducing an equatorially
superrotating jet on tidally locked exoplanets.

\begin{figure}
\vskip 10pt
\includegraphics[scale=0.6, angle=0]{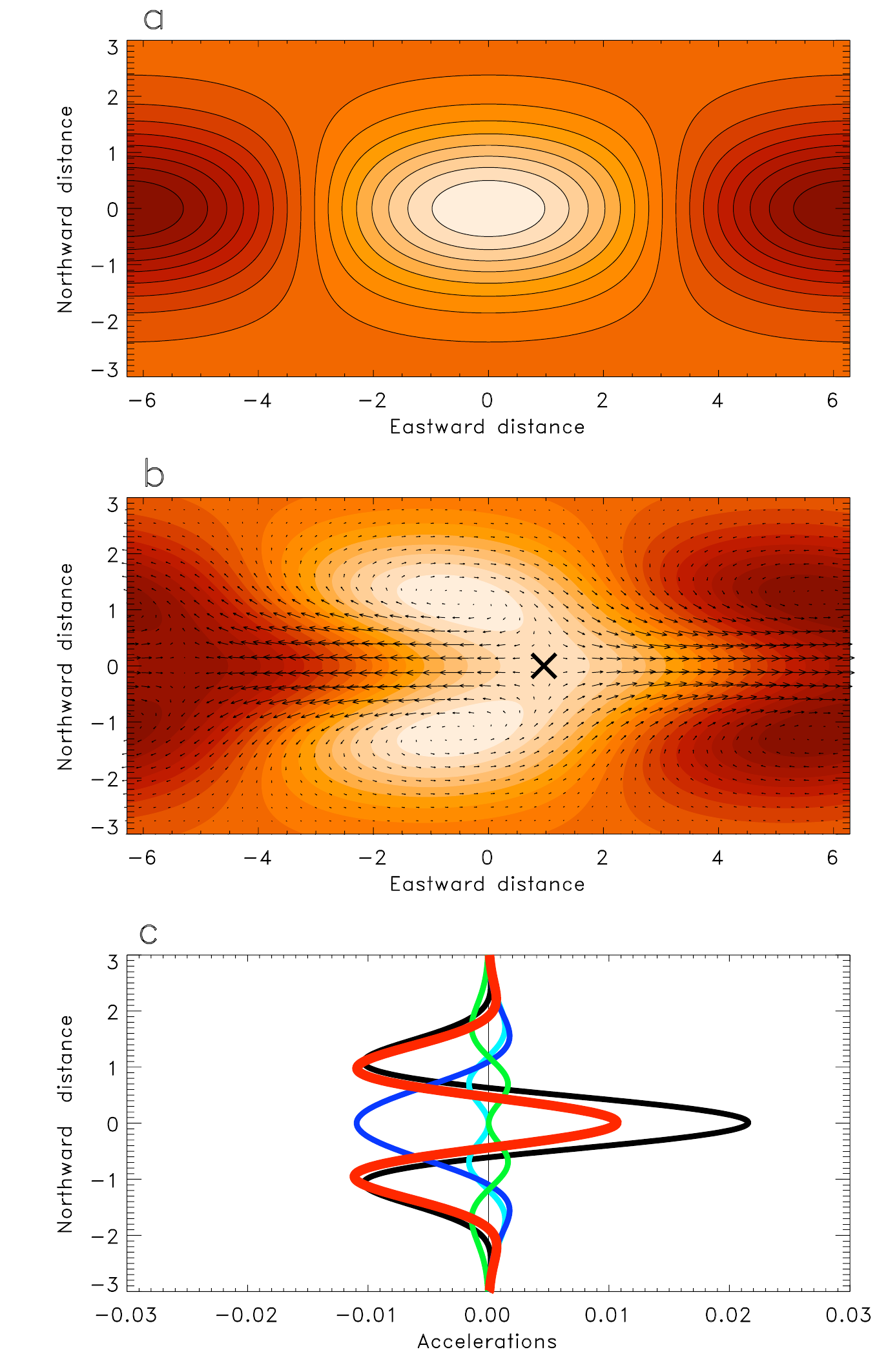}
\caption{An example linear, analytic solution for parameters relevant to
hot, tidally locked exoplanets.  ({\it a})  
Spatial structure of radiative-equilibrium height field, $h_{\rm eq}$
(orangescale and contours).
({\it b}) Height field (orangescale) and horizontal wind velocities
(arrows) for the linear, analytic solution forced by relaxation to the 
$h_{\rm eq}$ profile shown in panel ({\it a}) and with 
nondimensional zonal wavenumber $k=0.5$ and nondimensional radiative
and drag times $\tau_{\rm rad}=\tau_{\rm drag}=5$.  For a hot Jupiter
or hot super Earth,
these correspond to dimensional zonal wavelengths of a planetary
circumference and dimensional radiative and drag time constants
of $\sim$$10^5\rm\,sec$ (see Appendix~\ref{nondim}). In ({\it a}) and
({\it b}), the horizontal and vertical axes are dimensionless
eastward and northward distance, respectively; one unit of distance
corresponds to a dimensional distance of one Rossby deformation radius,
$(\sqrt{gH}/\beta)^{1/2}$.  The $\times$ marks the longitude along
the equator where $h$ reaches a maximum and the eddy zonal wind
changes sign. ({\it c}) Zonal (east-west)
accelerations of the zonal-mean flow implied by the linear solution.
The black and dark blue curves give the accelerations due to
horizontal and vertical eddy transport (terms II and III, respectively,
in Eq.~\ref{tem}).  The light-green and cyan curves show friction (term IV)
and the effect of the mean-meridional circulation (term I), respectively. 
The red curve shows the sum of all terms.  The numerical
values adopt a forcing amplitude $\Delta h_{\rm eq}/H=1$.  For
this value, the nondimensional peak winds approach 0.5, corresponding
to speeds of $\sim$$1\km\sec^{-1}$ on a hot Jupiter.}
\label{matsuno}
\end{figure}

Appendix \ref{gill-solutions} describes the solution method of
Eqs.~(\ref{u-mom-nd})--(\ref{h-nd}) and presents the 
solution for the specific case where the forcing consists solely 
of the $S_0$ term varying sinusoidally in longitude, i.e.,
$S(x,y)=\hat S_0 e^{ikx}\psi_0(y)$, where $\hat S_0$ is a constant.  
Figure~\ref{matsuno} 
shows an example for parameter values typical of a hot Jupiter or
hot super Earth (zonal wavelengths associated with the day-night heating 
contrast of a planetary circumference and 
radiative time constants of order $10^5\rm\,sec$).  For this example,
the drag time constant is taken equal to the radiative time constant.
Figure~\ref{matsuno}a shows the radiative-equilibrium height field and 
Fig.~\ref{matsuno}b presents the steady state height and velocity fields.

The solutions exhibit several important features.  Although
the radiative-equilibrium height field 
is symmetric in longitude about the substellar point (Fig.~\ref{matsuno}a),
the actual height field deviates significantly from
radiative equilibrium and exhibits considerable dynamical structure
(Fig.~\ref{matsuno}b).  Two fundamental types of behavior are 
present.  First, at mid-to-high latitudes
($|y|\sim1$--3 in the figure), the flow exhibits vortical behavior.
The dayside contains an anticyclone in each hemisphere, manifesting 
as a pressure high (i.e., local maximum of the height) around which winds flow
clockwise in the northern hemisphere and counterclockwise in the
southern hemisphere; the nightside contains a cyclone
in each hemisphere, manifesting as a pressure low around which winds
flow counterclockwise in the northern hemisphere and clockwise
in the southern hemisphere.  Second, at low latitudes ($|y| \lesssim 1$), 
the flows are nearly east-west; they diverge from a point east of the 
substellar longitude (marked with a cross in Fig.~\ref{matsuno}b)
and converge toward a point east of the antistellar 
longitude.   

As discussed by \citet{gill-1980}, these features can be interpreted
in terms of forced, damped, steady equatorial wave modes.  The mid-to-high
latitude feature described above is dynamically analogous to that of 
an $n=1$ equatorially trapped Rossby wave, which
exhibits cyclones and anticyclones---alternating in longitude---that
peak off the equator \citep[see][Fig.~4c for an example of the flowfield
in this mode]{matsuno-1966}.  The low-latitude feature discussed above
is dynamically analogous to a superposition of the $n=1$ Rossby
wave and the equatorial Kelvin wave, which is a fundamental equatorially 
trapped wave mode with strong zonal winds but very weak meridional winds and
whose amplitude is symmetric about, and peaks at, the equator
(see, e.g., \citet{holton-2004} or \citet{andrews-etal-1987}).
Both of these wave modes exhibit winds that are primarily
east-west at the equator; in the example shown in Fig.~\ref{matsuno},
the Kelvin component dominates over the $n=1$ Rossby component
at the equator.

\begin{figure*}
\vskip -15pt
\includegraphics[scale=0.9, angle=0]{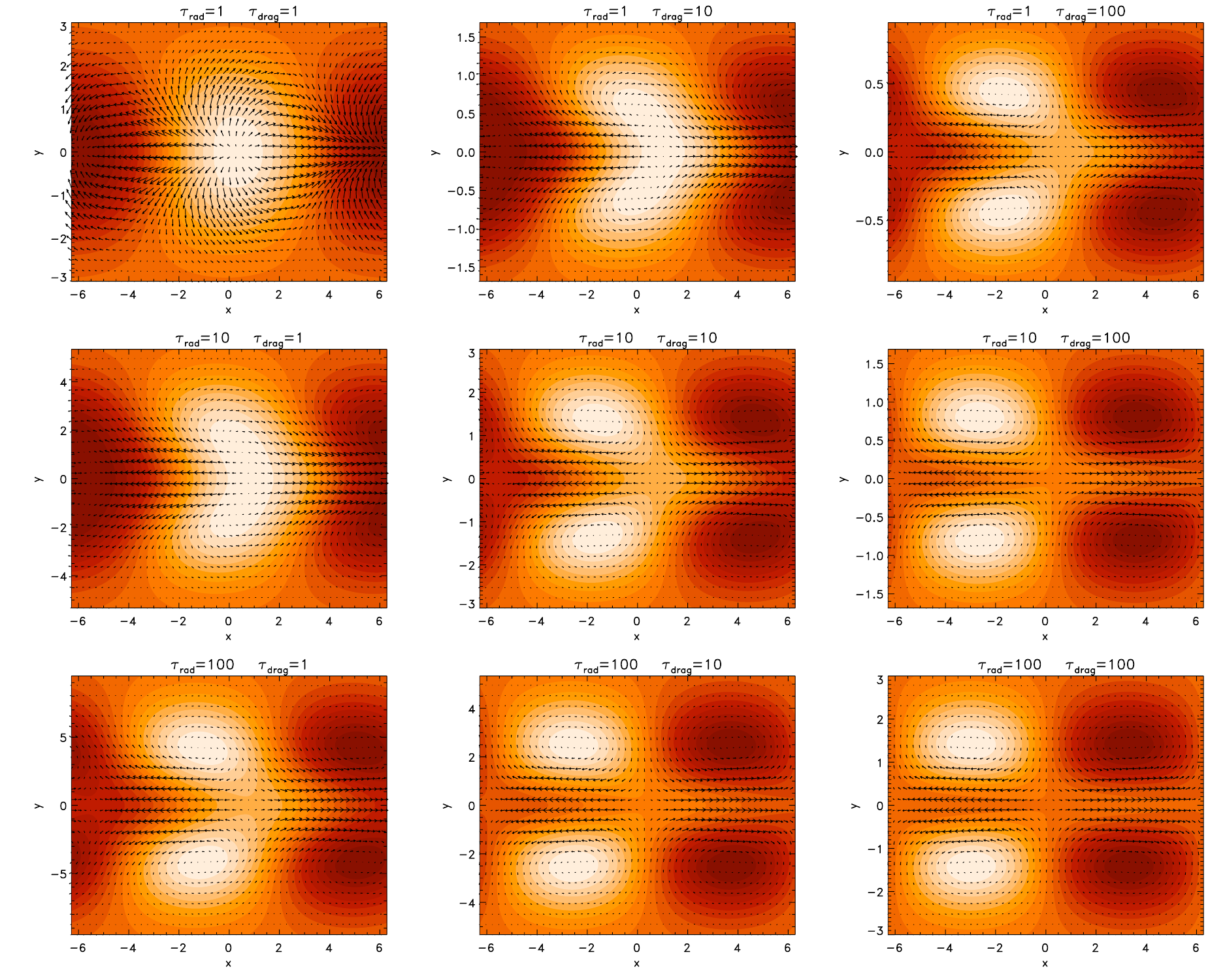}
\caption{Analytic solutions of linearized shallow-water equations 
(Eqs.~\ref{u-mom-nd}--\ref{h-nd}), as presented in 
Appendix~\ref{gill-solutions}, for dimensionless zonal
wavenumber $k=0.5$ and dimensionless radiative and drag
time constants of 1, 10, and 100, corresponding roughly
to dimensional time constants of 6 hours, 3 Earth days,
and one month for parameters appropriate to hot Jupiters
(see Appendix~\ref{nondim}).  All of these cases are forced by Newtonian
relaxation of the height field toward a distribution analogous
to that in Fig.~\ref{matsuno}a.}
\label{cases}
\end{figure*}

Equations~(\ref{u-mom-nd})--(\ref{h-nd}) indicate that this is
a problem governed by two parameters: the radiative time constant and
the drag time constant.  We now examine how the behavior depends on 
their values.  Figure~\ref{cases} shows linear
solutions, as presented in Appendix~\ref{gill-solutions}, 
for dimensionless radiative 
time constants of 1, 10, and 100 (top, middle, and bottom rows, respectively) 
and drag time constants of 1, 10 and 100 (left, middle, and right
columns, respectively).  For parameters appropriate to hot Jupiters
(rotation period of 3 Earth days and $gH\approx 4\times10^6\m^2\sec^{-2}$),
these dimensionless values correspond to dimensional 
time constants of $\sim$$3\times10^4$, $3\times10^5$, and 
$3\times10^6\rm\,sec$, respectively (see Appendix~\ref{nondim}).  
When the radiative and drag time constants are short (upper left 
corner of Fig.~\ref{cases}), the maximum and minimum thermal (height)
perturbations lie on the equator and are close to the substellar
and antistellar points; in this limit, the height field is close
to radiative equilibrium (compare top left of Fig.~\ref{cases}
with Fig.~\ref{matsuno}a)\footnote{Appendix~\ref{gill-solutions} 
demonstrates formally that, in the limit of either time constant
going to zero, the height field converges to the radiative-equilibrium 
height field.}, and distinct Rossby-wave gyres do not appear.
When the radiative and drag time constants have intermediate
values (middle of Fig.~\ref{cases}), cyclones and anticyclones
become visible and---as in Fig.~\ref{matsuno}b---exhibit height
extrema that are phase shifted westward of the extrema in radiative
equilibrium.  Similarly, the height extrema along the equator
become phase shifted eastward relative to radiative equilibrium;
thickness variations along the equator become modest relative
to those in midlatitudes.  When the radiative and drag time constants are 
long (lower right corner of Fig.~\ref{cases}), the height field
becomes dominated by the off-equatorial anticyclones and cyclones,
with minimal variation of height at the equator.  
In the limits $\tau_{\rm rad}\to\infty$ and $\tau_{\rm drag}\to\infty$,
the solution becomes flat at the equator and is
symmetric in longitude about the $x=0$ axis 
(a point demonstrated explicitly in Appendix~\ref{gill-nodrag}); 
Fig.~\ref{cases} shows that this limit is almost reached even for 
$\tau_{\rm rad}$ and $\tau_{\rm drag}$ of 100.  

Much of the behavior in Fig.~\ref{cases} can be understood
in terms of the zonal propagation of equatorially trapped
Rossby and Kelvin modes.  Kelvin waves exhibit eastward group
propagation while long-wavelength, equatorially trapped 
Rossby waves exhibit westward group propagation.  When $\tau_{\rm rad}$
and $\tau_{\rm drag}$ are very short (upper left corner of Fig.~\ref{cases}), 
the damping is so strong that the waves are unable to propagate zonally.
As a result, the height is close to the radiative equilibrium height
field.   When the two time constants have intermediate
values, the propagation produces an eastward phase 
shift of the height field at the equator (the Kelvin component) and a 
westward phase shift of the height field in the off-equatorial cyclones 
and anticyclones (the Rossby component)---exactly as seen in Fig.~\ref{matsuno}b
and the middle of Fig.~\ref{cases}.  As the two time constants become
very long, the westward phase offset of the off-equatorial cyclones
and anticyclones achieves maximal values of $90^{\circ}$.  At the equator,
however, the height variations go to zero; this is explained by the
fact that Coriolis forces are zero at the equator, so the linearized
force balance is between pressure-gradient forces and drag.  Weak
drag requires weak pressure-gradient forces and hence a flat layer
at the equator.

Now, the key point of our paper is that these linear solutions
have major implications for the development of equatorial superrotation
on tidally locked exoplanets.   As can be seen in Figs.~\ref{matsuno}b and
\ref{cases}, the wind vectors exhibit an
overall tilt from northwest-to-southeast in the northern hemisphere
and southwest-to-northeast in the southern hemisphere.  This pattern,
which resembles a chevron centered at the equator and pointing east, is
particularly strong when the radiative and drag time constants are
short, but occurs in all the cases shown.  This structure implies
that, on average, equatorward moving air has faster-than-average eastward
wind speed while poleward moving air has slower-than-average eastward 
wind speed, so that  
$\overline{u'v'} < 0$ in the northern hemisphere and $\overline{u'v'} > 0$ 
in the southern hemisphere.  As shown schematically in 
Fig.~\ref{qualitative-scenario}, this is exactly the
type of pattern that causes a flux of eastward eddy momentum
to the equator and can induce equatorial
superrotation.  Since momentum is being removed from the mid-latitudes,
one would expect westward zonal-mean flow to develop there.

The physical mechanism responsible for producing these phase
tilts are twofold. First, the differential wave propagation discussed
above: this propagation causes 
an eastward phase shift of the height field in the Kelvin waves 
and a westward shift of the height field in the Rossby waves
relative to the radiative-equilibrium height field.
Because the Rossby wave lies on the poleward flanks of the
Kelvin wave, the result is a chevron pattern
where the height contours tilt northwest-southeast
in the northern hemisphere and southwest-northeast in the
southern hemisphere.  To the extent that velocity vectors
approximately parallel the geopotential contours (as they
tend to do away from the equator when drag is weak or moderate),
this will generate tilts in the velocities such
that $\overline{u'v'}<0$ in the northern hemisphere and 
$\overline{u'v'}>0$ in the southern hemisphere.

The second mechanism for generating the velocity tilts
needed for equatorial superrotation is simply the three-way
force balance between Coriolis, drag, and pressure-gradient
forces.  Because drag acts opposite to the velocity, and
Coriolis forces are perpendicular to the velocity, this 
three-way force balance requires the velocities to be 
rotated clockwise of $-\nabla\eta$ in the northern hemisphere
and counterclockwise of $-\nabla\eta$ in the southern hemisphere.
Given the expected day-night gradients in $\eta$, this 
balance implies that the velocities will tend to tilt
northwest-southeast in the northern hemisphere and southwest-northeast
in the northern hemisphere.  We demonstrate this fact
explicitly with an analytic solution in the limit of $\tau_{\rm rad}\to0$
in Appendix~\ref{zero-taurad}; even when the height field is
nearly in radiative equilibrium and hence exhibits no overall phase
tilts, the velocities themselves develop tilts such that
$\overline{u'v'}<0$ in the northern hemisphere and $\overline{u'v'}>0$ 
in the southern hemisphere (see Fig.~\ref{analyt-zero-taurad}).
The calculation in the limit $\tau_{\rm rad}\to0$ is particularly
interesting because, in this limit, there is no zonal propagation
of the Kelvin and Rossby waves: the radiative damping is infinitely
strong and the zonal phase shift of the height field (relative
to radiative equilibrium) is zero.  This is 
the dominant mechanism for the velocity tilts
in the top-left panel of Fig.~\ref{cases}.

To demonstrate explicitly how superrotation would emerge from
these standing-wave patterns, we analyze the zonal
accelerations associated with these linear solutions.  
Decomposing variables into their zonal means 
(denoted by overbars) and deviations therefrom (denoted with primes)
and zonally averaging the zonal-momentum equation (Eq.~\ref{momentum})
leads to \citep[e.g.,][]{thuburn-lagneau-1999}
\begin{eqnarray}
\nonumber
{\partial\overline{u}\over\partial t}=\underbrace{\overline{v}^*\left[f - 
{\partial \overline{u}\over\partial y}\right]}_{I}
\nonumber
\underbrace{-{1\over \overline{h}}{\partial\over\partial y}
[\overline{(hv)'u'}]}_{II}\\
 + \underbrace{\left[{1\over\overline{h}}
\overline{u'Q'} + \overline{R_u}^*\right]}_{III} 
\underbrace{-{\overline{u}^*\over \tau_{\rm drag}}}_{IV} 
- {1\over\overline{h}}{\partial(\overline{h'u'})\over\partial t}
\label{tem}
\end{eqnarray}
where $a$ is the planetary radius and $\overline{A}^*\equiv \overline{hA}/
\overline{h}$ denotes the thickness-weighted zonal average of any quantity $A$.
Eq.~(\ref{tem}) is the shallow-water version of the 
Transformed Eulerian Mean (TEM) momentum equation, analogous to that in
the isentropic-coordinate form of the primitive equations 
\citep[see][Section 3.9]{andrews-etal-1987}.  On the right-hand side,
terms I, II, and III represent accelerations due to (i) 
momentum advection by the mean-meridional circulation,
(ii) the convergence of the meridional flux of zonal eddy momentum, and (iii)
correlations between the regions of eddy zonal flow and eddy mass source
(essentially vertical eddy-momentum transport).  Within this term,
the quantity $R_u$ is the zonal component of ${\bf R}$ 
(equal to $-Qu/h$ when $Q>0$ and 0 when $Q<0$).  Term IV is
frictional drag.   The final term represents the time rate of
change of the eddy momentum.  In the linear limit,
all the terms on the right side of Eq.~(\ref{tem}) have vanishingly
small amplitude and, in this case, the solutions in
Figs.~\ref{matsuno}--\ref{cases}
represent true steady states.  At any finite amplitude, however,
terms I--IV are nonzero and would cause generation of a zonal-mean
zonal flow.

Figure~\ref{matsuno}c depicts these terms for the example solution
presented in Fig.~\ref{matsuno}b.  As expected, horizontal convergence 
of eddy momentum, term II, causes a strong eastward acceleration at the 
equator and westward acceleration in the midlatitudes ({\it black
curve}).  On the other hand, the acceleration associated with
vertical eddy-momentum transport, term III, is strong and westward
at the equator ({\it blue}), implying downward transport of eddy
momentum at the equator. The remaining terms---the mean-meridional 
circulation (term I, {\it cyan}) and mass-weighted friction
(term IV, {\it light green})---are small at the equator.
The two eddy terms partially
cancel at the equator, but the acceleration due to horizontal eddy
momentum convergences exceeds that due to vertical eddy momentum
convergences, leading to a net eastward acceleration at the equator
and westward acceleration in midlatitudes ({\it red curve}).

\begin{figure*}
\vskip 10pt
\includegraphics[scale=0.9, angle=0]{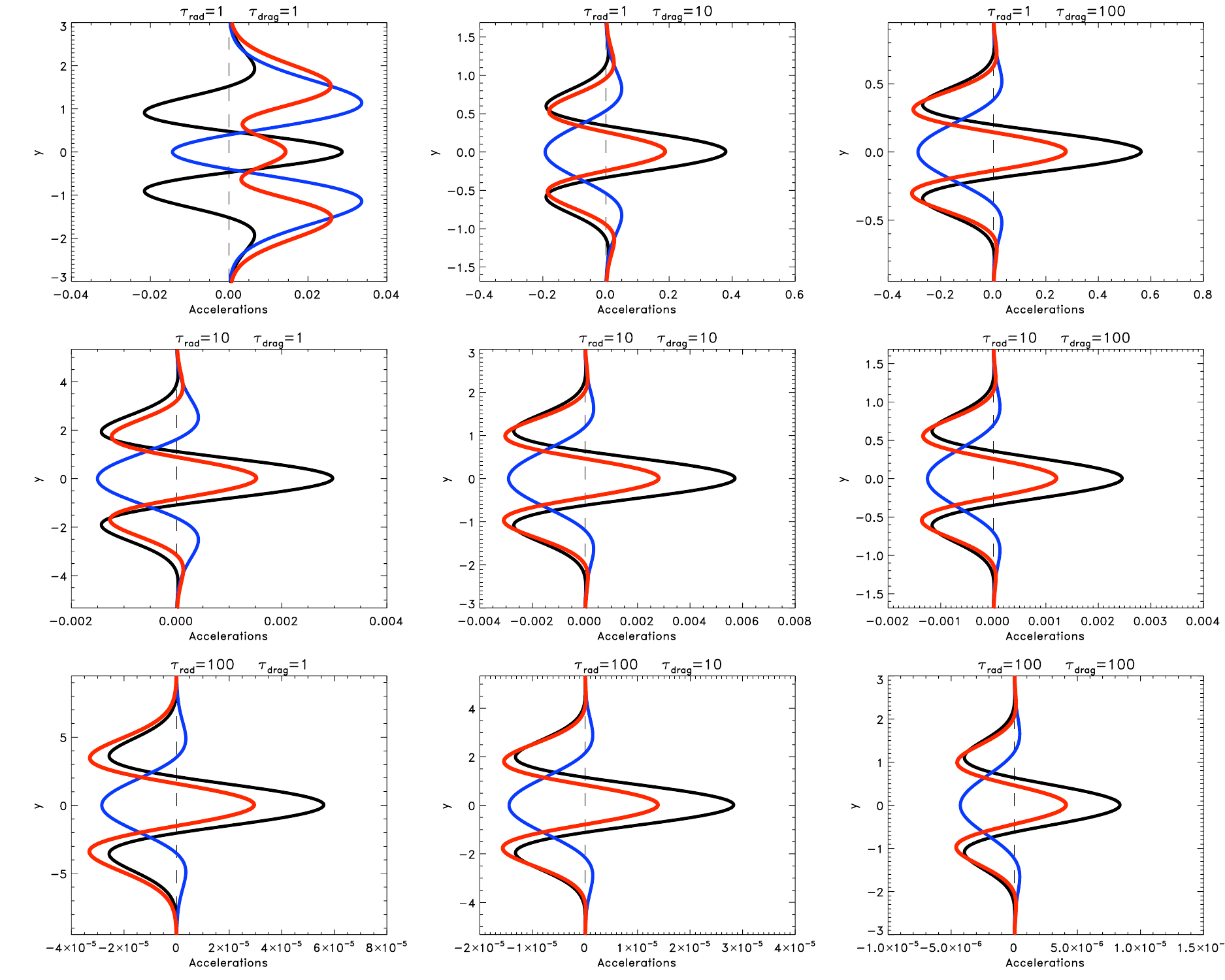}
\caption{Accelerations of the zonal-mean zonal flow for the same
solutions as in Fig.~\ref{cases}.  Black curve is acceleration
due to horizontal eddy-momentum convergence (term II in Eq.~\ref{tem}), and
blue curve is acceleration due to vertical eddy-momentum
transport (term III).  Red curve is sum of terms I, II, III, and IV.
Numerical values plotted here assume that $\hat S_0/\tau_{\rm rad}$ (which
is essentially equivalent to $\Delta h_{\rm eq}/H$) equals one.}
\label{cases-accel}
\end{figure*}

Remarkably, despite the wide
range of morphologies that occur when $\tau_{\rm rad}$ and
$\tau_{\rm drag}$ are varied (Fig.~\ref{cases}), all the solutions exhibit
an equatorward flux of eddy momentum and a net eastward acceleration
at the equator.  This is shown in Fig.~\ref{cases-accel}, which
presents the two eddy acceleration terms from Eq.~(\ref{tem}) for each of
the cases shown in Fig.~\ref{cases}. 
These solutions therefore suggest that superrotation 
at the equator and westward mean flow in the midlatitudes should
occur at essentially any value of the control parameters.

The patterns of spatial velocity and mass source/sink illuminate
the physical origin of the westward equatorial acceleration caused by the
vertical eddy exchange.   The solutions show that the
longitudes of zero zonal wind at the equator lie east of the
mass-source extrema (Fig.~\ref{matsuno}b), a feature also clearly visible
in the steady, linear calculations of \citet[][Fig.~9]{matsuno-1966} 
and \citet[][Fig.~1]{gill-1980}.  Because of this shift,
equatorial mass sources (sinks) occur predominantly in regions of 
westward (eastward) eddy zonal flow.  On average, therefore,
the mass sink regions transport air with eastward column-integrated
eddy momentum out of the layer.  The mass source regions
transport air with no relative zonal momentum from the quiescent
abyssal layer into the upper layer; this process conserves the
local, column-integrated relative momentum ${\bf v}h$ of the upper layer.
Thus, when zonally averaged, vertical exchange at the equator removes 
momentum from the layer, leading to $(\overline{u'Q'} + \overline{R_u}^*)<0$
and contributing a westward acceleration (blue curve in Fig.~\ref{matsuno}c).

The above argument, however, does not determine which of
the two eddy terms (II and III in Eq.~\ref{tem}) dominates.
To determine which is larger---and hence whether the net equatorial
eddy acceleration is eastward or westward---we write the zonally
averaged zonal momentum equation in the form
\begin{equation}
{\partial\overline{u}\over \partial t}=\overline{v'\zeta'} +
\overline{v}(f + \overline{\zeta}) - {\overline{u}\over\tau_{\rm drag}}
+ \overline{R_u}.
\label{eulerian-mean}
\end{equation}
where $\zeta$ is the relative vorticity.
For the case where the forcing is symmetric about the equator, 
the solutions are symmetric about the equator in $u$ and $h$ but
antisymmetric about the equator in $v$ and $\zeta$.  As a result, the meridional velocity
and relative vorticity are zero at the equator, so the terms
$\overline{v}(f+\zeta)$ and $\overline{v'\zeta'}$ vanish there.
Therefore, 
\begin{equation}
{\partial\overline{u}\over\partial t}=-{\overline{u}\over\tau_{\rm drag}}
+\overline{R_u}\qquad\qquad {\rm at}\;y=0.
\label{balance}
\end{equation}
Essentially, at the equator,
$\overline{R_u}$ is the mismatch between the accelerations
caused by horizontal and vertical eddy-momentum fluxes.  The
analytic solutions, which assume $\overline{u}=0$, 
show that $u$ is predominantly westward in regions 
where $Q>0$, which therefore implies that $\overline{R_u}>0$. 
From Eq.~(\ref{balance}), the net eddy-induced acceleration
is therefore eastward.  
This explains, in a general way, the sign of the net eddy
accelerations
at the equator in Fig.~\ref{cases-accel}.  Of course, once a 
zonal-mean flow ($\overline{u}\neq 0$) 
develops, the magnitude of $\overline{R_u}$ changes and 
the friction term becomes important in Eq.~(\ref{balance}); eventually
these terms balance and allow a steady state to be achieved.  We
discuss the possible steady states in light of this equation in
\S\ref{nonlinear-sw}.

We have so far emphasized the {\it spatial patterns} of the circulation, 
but it is also interesting to examine the {\it magnitudes} of
the velocities predicted by our linear solutions.  When the day-night 
difference in the
radiative-equilibrium height is comparable to the mean value
and the radiative time constant is a few days or less (as expected for 
the strongly forced conditions on hot Jupiters),
the winds shown in Fig.~\ref{matsuno}b and Fig.~\ref{cases} 
reach nondimensional speeds of order unity.  For a hot Jupiter,
with typical $g=20\m\sec^{-2}$ and $H=200\km$, this corresponds to dimensional 
speeds of $\sim$$2\rm\,km\,sec^{-1}$.  
To within a factor of a few, this is similar to the speeds obtained 
in fully nonlinear three-dimensional atmospheric circulation models of 
hot Jupiters \citep{showman-guillot-2002, cooper-showman-2005,
showman-etal-2008a, showman-etal-2009, dobbs-dixon-lin-2008,
dobbs-dixon-etal-2010, menou-rauscher-2009, rauscher-menou-2010,
thrastarson-cho-2010}.  For a tidally locked, Earth-like
planet in the habitable zone of an M-dwarf, with $g=10\m\sec^{-2}$,
$H=10\km$, and an Earth-like radiative time constant of $\sim$10 days
(corresponding to dimensionless time constants of 10--100), the solutions
then yield nondimensional speeds of $\sim0.02$--0.1.  This corresponds
to dimensional speeds of up to a few tens of $\m\sec^{-1}$, similar to
speeds obtained in models of tidally locked 
terrestrial planets \citep{joshi-etal-1997, heng-vogt-2011,
merlis-schneider-2011}.

\subsection{Nonlinear solutions}
\label{nonlinear-sw}

Next, we relax the small-amplitude and Cartesian constraints to
demonstrate how nonlinearity and full spherical geometry
affect the solutions, 
and we show how the wave-induced accelerations
interact with the mean flow to generate an equilibrated
state exhibiting equatorial superrotation.  To do so, we solve 
the fully nonlinear forms of Eqs.~(\ref{momentum})--(\ref{R}) 
in global, spherical geometry, using a radiative-equilibrium
thickness given by
\begin{equation}
h_{\rm eq}=H + \Delta h_{\rm eq}\cos\lambda\cos\phi
\label{heq}
\end{equation}
where $H$ is the mean thickness, $\Delta h_{\rm eq}$ is the day-night
contrast in radiative-equilibrium thickness, and the substellar 
point is at longitude $0^{\circ}$ and latitude $0^{\circ}$.  The
planet is assumed to be synchronously rotating, so that the pattern
of $h_{\rm eq}(\lambda,\phi)$ remains fixed in time.  For concreteness,
we adopt planetary parameters appropriate to a hot Jupiter, although we
expect qualitatively similar solutions to apply to super Earths. 
For a typical gravity of $20\rm\,m\,sec^{-2}$ and scale height of
$200\rm\,km$ appropriate to hot Jupiters, we might expect
$gH=4\times10^6\rm\,m^2\,sec^{-2}$, and we adopt this value for all
our runs.  (Note that $g$ and $H$ do not need to be specified
independently.)  We also take $\Omega=3.2\times10^{-5}\rm\,sec^{-1}$
and $a=8.2\times10^7\rm\,m$, corresponding to rotation period
and planetary radius of 2.3 Earth days and 1.15 Jupiter radii,
respectively, similar to the values for HD 189733b.

We reiterate that the equations represent a two layer system
with an active layer overlying a quiescent, infinitely deep
lower layer.  Because of coupling between the layers (specifically,
mass exchange in the presence of heating/cooling), the solutions
readily reach a steady state for any value of the 
drag time constant, including the limit where drag is excluded
entirely in the upper layer ($\tau_{\rm drag}\to\infty$).  
This in fact is a simple
representation of the situation in many full 3D GCMs of Solar System 
atmospheres, including Earth, which often have strong frictional drag near the 
surface, little-to-no friction in the upper layers, and yet
easily reach a steady configuration throughout
{\it all} the model layers.  In our case, we find that,
when drag is strong, the solutions reach steady states in
runtimes $\lesssim 10\tau_{\rm drag}$.  In the case where
drag is turned off, the time to reach steady state is 
determined by the magnitude of momentum and energy exchange
between the layers (e.g., by the magnitude of the ${\bf R}$ term), 
and is generally $\lesssim 10\tau_{\rm rad}
|H/\Delta h|$, where $|\Delta h|/H$ is a characteristic
value of the fractional height variations in the active layer.
All solutions shown here are equilibrated and steady.

We solve Eqs.~(\ref{momentum})--(\ref{R}) using the Spectral Transform 
Shallow Water Model (STSWM) of \citet{hack-jakob-1992}. Rather
than integrating the equations for $u$ and $v$, the code
solves the momentum equations in a vorticity-divergence form.  The
initial condition is a flat layer of geopotential $gH$ at rest;
the equations are integrated using a spectral truncation of 
T170, corresponding to a resolution of $0.7^{\circ}$ in longitude
and latitude (i.e., a global grid of $512\times256$ in longitude
and latitude).  A $\nabla^6$ hyperviscosity is applied to each
of the dynamical variables to maintain numerical stability.
The code adopts the leapfrog timestepping scheme and applies
an Asselin filter at each timestep to suppress the computational
mode.   These methods are standard practice; for further
details, the reader is referred to \citet{hack-jakob-1992}.

\begin{figure}
\vskip 10pt
\includegraphics[scale=0.45, angle=0]{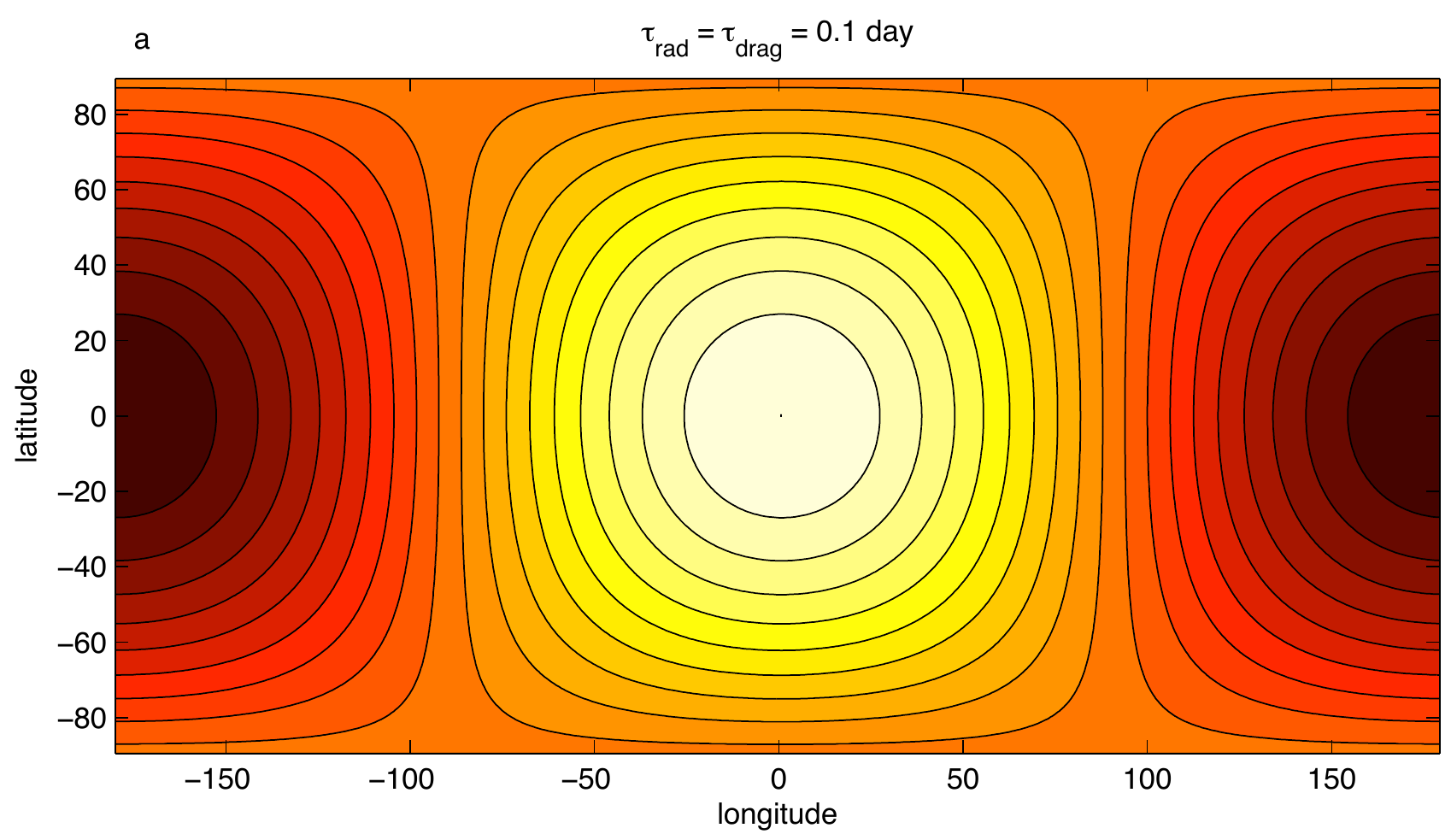}\\
\includegraphics[scale=0.45, angle=0]{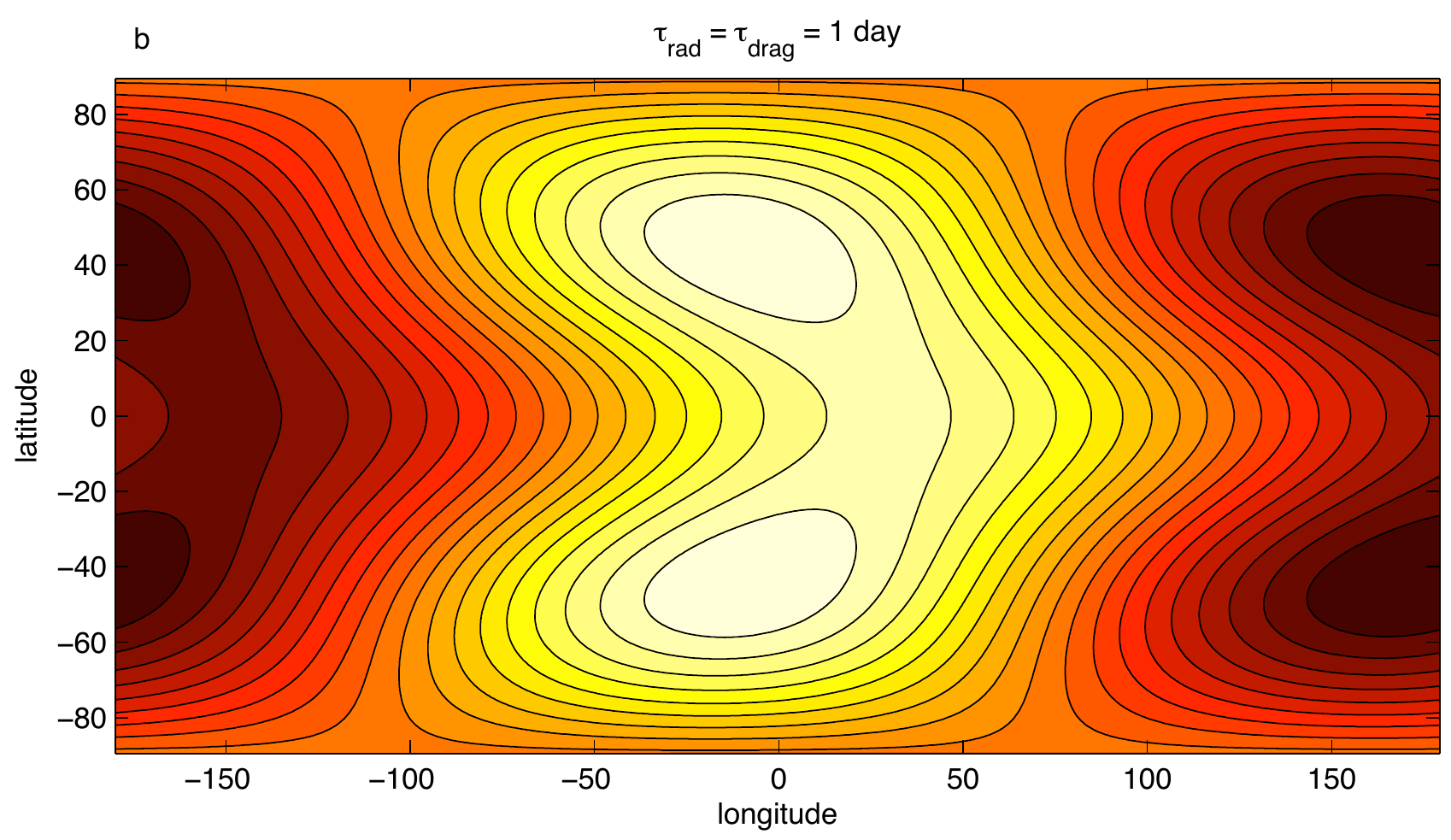}\\
\includegraphics[scale=0.45, angle=0]{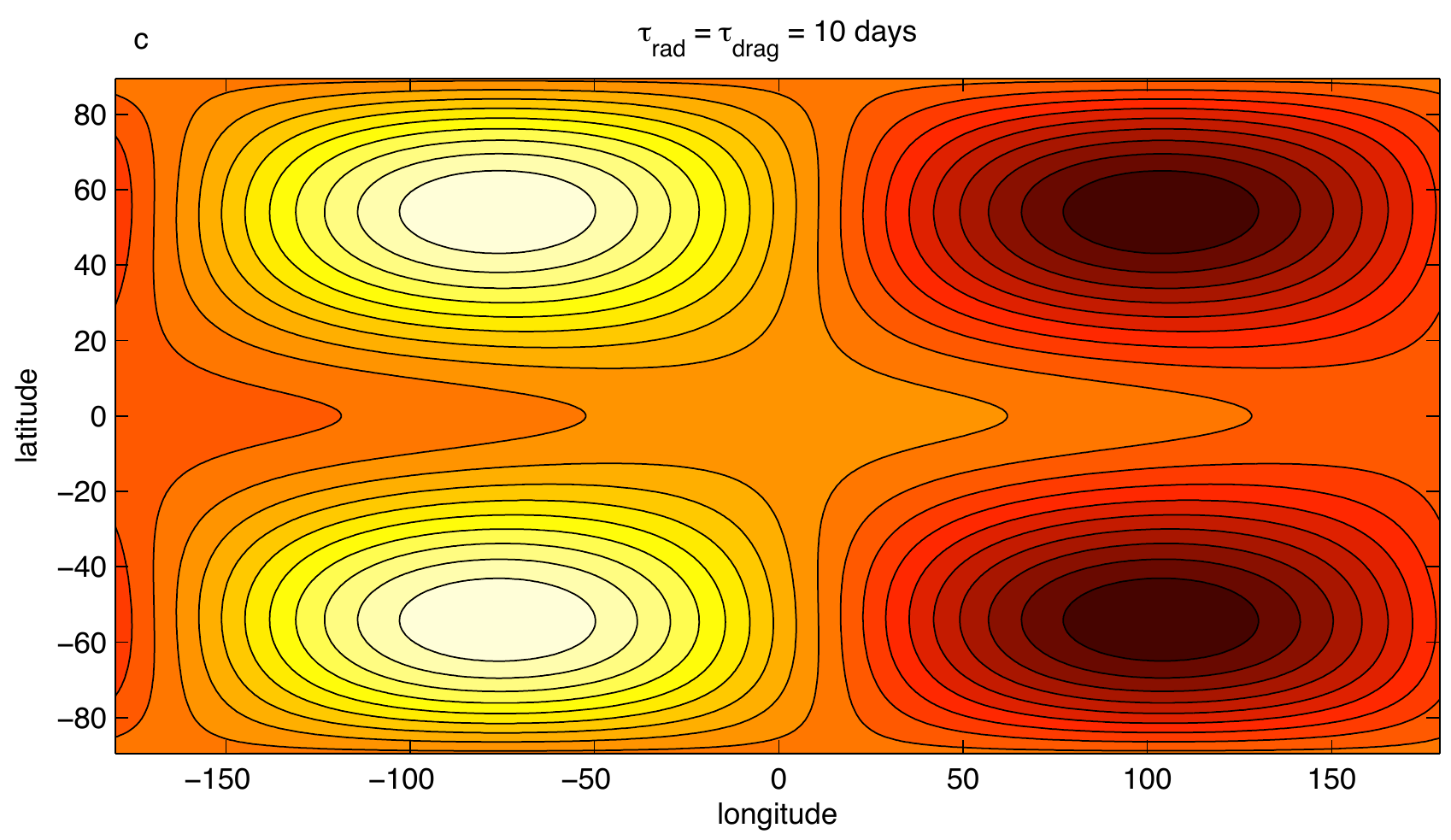}
\caption{Geopotential $gh$ (orangescale and contours) for the equilibrated
(steady-state) solutions
to the shallow-water equations (Eqs.~\ref{momentum}--\ref{R}) 
in full spherical geometry assuming equal radiative and drag
time constants of 0.1, 1, and 10 days in {\it (a), (b),} 
and {\it (c)}, respectively.  White is thick and dark is thin.
Although the equations solved are fully nonlinear, the
forcing amplitude is small here ($\Delta h_{\rm eq}/H=0.01$)
so that the solutions in these panels are essentially linear.}
\label{stswm-equal-tau}
\end{figure}

To facilitate comparison with the analytic theory in \S\ref{linear-sw},
we first describe the solutions at very low amplitude where the
behavior is linear.  Figure~\ref{stswm-equal-tau} shows
the geopotential (i.e., $gh$) for equal radiative and drag time constants
of 0.1, 1, and 10 (Earth) days, respectively.  Qualitatively,
the numerical solutions in spherical geometry bear a striking resemblance
to the analytic solutions on a $\beta$ plane.  At time constants 
of a fraction of day, the geopotential maxima occur on the equator, and 
for time constants of 0.1 day ({\it a}), the geopotential 
resembles the radiative-equilibrium solution, with wind
flowing from the substellar point to the antistellar point.
Longer time constants (1 day, panel {\it b}) allow zonal
energy propagation of the Kelvin and Rossby waves, leading to an
eastward phase shift of the geopotential at the equator
and a westward phase shift at high latitudes ($\sim$40--$90^{\circ}$).
The result is contours of geopotential that develop
northwest-southeast tilts in the northern hemisphere and
southwest-northeast tilts in the southern hemisphere.
When the time constants are
long (10 days, panel {\it c}) off-equatorial cyclones and
anticyclones dominate the geopotential, with only weak 
geopotential variations along the equator.  These vortices
are oval in shape, exhibiting no overall phase tilt, though
the regions close to the equator do develop phase tilts 
(westward-poleward to easward-equatorward).   The momentum
fluxes cause a prograde eddy acceleration (and superrotation)
at the equator for all these cases. All of these features are 
also shared by the analytic solutions (Fig.~\ref{cases}).

\begin{figure}
\vskip 10pt
\includegraphics[scale=0.45, angle=0]{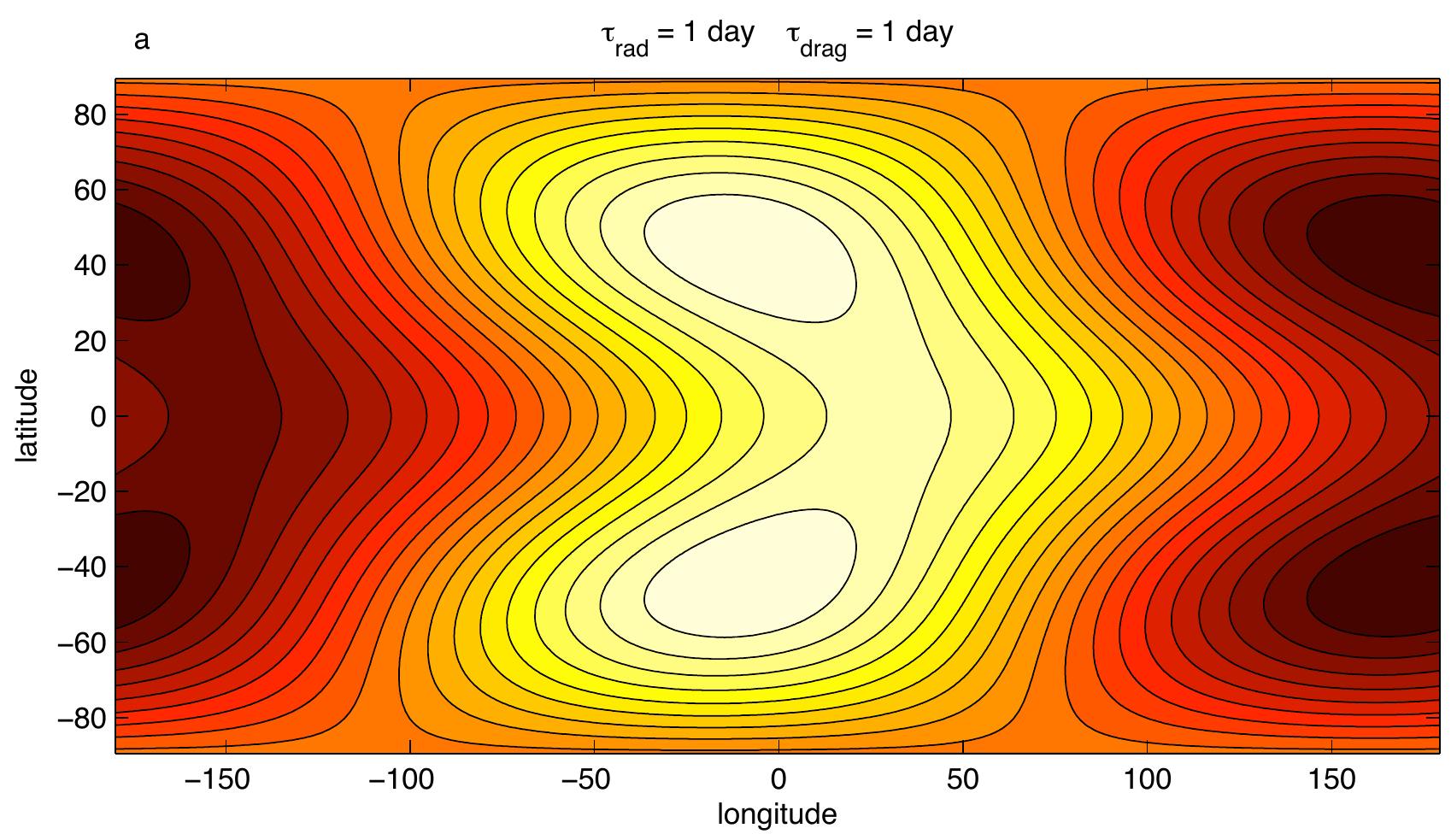}\\
\includegraphics[scale=0.45, angle=0]{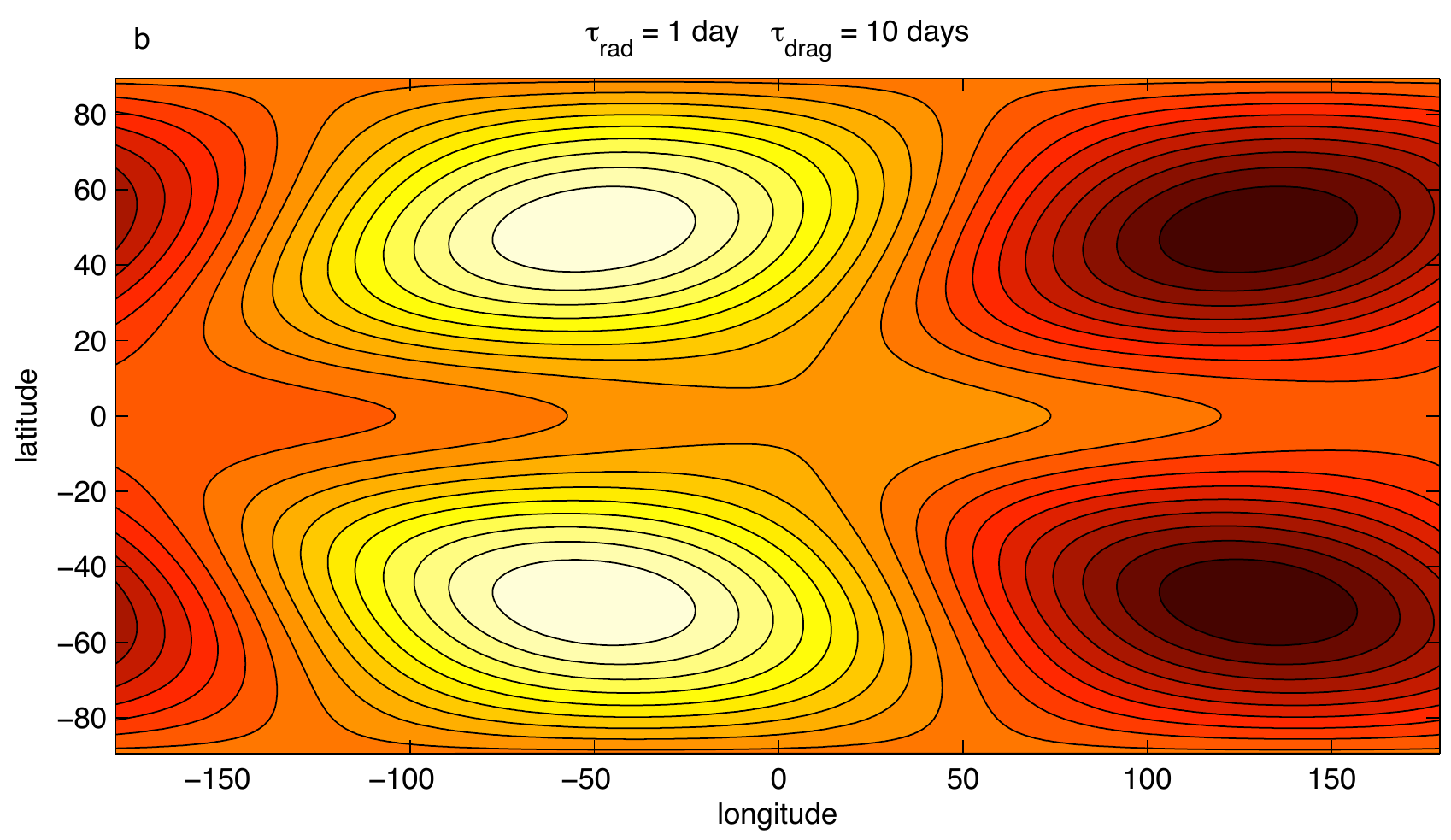}\\
\includegraphics[scale=0.45, angle=0]{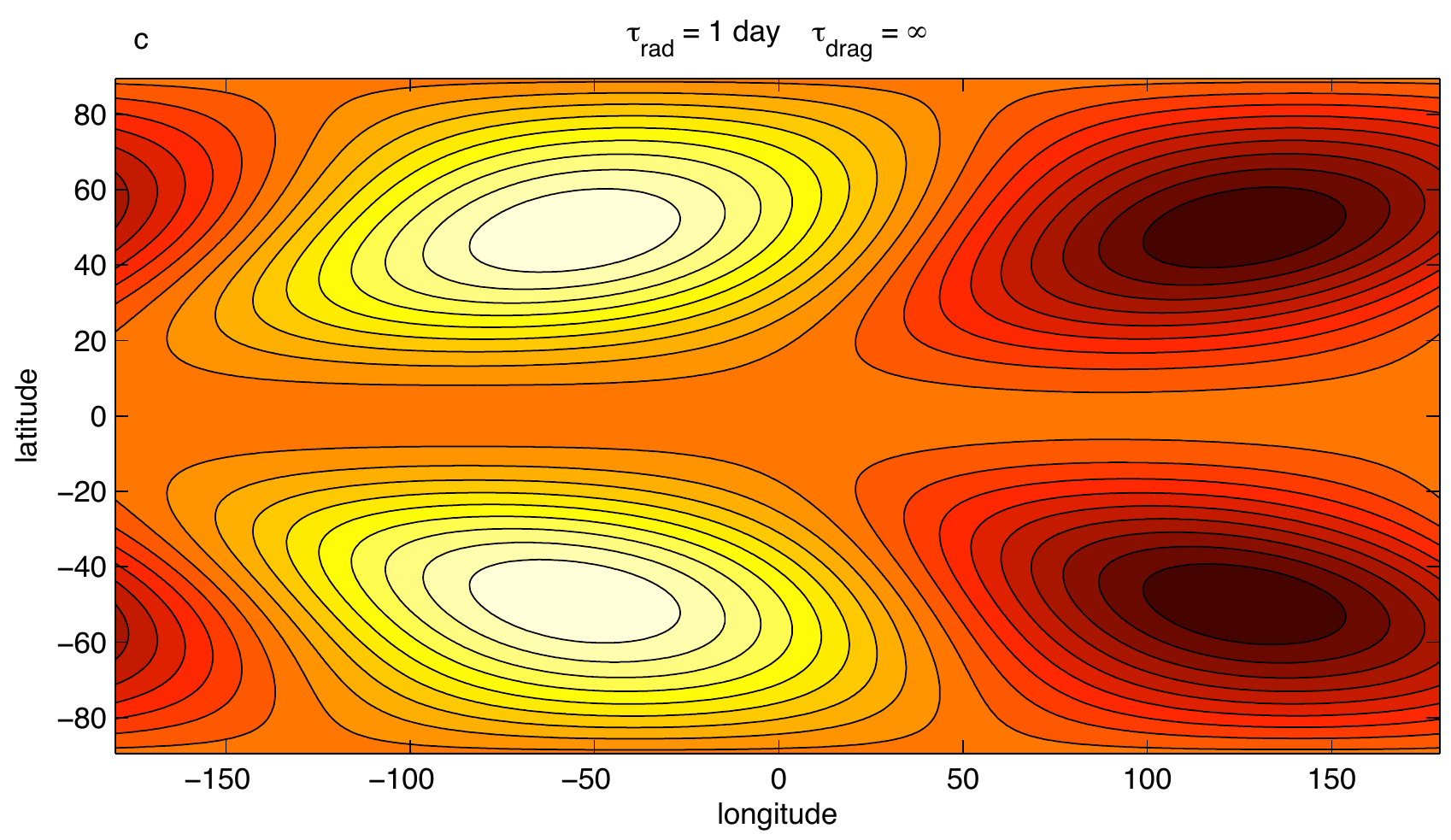}
\caption{Geopotential $gh$ (orangescale and contours) for the equilibrated
(steady-state) solutions to the shallow-water equations 
(Eqs.~\ref{momentum}--\ref{R}) in full spherical geometry illustrating 
the effect of varying $\tau_{\rm drag}$ in the linear limit.  All of the
cases depicted have $\tau_{\rm rad}=1\,\rm day$.  The drag time constant
is 1 day, 10 days, and infinite in the top, middle, and bottom panels,
respectively.   White is thick and dark is thin.
As in Fig.~\ref{stswm-equal-tau}, $\Delta h_{\rm eq}/H=0.01$ here.}
\label{stswm-vary-drag}
\end{figure}

We now explore how the solutions change when the radiative and
frictional time scales are different.
In the linear limit, the latitudinal width of the region exhibiting
prograde phase tilts (i.e., northwest-to-southeast in the northern
hemisphere and southwest-to-northeast in the southern hemisphere)
contracts toward the equator when the drag time constant greatly
exceeds the radiative time constant.  This is illustrated in
Fig.~\ref{stswm-vary-drag}, which shows the equilibrated (steady-state) 
solutions for $\tau_{\rm rad}=1\rm\,day$ and $\tau_{\rm drag}/\tau_{\rm rad}
=1$
({\it top}), 10 ({\it middle}), and infinite ({\it bottom}).
When the time constants are equal, the entire northern (southern)
hemisphere exhibits northwest-to-southeast (southwest-to-northeast) 
phase tilts.  When $\tau_{\rm drag}/\tau_{\rm rad}=10$, these
phase tilts are confined within $\sim$$20^{\circ}$ latitude of
the equator, and for $\tau_{\rm drag}\to\infty$, the width shrinks toward
zero.  This behavior is explained by the analytic
theory in \S\ref{linear-sw}.  As shown in 
Eq.~(\ref{parabolic-cylinder-functions}),
the parabolic cylinder functions comprising the latitudinal structure
exhibit a characteristic latitudinal width of $L(\tau_{\rm rad}/
\tau_{\rm drag})^{1/4}$, where $L=(\sqrt{gH}/\beta)^{1/2}$ is 
the equatorial Rossby deformation radius; these functions thus
collapse toward the equator as $\tau_{\rm drag}/\tau_{\rm rad}$ becomes
infinite.\footnote{The numerical solutions show that the region of prograde
phase tilts does not become precisely zero as $\tau_{\rm drag}$ becomes 
infinite because of the ${\bf R}$ term.  As shown in Eq.~(\ref{momentum2}),
${\bf R}$ plays a role analogous to drag, and the {\it effective}
drag time constant (one over the quantity in square brackets in 
Eq.~\ref{momentum2}) has a characteristic 
magnitude $\tau_{\rm rad} h/(h_{\rm eq}-h)$.  This suggests that,
in the absence of drag, the region of prograde phase tilts exhibits
a latitudinal width of order $L[(h_{\rm eq}-h)/h]^{1/4}$.  This goes
to zero in the limit of zero amplitude but is nonzero at any finite
amplitude.} Poleward of this region, the solutions exhibit 
phase tilts of the opposite direction
(northeast-to-southwest in the northern hemisphere
and southeast-to-northwest in the southern hemisphere).  
Appendix~\ref{gill-nodrag} gives the explanation for this reversal in 
phase tilts; the low-amplitude, full spherical numerical solutions
at $\tau_{\rm drag}/\tau_{\rm rad}\gg1$ strongly resemble analytic 
solutions in the absence of drag, presented in Appendix~\ref{gill-nodrag} 
(compare Figs.~\ref{stswm-vary-drag}c and \ref{analyt-nodrag}).

\begin{figure}
\vskip 10pt
\includegraphics[scale=0.53, angle=0]{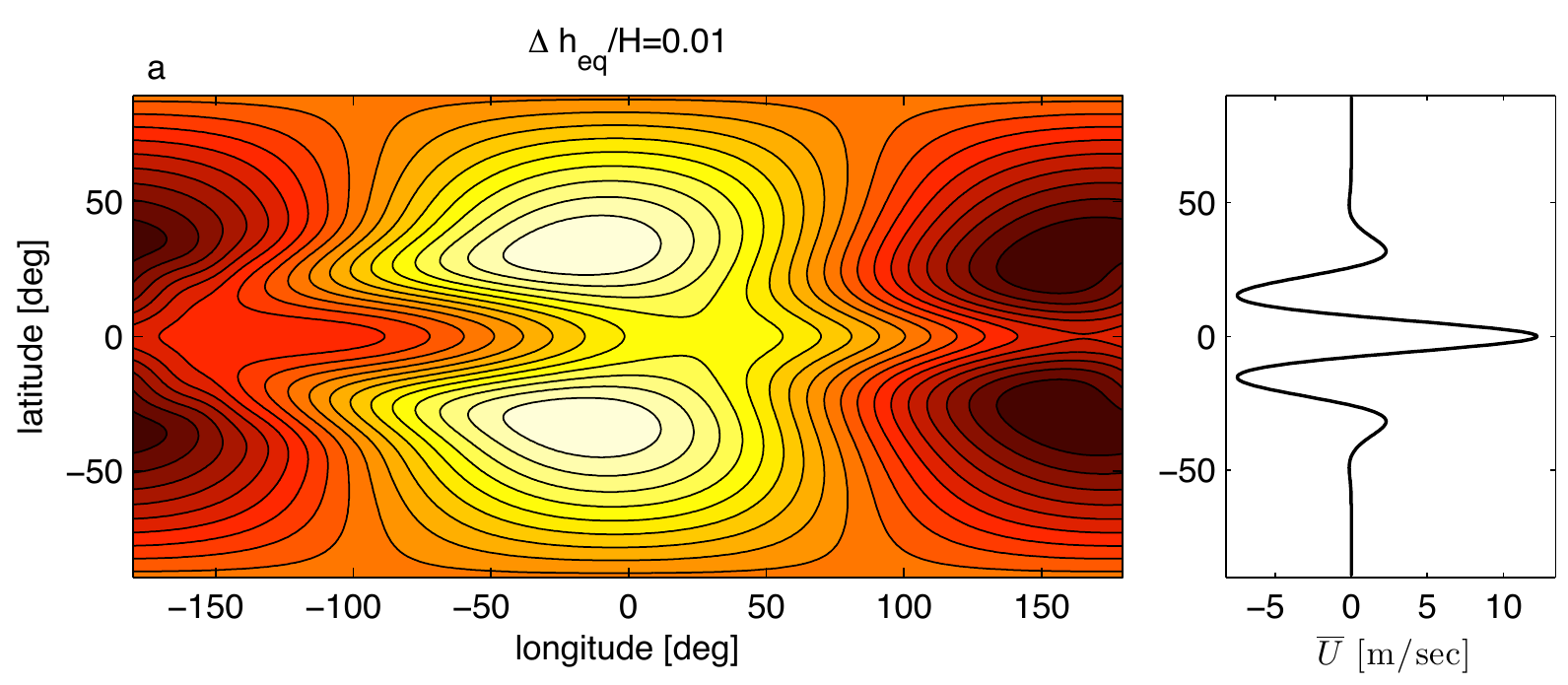}\\
\includegraphics[scale=0.53, angle=0]{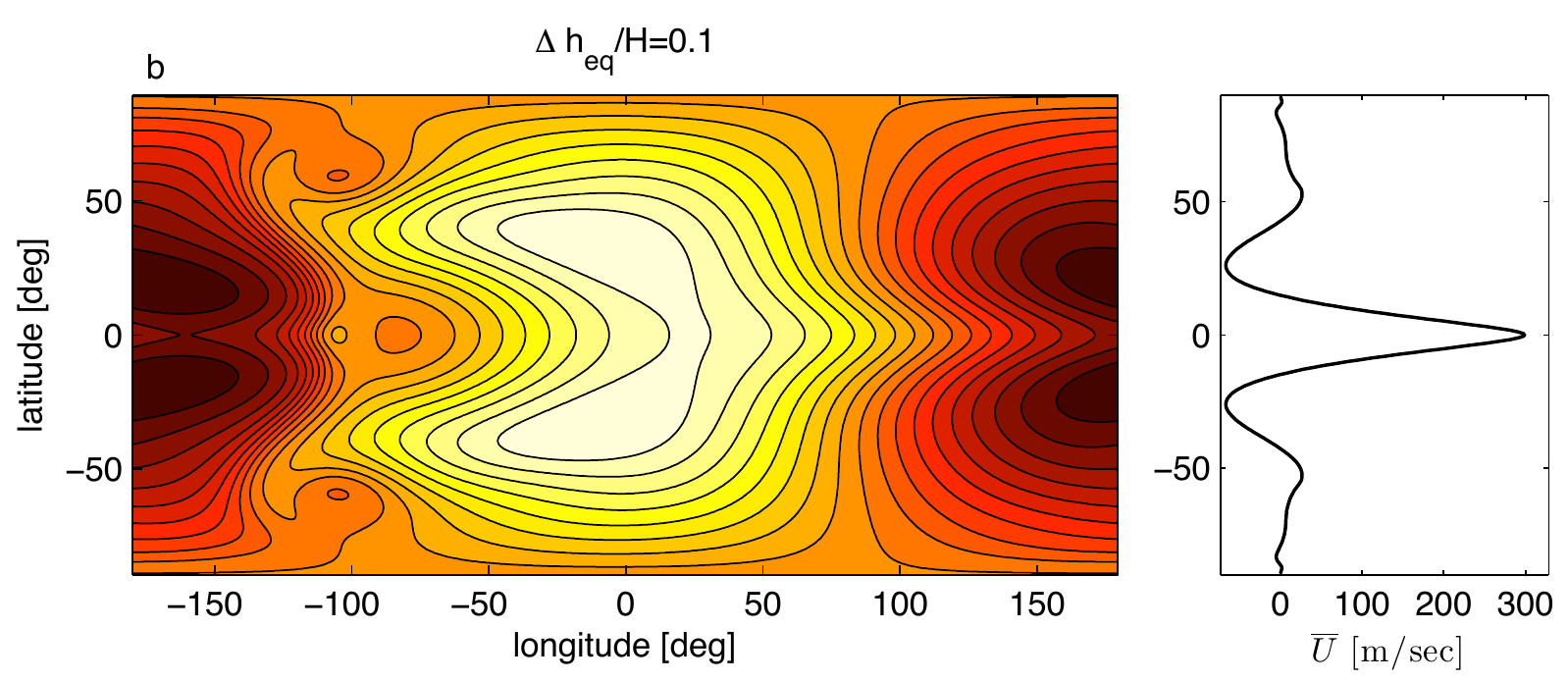}\\
\includegraphics[scale=0.53, angle=0]{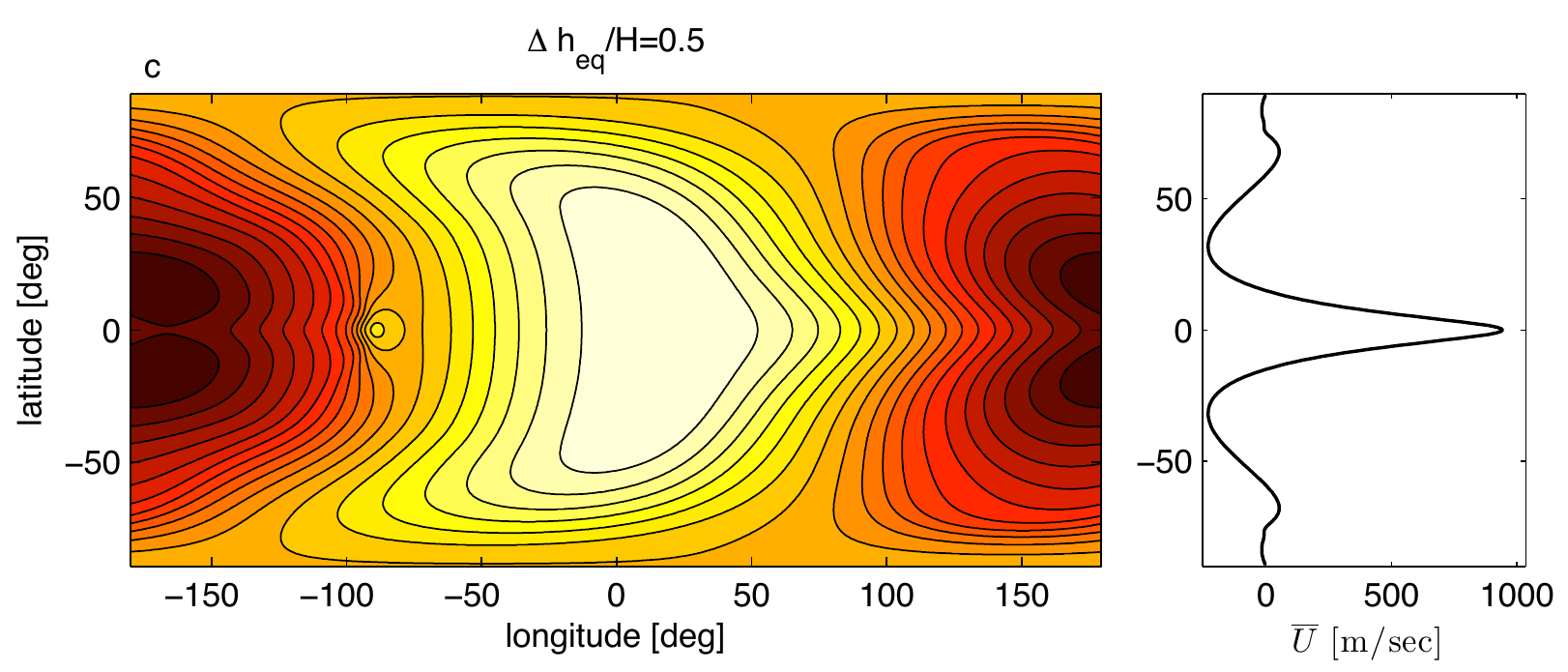}
\caption{Geopotential $gh$ (orangescale and contours) for the equilibrated
solutions to the shallow-water equations 
(Eqs.~\ref{momentum}--\ref{R}) in full spherical geometry illustrating 
the effect of nonlinearity.  All of the
cases depicted have $\tau_{\rm rad}=0.1\,$ day and $\tau_{\rm drag}=10\,$
days.  The forcing amplitude is $\Delta h_{\rm eq}/H=0.01$ 0.1, and 0.5
in ({\it a}), ({\it b}), and ({\it c}), respectively.  
 White is thick and dark is thin.  Range of $gh$ values is 
3.97--$4.03\times10^6\m^2\sec^{-2}$, 3.6--$4.3\times10^6\m^2\sec^{-2}$, and 
2.1--$5.5\times10^6\m^2\sec^{-2}$ from top to bottom, respectively.
The solution in ({\it a}) is steady;
those in ({\it b}) and ({\it c}) exhibit small-scale eddy generation
and time variability (most evident in panel {\it b}), although the
global properties (total kinetic energy, potential energy, and equatorial
jet speed) are essentially constant in time.}
\label{stswm-nonlinearity}
\end{figure}

Nonlinearity alters the solutions in several important ways,
which we illustrate in Fig.~\ref{stswm-nonlinearity}, showing
a sequence of solutions 
for $\tau_{\rm rad}=0.1\rm\,day$, $\tau_{\rm drag}=10\rm\,days$
and, from top to bottom, amplitudes $\Delta h_{\rm eq}/H$ of 0.01, 
0.1, and 0.5, respectively.  This choice of time constants
is representative of the regime of strong radiative forcing
and weak drag that may be appropriate to typical hot Jupiters. 
Increasing the forcing amplitude
(i.e., increasing $\Delta h_{\rm eq}/H$ while holding $\tau_{\rm rad}$
and $\tau_{\rm drag}$ constant) of course leads to increased wind speeds and
day-night geopotential variations; for the parameters in 
Fig.~\ref{stswm-nonlinearity}, the zonal-mean zonal wind speed at
the equator ranges from $\sim$$10\rm\,m\,sec^{-1}$ at the lowest
amplitude to almost $1000\rm\,m\,sec^{-1}$ for the highest amplitude shown.
Moreover, beyond a critical value of $\Delta h_{\rm eq}/H$ (depending on 
the values of $\tau_{\rm rad}$ and $\tau_{\rm drag}$), the solutions begin
to deviate {\it qualitatively} from the linear solutions. 

First, nonlinearity allows greater geopotential variations to
occur along the equator, such that at extreme forcing amplitude
the geopotential extrema can in some cases occur
along the equator when they otherwise would not.  In the linear
limit, the zonal force balance at the equator in the steady state 
is between the pressure-gradient force and drag (cf Eq.~\ref{u-mom-linear});
therefore, when drag is weak, the pressure-gradient force must likewise
be small, implying that minimal variations of geopotential occur
along the equator.  This restriction does not apply at higher latitudes
(where the Coriolis force can balance the pressure-gradient force),
so for very weak drag the thickness extrema generally occur off the
equator (as can be seen in the lower right portion of Fig.~\ref{cases};
Fig.~\ref{stswm-equal-tau}{\it b} and {\it c}; Fig.~\ref{stswm-vary-drag},
and Fig.~\ref{stswm-nonlinearity}{\it a}).  At large forcing amplitude,
however, the momentum advection term ${\bf v}\cdot{\nabla h}$ and the
${\bf R}$ terms become
important and can balance the pressure-gradient force, allowing
significant zonal pressure gradients---and hence significant variations
in thickness---to occur along the equator.  For the parameters
in Fig.~\ref{stswm-nonlinearity}, the thickness variations peak at
the equator when the forcing amplitude is sufficiently large ({\it bottom
panel}).

Second, at high amplitude, the phase tilts of wind and geopotential
tend to be from northwest-to-southeast (southwest-to-northeast) throughout
much of the northern (southern) hemisphere---as in 
Fig.~\ref{stswm-nonlinearity}{\it b} and \ref{stswm-nonlinearity}{\it c}---
even when the phase tilts are in the opposite direction at low
amplitude (as in Fig.~\ref{stswm-nonlinearity}{\it a}).  This effect can
be directly attributed to the term ${\bf R}$ in the momentum equations.  As
shown in Eq.~(\ref{momentum2}), ${\bf R}$ plays a role analogous to
drag.  When true drag is weak or absent, the {\it effective} drag time
constant (one over the quantity in square brackets in Eq.~\ref{momentum2})
has a characteristic magnitude $\tau_{\rm rad}h/(h_{\rm eq}-h)$.  At
large forcing amplitude, $h/(h_{\rm eq}-h) \sim 1$, and in that case
the {\it effective} drag time constant is comparable to $\tau_{\rm rad}$.
The linear solutions show that prograde phase tilts dominate over
much of the globe when the radiative and drag time constants are comparable,
but when the drag time constant greatly exceeds the radiative time constant,
the phase tilts are in the opposite direction 
(see Fig.~\ref{stswm-vary-drag}).  In Fig.~\ref{stswm-nonlinearity},
$\tau_{\rm drag}/\tau_{\rm rad}=100$, but the ratio of the {\it effective}
drag time constant to the radiative time constant decreases from
top to bottom and reaches $\sim$1 in the bottom panel, explaining
the transition in the phase tilts from Fig.~\ref{stswm-nonlinearity}{\it a}
through \ref{stswm-nonlinearity}{\it c}.  Through
the momentum fluxes that accompany these phase tilts, the equatorial
jet becomes broader and more dominant with increasing nonlinearity.

The momentum balance in the equatorial jet can achieve steady
state in two ways.  In steady state, Eq.~(\ref{balance}) becomes
\begin{equation}
{\overline{u}\over\tau_{\rm drag}}=\overline{R_u} \qquad\qquad 
({\rm at \;the\; equator}).
\label{steady-state}
\end{equation}
As described previously, 
when the zonal-mean zonal winds are weak, the zonal wind $u$ 
is predominantly westward
in regions where $Q>0$, so that $\overline{R_u}>0$ (\S\ref{linear-sw}
and Fig.~\ref{cases-accel}).  This implies an eastward eddy acceleration of
the zonal-mean zonal winds at the equator, which induces equatorial
superrotation.  In the first type of 
steady state, corresponding to a regime of strong friction (short
drag time constant), this superrotation implies a strong westward 
acceleration due to friction ($-\overline{u}/\tau_{\rm drag}$).  Steady
state occurs when the zonal-mean equatorial jet becomes strong enough for
the friction to balance the eastward eddy-induced acceleration at 
the equator.  We call this the ``high Prandtl number'' regime.
In the second type of steady state, which we call the ``low Prandtl number''
regime, the friction is
sufficiently weak that the term 
$-\overline{u}/\tau_{\rm drag}$ is unimportant in the momentum balance.  
Because of the eastward eddy acceleration, the zonal-mean zonal winds
can build to high speed.  Once they do, they change the nature of
$\overline{R_u}$---the larger $\overline{u}$ becomes, the smaller the
extent to which $u<0$ in the region $Q>0$, as necessary
for $\overline{R_u}>0$.  Eventually, for sufficiently
large $\overline{u}$,  the quantity $\overline{R_u}$ goes to zero
at the equator.  The equatorial jet thus achieves steady state.

\begin{figure*}
\vskip 10pt
\includegraphics[scale=0.9, angle=0]{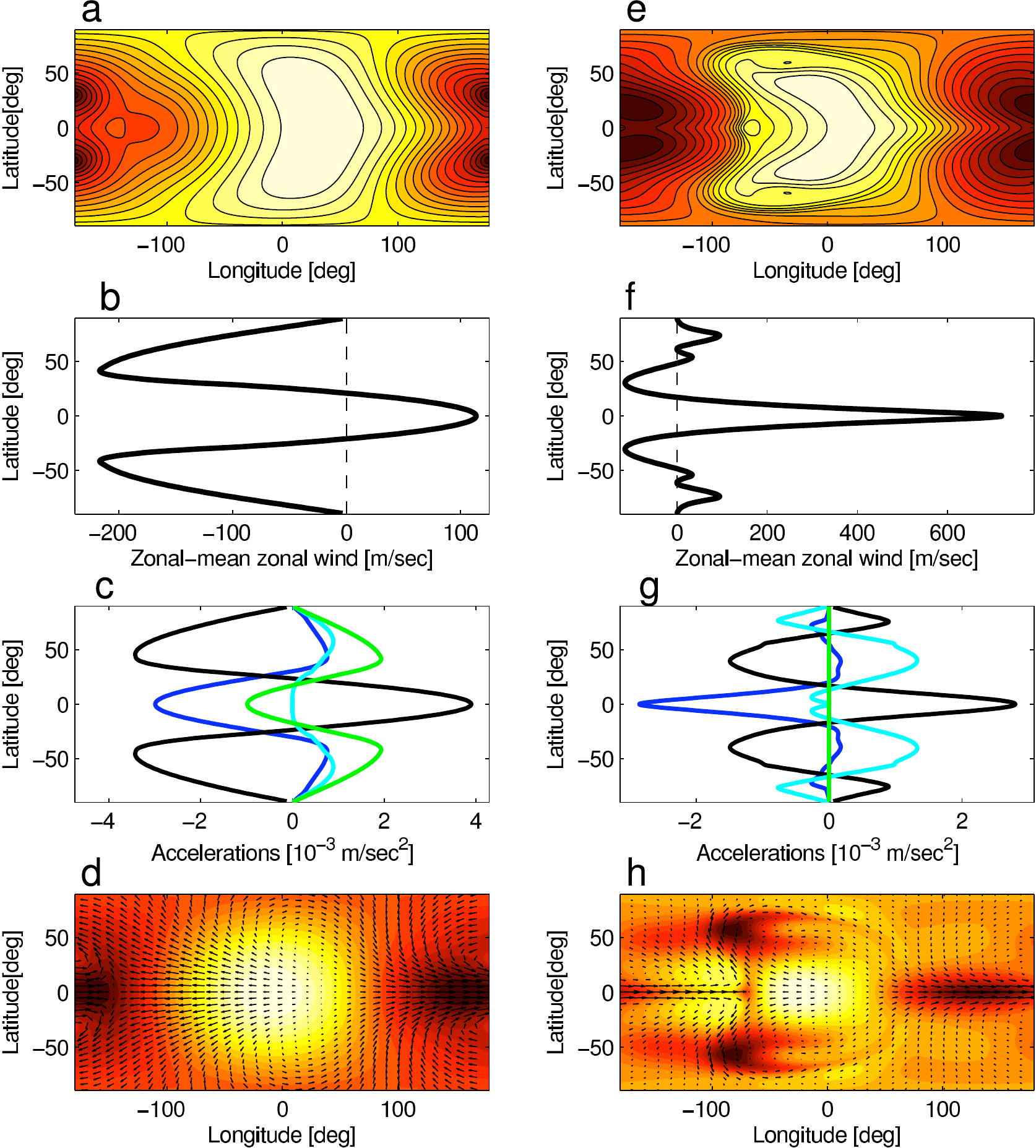}
\caption{Fully nonlinear shallow-water solutions with 
$\tau_{\rm rad}=1\rm\,day$, $\tau_{\rm drag}=1\rm
\,day$, and $\Delta h_{\rm eq}/H=1.25$ ({\it left column}) and 
$\tau_{\rm rad}=0.1\rm\,day$, no drag ($\tau_{\rm drag}\to \infty$),
and $\Delta h_{\rm eq}/H=0.2$ ({\it right column}).  ({\it a, e}) Geopotential
$gh$ (orangescale and contours, with thick and thin regions in white
and black, respectively).  Scale runs from $1.1$--$5.4\times10^6\rm\,m^2
\,sec^{-2}$ in top panel and
$3.3$--$4.6\times10^6\rm\,m^2\,sec^{-2}$ in bottom panel. 
({\it b, f}) Zonal-mean zonal wind $\overline{u}$.
({\it c, g}) Accelerations of the zonal-mean zonal wind in the equilibrated
state.  From Eq.~(\ref{tem2}), curves are terms I ({\it cyan}), II ({\it black}), III ({\it dark blue}),
IV ({\it light green}).   ({\it d, h}) Mass source $gQ$ at the equator (orangescale, white being
positive and dark being negative values) and winds (arrows).  Mass source scale runs from $-41.3$ 
to $45.1\m^2\sec^{-3}$ (left) and $-38.1$ to $33.2\m^2\sec^{-3}$ (right).}
\label{stswm-diag}
\end{figure*}

Figure~\ref{stswm-diag}
shows examples of each of these regimes illustrating how the 
momentum balance occurs.  The left column presents an example with strong
drag ($\tau_{\rm rad}=\tau_{\rm drag}=1\rm\,day$) and the right column
presents an example with weak drag 
($\tau_{\rm rad}=0.1\rm\,day$ and $\tau_{\rm drag}=\infty$); both are
equilibrated and steady. 
These are high-amplitude cases, so the thickness 
({\it top row, a} and {\it e}) exhibits large fractional 
variations, and the phase tilts exhibit an overall trend of
northwest-to-southeast (southwest-to-northeast) in the northern
(southern) hemisphere, as explained in previous discussion
(cf Fig.~\ref{stswm-nonlinearity}).  These tilts indicate transport of
eddy momentum from midlatitudes to the equator; as a result,
the zonal-mean zonal winds are eastward at the equator and westward 
in the midlatitudes ({\it b} and {\it f}). Interestingly, however,
the relative strengths of the equatorial
and midlatitude jets differ and reflect the range of possible variation.
Panels ({\it c}) and ({\it g}) show the terms in the zonal-mean momentum
equation, just the spherical equivalent of Eq.~(\ref{tem}):
\begin{eqnarray}
\nonumber
{\partial\overline{u}\over\partial t}=\underbrace{\overline{v}^*\left[f - 
{1\over a\cos\phi}{\partial (\overline{u}\cos\phi)\over\partial \phi}\right]}_{I}\\
\nonumber
\underbrace{-{1\over \overline{h}a\cos^2\phi}{\partial\over\partial\phi}
[\overline{(hv)'u'}\cos^2\phi]}_{II}\\
 + \underbrace{\left[{1\over\overline{h}}
\overline{u'Q'} + \overline{R_u}^*\right]}_{III} 
\underbrace{-{\overline{u}^*\over \tau_{\rm drag}}}_{IV} 
- {1\over\overline{h}}{\partial(\overline{h'u'})\over\partial t}.
\label{tem2}
\end{eqnarray}
  As expected, horizontal convergence of eddy
momentum, term II, causes a strong eastward acceleration at the 
equator and westward acceleration in midlatitudes ({\it black curves}).
The vertical eddy-momentum transport,
term III ({\it dark blue curves}), causes a westward acceleration
at the equator that counteracts the eastward acceleration
due to horizontal eddy-momentum convergence.  In the case of
strong drag (Fig.~\ref{stswm-diag}{\it c}), the cancellation is
imperfect, leading to a net eddy-induced acceleration that is
eastward at the equator---as predicted by the linear, analytic theory 
in \S\ref{linear-sw} (compare to Fig.~\ref{matsuno}c).  
A superrotating equatorial jet therefore
emerges and only reaches steady state when the jet becomes
sufficiently strong that the zonal-mean drag on the jet, $-\overline{u}^*/
\tau_{\rm drag}$, balances the eastward acceleration at the equator
(Fig.~\ref{stswm-diag}{\it c}).  On the other hand, when drag is 
absent, the superrotation induced by the eddy fluxes 
becomes quite strong (Fig.~\ref{stswm-diag}{\it f}).
This mean flow alters the eddy fluxes, causing them to self-adjust
to an equilibrium where the accelerations at the equator due to horizontal
and vertical momentum fluxes cancel, leading to no net eddy-induced
acceleration at the equator in steady state (Fig.~\ref{stswm-diag}{\it g}).

\begin{figure}
\vskip 10pt
\includegraphics[scale=0.6, angle=0]{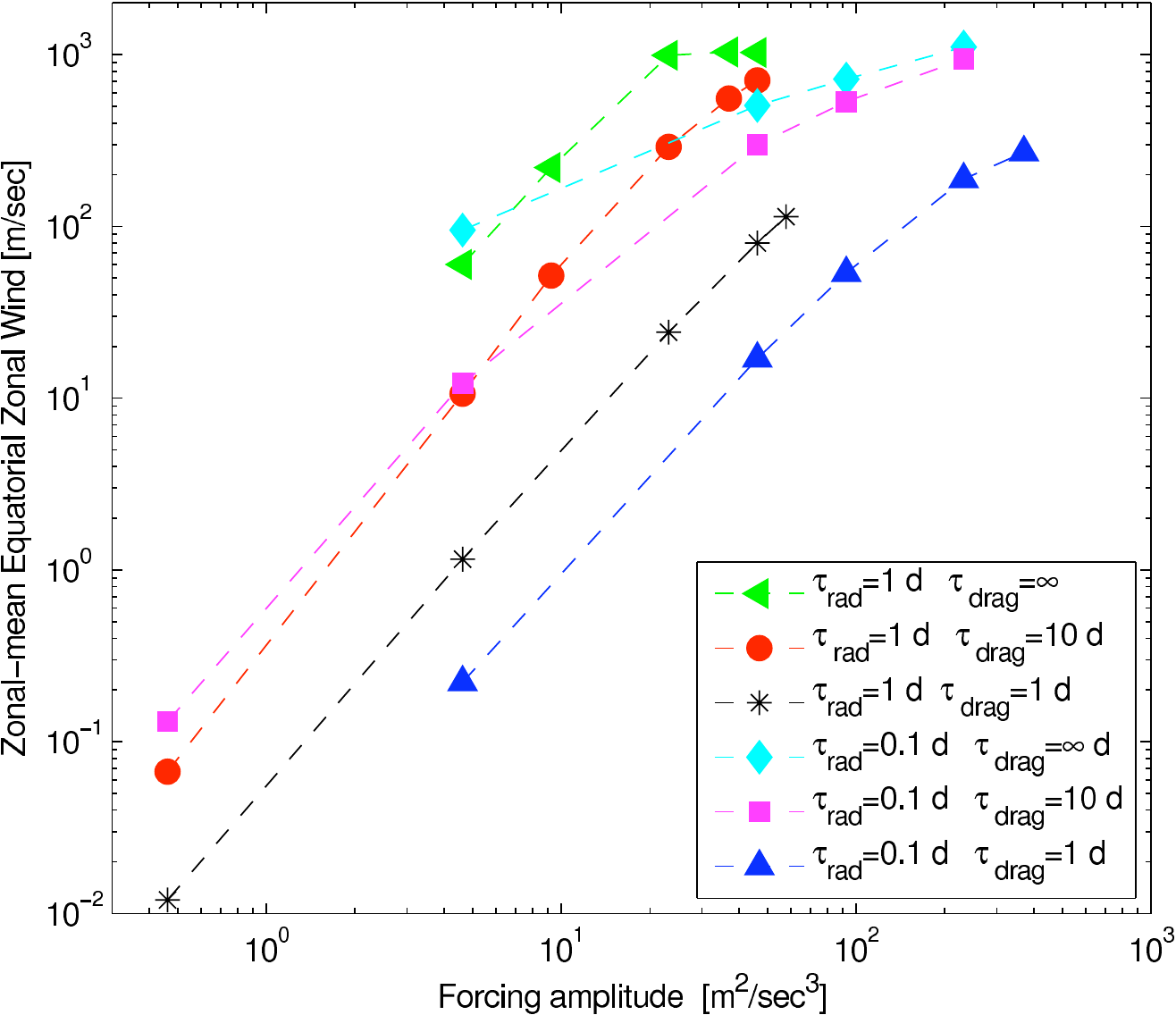}
\caption{Equilibrated (steady-state) equatorial jet speed from the 
nonlinear shallow-water solutions for a variety of forcing amplitudes,
radiative time constants, and drag time constants.  The jet is eastward
(i.e., superrotating) in all cases.  In most
cases, the zonal-mean jet speed scales as forcing amplitude squared 
at low amplitude but exhibits a flatter dependence at high amplitude.
Here, forcing amplitude is defined as $g\Delta h_{\rm eq}/\tau_{\rm rad}$.}
\label{stswm-ueq}
\end{figure}

We find that equatorial superrotation occurs at 
all forcing amplitudes, even arbitrarily small amplitudes where the 
solutions behave linearly.  This
is illustrated in Fig.~\ref{stswm-ueq}, which shows the equilibrated 
equatorial zonal-mean zonal wind versus forcing amplitude
for solutions with a range of $\tau_{\rm rad}$ and $\tau_{\rm drag}$ 
combinations. We therefore conclude that the mechanism 
for generating equatorial superrotation described here
has no inherent threshold. Nevertheless, other processes---not included in 
the shallow-water model---can in some cases overwhelm the desire of
the day-night forcing to trigger superrotation, particularly when the
day-night forcing is weak.  This occurs for example in the cases
examined by \citet{suarez-duffy-1992} and \citet{saravanan-1993},
where superrotation only developed for forcing amplitudes exceeding
a threshold value.  In their case, the tropical wave forcing only
triggers superrotation when it attains sufficiently great amplitudes
to overcome the westward torques provided by midlatitude eddies
propagating into the tropics.  These issues are discussed further in 
\S\ref{discussion}.

In all the cases shown in Fig.~\ref{stswm-ueq} where
$\tau_{\rm drag}$ is finite, the zonal-mean speed of equatorial
superrotation scales
with the square of the forcing amplitude when the forcing amplitude
is sufficiently small.  This
in fact is the expected low-amplitude behavior in the high-Prandtl-number
regime described above: at low amplitude, the solutions become linear,
such that the velocities, 
height perturbations, and mass source/sink scale with the 
forcing amplitude.  Because $\overline{R_u}$ scales as 
the product of the mass source/sink and the velocities at low
amplitude, it is therefore quadratic in the forcing amplitude.  
In the frictional regime, Eq.~(\ref{steady-state}) implies that
$\overline{u}$ at the equator is simply $\tau_{\rm rad}\overline{R_u}$,
and therefore $\overline{u}$ itself is quadratic in the forcing
amplitude.  This behavior breaks down when the solutions become
sufficiently high amplitude, as can be seen in Fig.~\ref{stswm-ueq}.
The low-Prandtl-number regime is more complex and can lead to a variety
of scaling behaviors depending on the parameters.

The flow in the shallow-water models differ from that in 
3D models in one major respect.  In many three-dimensional
models of hot Jupiters, eastward equatorial flow occurs not 
only in the zonal mean but 
at {\it all longitudes}, at least over some range of pressures.
In contrast, although the shallow-water models described here
all exhibit eastward {\it zonal-mean} flow at the equator, the
zonal wind at the equator is always westward over some range 
of longitudes.   This can be seen as follows: $\overline{R_u}$ is
essentially the mismatch in equatorial zonal acceleration between
horizontal and vertical eddy-momentum transport and in steady state,
when $\overline{u}>0$,
will be greater than or equal to zero.  From the definition of
$R_u$, this implies westward equatorial flow at some longitudes. 
This trait probably arises because the meteorologically active
atmosphere has here been resolved with only one layer overlying
a deep interior; in future work, it would be interesting
to explore models that represent the flow with two or 
more layers overlying
a quiescent interior to see whether they can develop 
equatorial flow that is eastward at all longitudes.

\section{Three-dimensional model of equatorial superrotation}
\label{3d}

Here, we show how
the basic mechanism for generating equatorial 
superrotation identified in \S\ref{sw} occurs also in three dimensions
under realistic conditions.  To do so, we analyze the three-dimensional 
model of HD 189733b presented in \citet{showman-etal-2009}.
\citet{showman-etal-2009} coupled the dynamical core of the MITgcm
\citep{adcroft-etal-2004}, 
which solves the primitive equations of meteorology in global, spherical 
geometry, using pressure as a vertical coordinate, 
to the state-of-the-art, non-gray radiative transfer scheme of 
\citet{marley-mckay-1999}, which solves the multi-stream radiative
transfer equations using the correlated-k method to treat the
wavelength dependence of the opacities.  This coupled model, dubbed the
Substellar and Planetary Atmospheric Circulation and Radiation 
(SPARC) model, is to date the only GCM to include realistic radiative
transfer for hot Jupiters.  Gaseous opacities were calculated assuming local
chemical equilibrium for
a specified atmospheric metallicity, assuming rainout of any condensates
(i.e. ignoring cloud opacity).  \citet{showman-etal-2009} presented 
synchronously rotating models of HD 189733b with one, five, and ten times 
solar metallicity and of HD 209458b with solar metallicity, along with
several models with non-synchronous rotation.  Their HD 189733b
models in particular compare favorably with a variety of observational
constraints \citep{showman-etal-2009, agol-etal-2010}, and here we
focus on their solar-metallicity, synchronously rotating HD 189733b
case.  This model adopts planetary radius and gravity
of $8.24\times10^7\rm\,m$ and $9.36\rm \,m\,sec^{-2}$. The
rotation rate is $3.3\times10^{-5}\rm\,sec$, corresponding to a rotation
period of 2.2 Earth days.

\begin{figure*}
\centering
\vskip 10pt
\includegraphics[scale=0.7, angle=0]{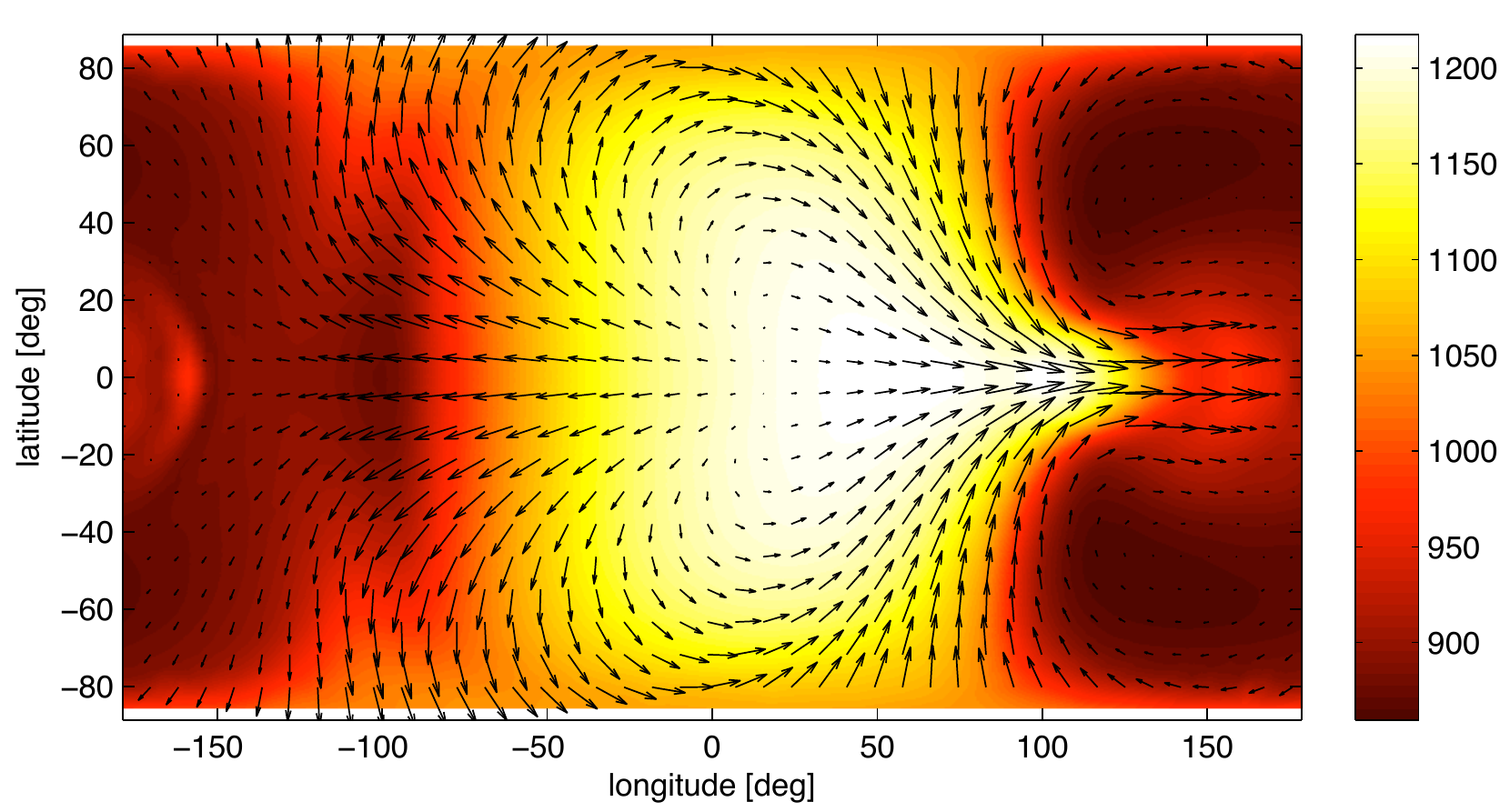}
\caption{Temperature (colorscale, in K) and winds (arrows) during the spin-up
phase of the three-dimensional, solar-metallicity
model of HD 189733b from \citet{showman-etal-2009}.  This shows 
the state at the 30-mbar pressure level---near 
the mean infrared photosphere---at a time of one Earth day, after
the day-night forcing has generated a global wave response but before
strong zonal-mean jets have developed.    Substellar point 
is at $0^{\circ}$ latitude, $0^{\circ}$ longitude.
Note the development of a ``Matsuno-Gill''-type pattern, 
which leads to an equatorward flux of eddy momentum
that pumps the equatorial jet. 
}
\label{3d-spinup-gill}
\end{figure*}

Figure~\ref{3d-spinup-gill} shows the velocity and temperature
structure at the 30-mbar level
during the spin-up phase of this model---after the
forcing has had sufficient time to trigger a global wave response but
before the equatorial jet has spun up to high speed.
The velocity pattern in the three-dimensional model (Fig.~\ref{3d-spinup-gill})
strongly resembles the standing 
Kelvin and Rossby-wave pattern described in \S\ref{sw}.  
The flow clearly exhibits the east-west divergence along the equator, emanating 
from a point near the substellar longitude, identified in \S\ref{sw}
as the standing Kelvin-wave response.  The longitude of peak divergence 
(i.e., the longitude at the equator where the zonal velocity changes sign) 
lies east of the substellar longitude, as expected from the 
analytic theory and nonlinear shallow-water runs in \S\ref{sw}.
Moreover, the flow exhibits the broad gyres in each hemisphere,
anticyclonic on the dayside and cyclonic on the nightside, identified
in \S\ref{sw} as the standing Rossby-wave response.  As predicted 
analytically, the velocities in these gyres exhibit a 
northwest-to-southeast (southwest-to-northeast) phase tilt in the
northern (southern) hemisphere.   These phase tilts
imply that $\overline{u'v'}$ is negative in the northern
hemisphere and positive in the southern hemisphere.  Eddy momentum
therefore fluxes from the midlatitudes to the equator, and it is
this flux that produces the superrotating equatorial jet 
(see Fig.~\ref{qualitative-scenario}).  The overall 
qualitative resemblance to the analytic calculation in Fig.~\ref{matsuno} is 
striking.

As in the shallow-water solutions, the three-dimensional
models exhibit a net downward eddy momentum flux at the equator
throughout the upper atmosphere where the radiative heating/cooling
is strong.  This momentum flux results from the fact that, at
the equator, (i) the Matsuno-Gill-type standing-wave patterns 
lead to net zonal eddy 
velocities that are predominantly westward on the dayside and
eastward on the nightside (see Fig.~\ref{3d-spinup-gill}),
and (ii) net radiative heating occurs on much of the dayside, leading 
to net upward velocities, whereas net radiative cooling occurs on the 
nightside, leading to net downward velocities.  Thus, at the equator,
upward velocities tend to be correlated with westward eddy velocities
and vice versa.  This transports eastward momentum downward and
causes a westward acceleration at the equator throughout the upper 
atmosphere, which counteracts the eastward equatorial acceleration caused by
latitudinal eddy-momentum transport---just as predicted by the 
analytic and numerical shallow-water solutions in \S\ref{sw} (see
Figs.~\ref{matsuno}, \ref{cases-accel}, and \ref{stswm-diag}).

To quantify the accelerations resulting from these momentum fluxes,
we consider the Eulerian-mean zonal-momentum equation in pressure coordinates.
By expanding the dynamical variables into
zonal-mean and deviation (eddy) components, and zonally averaging the
zonal-momentum equation, and adopting pressure as the vertical coordinate,
we obtain\footnote{An equation analogous to this, except using
log-pressure rather than pressure itself as the coordinate, can
be found in \citet[][Eq.~3.3.2a]{andrews-etal-1987}.}
\begin{eqnarray}
\nonumber
{\partial\overline{u}\over\partial t}=\overline{v}\left[f 
- {1\over a\cos\phi}{\partial(\overline{u}\cos\phi)\over
\partial \phi}\right] - \overline{\omega}{\partial\overline{u}\over\partial p}\\
-{1\over a\cos^2\phi}{\partial(\overline{u'v'}\cos^2\phi)\over\partial \phi}
-{\partial(\overline{u'\omega'})\over\partial p} + \overline{X}
\label{eulerian-mean}
\end{eqnarray}
On the righthand side, the terms describe the meridional momentum
advection by the zonal-mean circulation, vertical momentum advection
by the zonal-mean circulation, the meridional eddy-momentum convergence, the
vertical eddy-momentum convergence, and friction (represented 
generically by $\overline{X}$), respectively.
At the equator, the Coriolis term is zero.  Because of the approximate
symmetry of the flow about the equator, $\overline{v}$ and the meridional
gradient of $\overline{u}$ are small there,  so the mean-meridional
advection term is small at the equator.  The mean vertical-advection 
term also tends to be weak for the flow considered here, and the 
net zonal acceleration at the equator is
then determined primarily by a competition between the horizontal and vertical
eddy-momentum convergence terms (the analogs of terms 
II and III in Eq.~(\ref{tem2}) for the shallow-water system).

\begin{figure}
\vskip 10pt
\includegraphics[scale=0.7, angle=0]{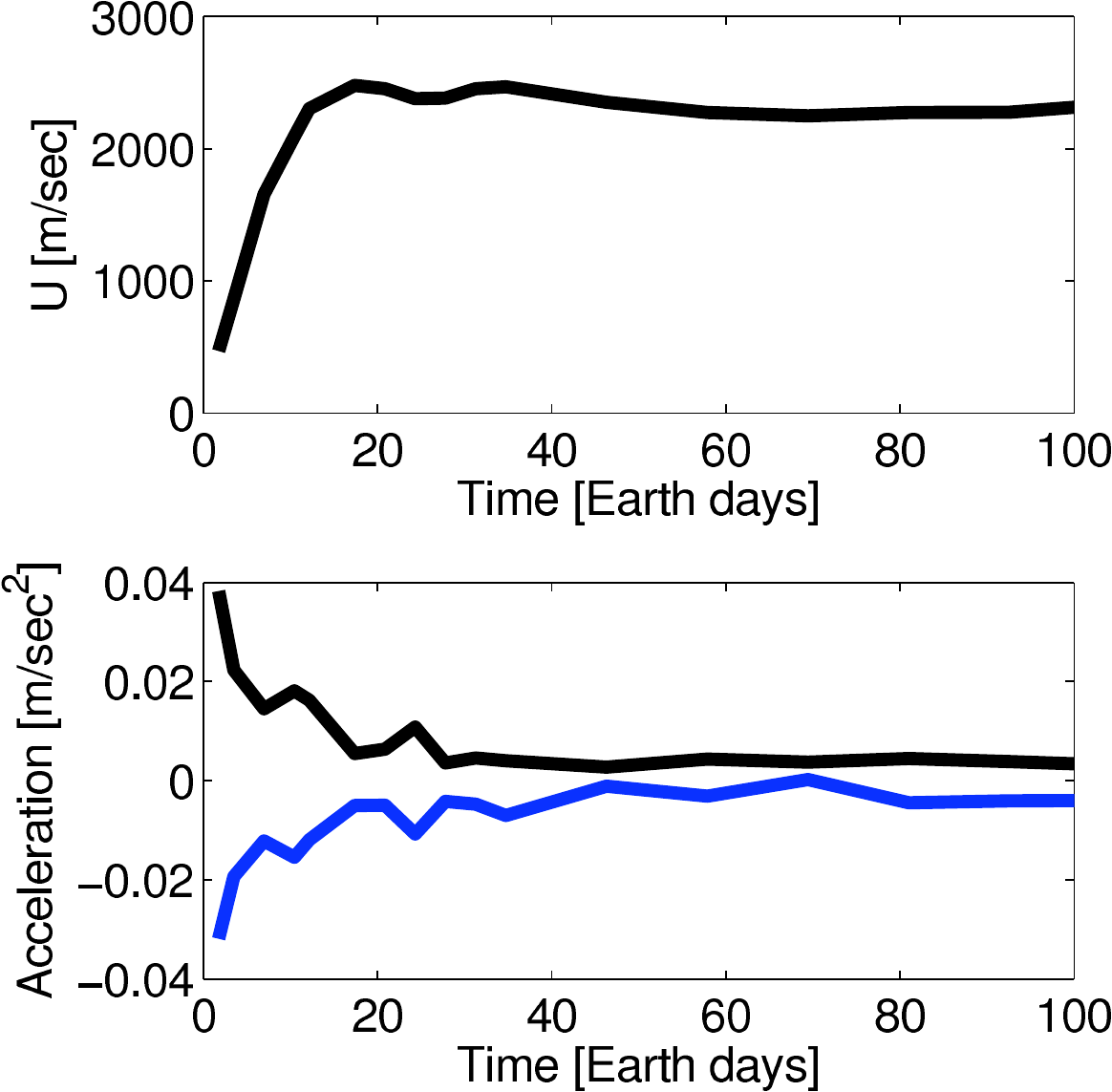}
\caption{Time evolution of the upper-atmospheric circulation in the three-dimensional
model of HD 189733b by \citet{showman-etal-2009}. ({\it Top:})
Zonal-mean zonal wind at the equator. ({\it Bottom:}) Zonal
accelerations due to latitudinal eddy-momentum convergence,
$-(a\cos^2\phi)^{-1}\partial(\overline{u'v'}\cos^2\phi)/\partial\phi$
({\it black curve}) and vertical eddy-momentum
convergence, $-\partial(\overline{u'\omega'})/\partial p$
({\it blue curve}). All quantities are shown in the upper atmosphere, averaged 
vertically from 30 mbar to the top of the model.}
\label{3d-time-evol}
\end{figure}

Figure~\ref{3d-time-evol} shows the time evolution of 
the zonal-mean zonal wind and the two eddy acceleration terms 
at the equator for the solar-metallicity model of HD 189733b 
from \citet{showman-etal-2009}.  These are vertical averages
through the top portion of the atmosphere where the radiative heating/cooling
is strong.  The zonal-mean zonal wind accelerates rapidly from the
initial rest state and approaches an equilibrium within $\sim$100 days
({\it top}).  As expected, the acceleration due to horizontal
eddy transport is eastward, while that due to
vertical eddy  transport is westward  ({\it bottom}).  Moreover,
as suggested by the linear and nonlinear shallow-water calculations,
the magnitude of the horizontal momentum convergence exceeds that
of the vertical momentum convergence during spin-up, 
so the net acceleration is eastward
at early times.  A superrotating equatorial jet therefore develops.
  As the jet speed builds,
the two acceleration terms weaken significantly, and the ratio of their
magnitudes approaches one.  As a result, the {\it net} acceleration
drops to zero, allowing the jet to equilibrate to a constant speed 
({\it top}). This model is in the same regime as the shallow-water
calculation presented in the right column of Fig.~\ref{stswm-diag}.

\begin{figure*}[ht]
\centering
\subfigure{
\includegraphics[scale=0.6]{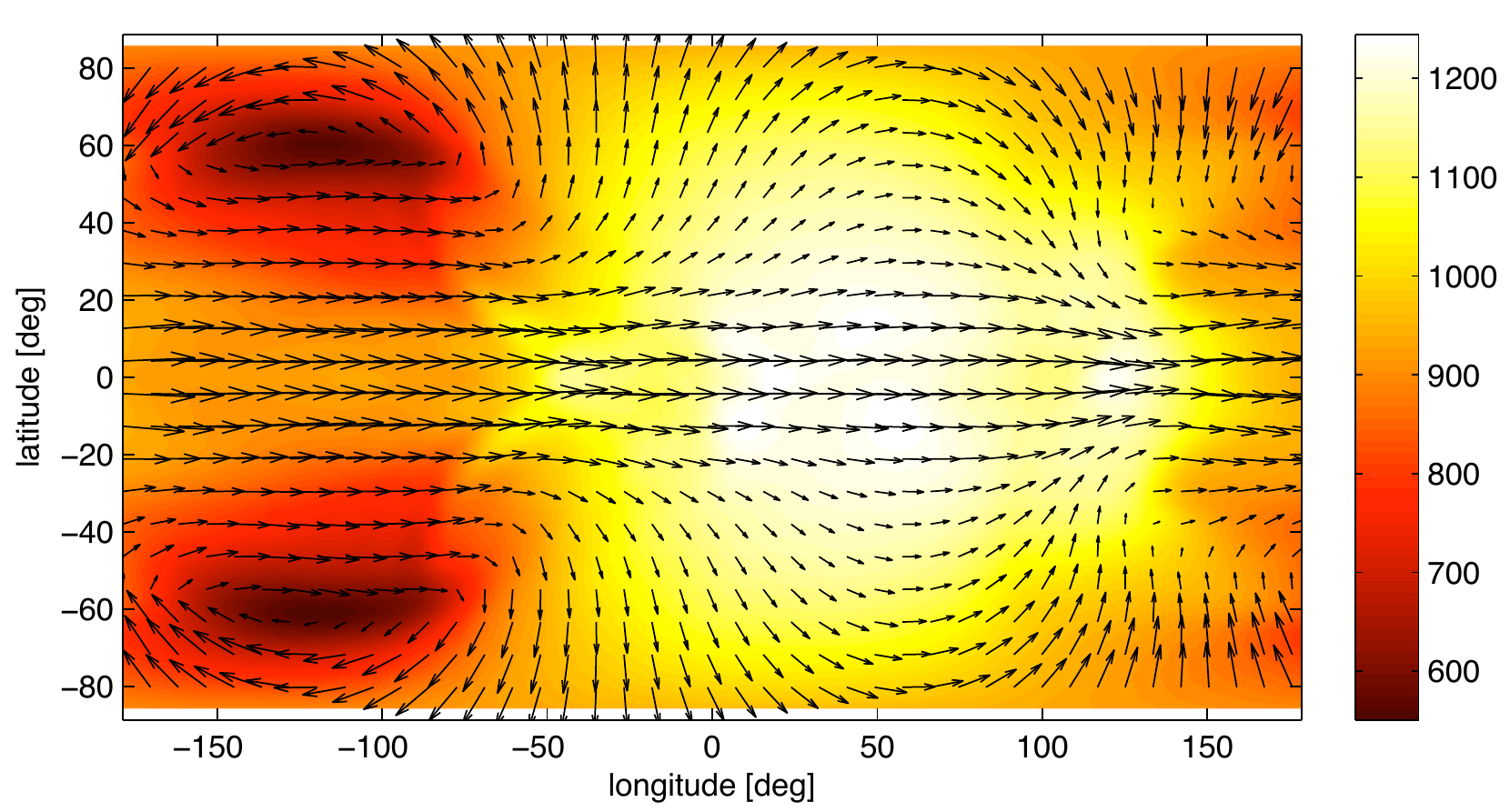}
}
\subfigure{
\includegraphics[scale=0.6]{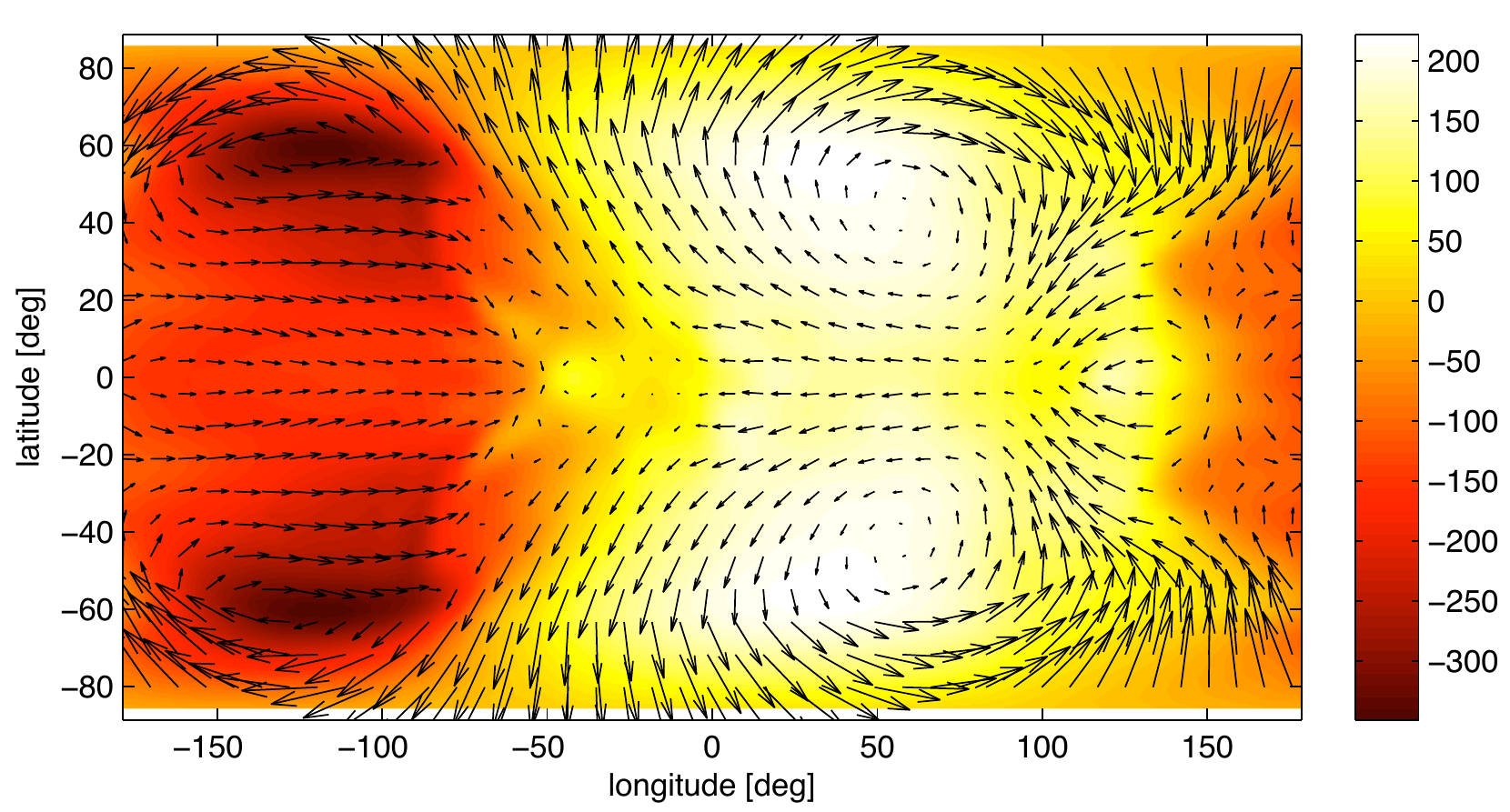}
}
\caption{{\it Top:} Temperature (colorscale, in K) and winds (arrows) for
the solar-metallicity model of HD 189733b from \citet{showman-etal-2009}.
{\it Bottom:} Eddy temperature $T'$ (colorscale, K) and eddy winds
$(u',v')$ (arrows) for the same model.  Both are shown at 30 mbar pressure,
near the mean infrared photosphere, after the winds at these levels
have reached steady state.}
\label{equil}
\end{figure*}

The weakening in time of the eddy accelerations seen in
Fig.~\ref{3d-time-evol} indicates that the mean flow, once
it forms, exerts a back-reaction on the eddies that alters their
structure.  The nature of these changes are illustrated in
Fig.~\ref{equil}.  The top panel shows the temperatures
and winds at 30 mbar pressure after the flow at this altitude has
become steady; the superrotating equatorial jet, eastward offset
of the hottest region from the substellar point, and other features
are evident as detailed in \citet{showman-etal-2009}.  The bottom
panel depicts the {\it eddy} temperature and {\it eddy} winds for the
same pressure and time---that is, $T'$ in colorscale and ($u'$,$v'$)
as arrows.  Several features are similar to those in
Fig.~\ref{3d-spinup-gill}: the eddy flow near the equator is approximately
zonal and exhibits a Kelvin-wave-like character, with predominantly 
eastward flow at some longitudes and westward flow at others; the
midlatitudes contain broad Rossby-wave gyres in each hemisphere,
anticyclonic on the dayside and cyclonic on the nightside.
Interestingly, however, the Kelvin-wave structure is shifted eastward,
and the midlatitude velocity structure differs significantly,
relative to that with weak mean flow (compare
Fig.~\ref{equil}b to Figs.~\ref{matsuno}b and
\ref{3d-spinup-gill}).  From longitudes of about
$-75^{\circ}$ to $+40^{\circ}$, the midlatitude velocity structure induces
equatorward momentum flux (i.e., $u'v'$ negative in the northern
hemisphere and positive in the southern hemisphere), but at longitudes
$\sim80$ to $150^{\circ}$ the flux is reversed (i.e., $u'v'$ positive
in the northern hemisphere and negative in the southern hemisphere).
Due to this cancellation, the magnitude of the zonally averaged flux
$\overline{u'v'}$ is significantly weaker in the equilibrated state
than during the spin-up phase, when the signs of the midlatitude
$u'v'$ add coherently at most longitudes (see Fig.~\ref{3d-spinup-gill}).

\begin{figure}
\vskip 10pt
\includegraphics[scale=0.6, angle=0]{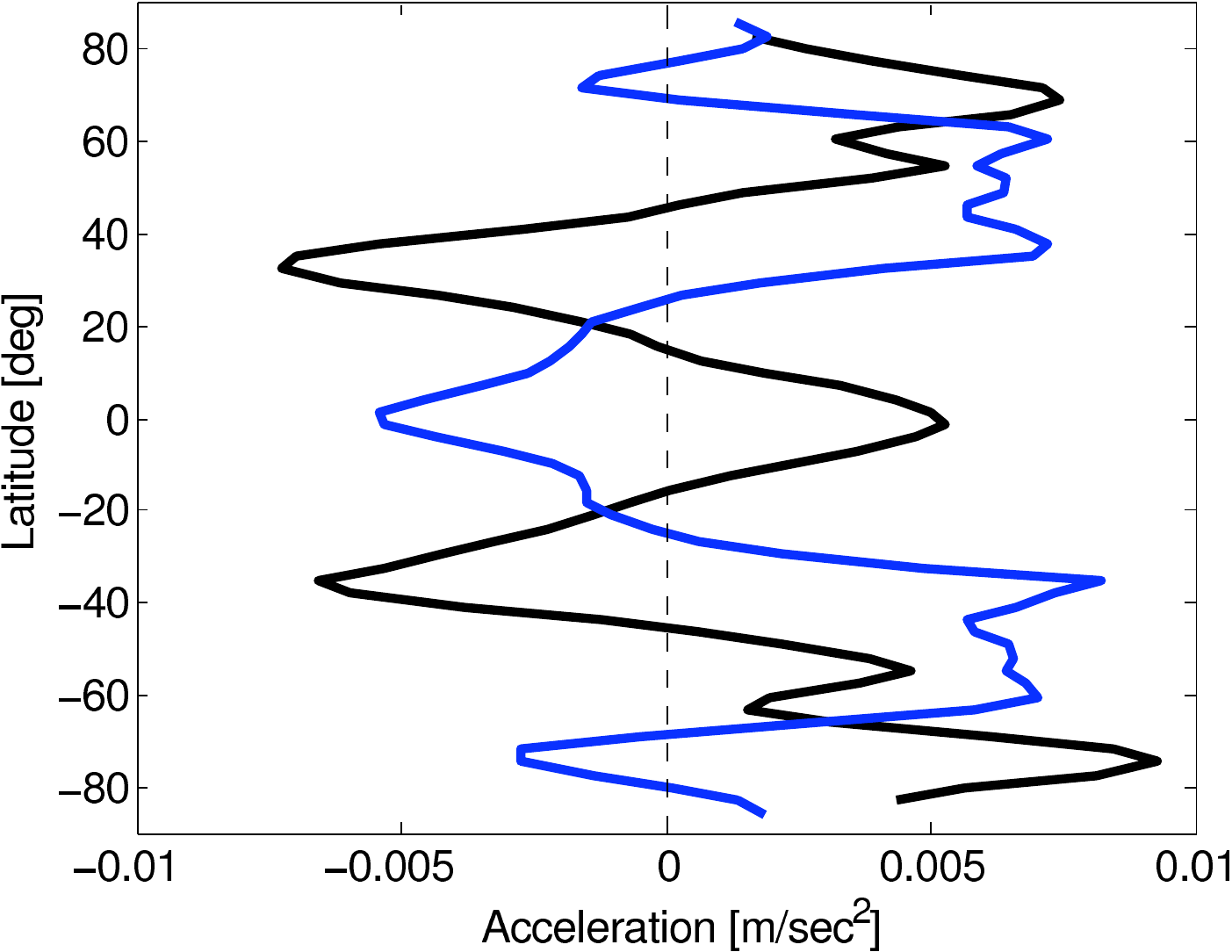}
\caption{Zonal-mean zonal accelerations from the solar-metallicity
model of HD 189733b from \citet{showman-etal-2009}.  Black and
blue show accelerations due to latitudinal and vertical
convergence of eddy momentum, respectively.  Both are vertically
averaged from 30 mbar to the top of the model at 0.2 mbar and temporally
averaged from 11 to 870 days.}
\label{3d-acceleration}
\end{figure}

The latitudinal pattern of zonal-mean
eddy accelerations in the upper atmosphere of the 3D
model, shown in Fig.~\ref{3d-acceleration}, exhibit a 
strong relationship to those from
the shallow-water calculations.  The comparison is most
apt to shallow-water calculations with short
radiative time constant ($\tau_{\rm rad}\sim0.1$--1 day), weak
frictional drag ($\tau_{\rm drag}\to\infty$), and large 
amplitude, as depicted for example in the right
column of Fig.~\ref{stswm-diag}.  
In the 3D model (Fig.~\ref{3d-acceleration}), 
the acceleration due to horizontal eddy-momentum 
convergence is eastward at the equator and westward in 
midlatitudes, in agreement with analytic theory and non-linear shallow-water 
solutions (compare black curve in 
Fig.~\ref{3d-acceleration} to Figs.~\ref{cases-accel}
and \ref{stswm-diag}g).  Near the poles,
the situation is more complex.  Poleward of $\sim$$50^{\circ}$ latitude, 
the acceleration in the upper atmosphere of the 3D run is eastward.
In the shallow-water solutions, the pattern of eddy
accelerations at high latitude depend on $\tau_{\rm rad}$, $\tau_{\rm drag}$,
and the forcing amplitude (e.g., compare Figs.~\ref{stswm-diag}c and 
\ref{stswm-diag}g
poleward of $\sim$$60^{\circ}$ latitude), but for the shallow-water 
cases most relevant to the 3D run shown here---such as the
right column in Fig.~\ref{stswm-diag}---the acceleration 
due to horizontal eddy convergence
becomes eastward at high latitudes (Fig.~\ref{stswm-diag}g), like
that in the 3D run.  In the 3D model, the acceleration due to 
vertical eddy-momentum convergence ({\it blue curve} in 
Fig.~\ref{3d-acceleration}) is westward at the equator
and eastward in the midlatitudes, again like that arising
in the shallow-water solutions, although a significant difference
is that the midlatitude eastward acceleration is weak in
the shallow-water runs but strong in the three-dimensional
run (relative to the magnitude of acceleration at the equator).

Of course, the standing eddy patterns and resulting zonal-wind
accelerations in 3D models depend on the strength
of radiative heating/cooling and drag, just as they do in the
shallow-water models.  For example, in the shallow-water solutions,
the Kelvin-wave structure and Rossby gyres are spatially distinct
when the radiative and/or drag time constants are long and the 
forcing amplitude is small but not when the time constants
are short or the forcing amplitude is large (e.g., compare the
upper left versus lower right of Fig.~\ref{cases}, the top
versus the bottom of Fig.~\ref{stswm-equal-tau}, and the
top versus the bottom of Fig.~\ref{stswm-nonlinearity}).
The 3D models shown here lie at an intermediate position along
this continuum, with Rossby and Kelvin-wave structures that
are visibly distinct, analogous for example to the shallow-water case in 
Fig.~\ref{matsuno}b.   3D models with very strong heating rates,
however, seem to exhibit eddy patterns lacking distinct Rossby
wave gyres, more analogous to the top-left case in Fig.~\ref{cases}
and the top case in Fig.~\ref{stswm-equal-tau}. 
Examples of models in this regime include the topmost part of the
atmosphere in the models of \citet{cooper-showman-2005, cooper-showman-2006},
\citet{koskinen-etal-2007},
\citet{rauscher-menou-2010} and the HD 209458b model of 
\citet{showman-etal-2009}.   In contrast, cases in the literature with 
more modest heating rates tend to exhibit distinct standing Rossby and 
Kelvin-wave structures; examples include \citet[][Fig.~5]{showman-guillot-2002},
\citet[][Figs.~1 and 12]{heng-vogt-2011}, the lower portion of 
some of the models of \citet[][Fig.~3b]{koskinen-etal-2007},
and several of the runs in \citet{thrastarson-cho-2010},
which exhibit a planetary-scale cyclone and anticyclone in each
hemisphere.

Despite differences of detail, the overall broad similarities 
described here between the 
3D and shallow-water models argues strongly that the mechanism
for equatorial jet maintenance that we have identified occurs
in both the shallow-water and 3D models.

\section{Discussion}
\label{discussion}
The development of an eastward equatorial 
jet---that is, equatorial superrotation---is a common feature emerging
from three-dimensional models of synchronously rotating hot Jupiters
and extrasolar terrestrial planets \citep{showman-guillot-2002, 
cooper-showman-2005, cooper-showman-2006,
showman-etal-2008a, showman-etal-2009, dobbs-dixon-lin-2008,
menou-rauscher-2009, rauscher-menou-2010, perna-etal-2010,
heng-etal-2010, joshi-etal-1997, merlis-schneider-2011, heng-vogt-2011}.  
\citet{showman-guillot-2002} first pointed out that, when
the radiative and advective time constants are similar,
this superrotation causes an eastward displacement of the hottest
regions from the substellar point---a phenomenon discovered on HD 189733b
five years later \citep{knutson-etal-2007b, knutson-etal-2009a}.
Despite its relevance, however, the dynamial mechanisms responsible for
generating the equatorial superrotation on tidally locked exoplanets
have not been previously identified.  

Here, we have shown that the equatorial superrotating jet
results from an interaction of the mean flow
with standing, planetary-scale Rossby and Kelvin waves generated by
the day-night thermal forcing.  The strong longitudinal variations 
in radiative heating---namely 
intense dayside heating and nightside cooling---trigger the formation 
of standing, planetary-scale equatorial Rossby and Kelvin waves; 
this is essentially a linear response when wind speeds are modest,
although nonlinearities affect the wave structure at high
amplitude.  The Kelvin waves straddle the equator while the
Rossby waves lie on their poleward flanks.  As a result of the 
differential zonal propagation---Kelvin waves propagating to the east and
long-wavelength Rossby waves to the west---as well as the multi-way 
force balance between pressure-gradient, Coriolis, advective,
and drag forces, the velocities develop tilts that resemble
an eastward-pointing chevron centered at the equator.  These velocity tilts
pump eastward momentum from high latitudes to the equator, thereby 
inducing equatorial superrotation.  In steady state, the zonal-mean 
equatorial jet speed near the photosphere is determined by a balance 
between this eastward, wave-induced acceleration and westward 
equatorial acceleration resulting from vertical eddy-momentum transport 
and/or drag.   We demonstrated the mechanism in a hierarchy of dynamical 
models---including linear, analytic shallow-water models, 
fully nonlinear shallow-water models, and state-of-the-art 
three-dimensional GCMs.  For conditions relevant
to hot Jupiters, such equatorial superrotation 
occurs over a wide range of radiative heating rates and drag time 
constants.   The consistency of the picture emerging
from this sequence of models with widely varying complexity is encouraging
and suggests that the mechanism is robust.

The mechanism identified here has several implications:
\begin{itemize}

\item It implies that the equatorial jet
results from a direct, essentially weakly nonlinear interaction
between the thermally forced waves and the mean flow at the planetary scale.
Eddy-eddy interactions, including the possibility of inverse
or forward energy cascades or other turbulent interactions,
may occur but are not essential to the basic
mechanism.  (This is analogous to the situation suggested by
\citet{ogorman-schneider-2007} for interaction of baroclinic
midlatitude eddies with the mean flow on Earth.)

\item  The wave-mean-flow interaction produces an equatorial jet whose 
latitudinal width is comparable to that of the Rossby waves, namely 
the equatorial Rossby deformation radius modified by radiative and 
frictional effects. For conditions typical of synchronously rotating 
hot Jupiters, this length is comparable to a planetary radius, explaining 
the broad scale of the equatorial jet obtained in most hot Jupiter models. 

\item When acting in isolation, this mechanism for generating
superrotation has no inherent forcing-amplitude threshold; it operates at
any forcing amplitude, unlike the behavior reported in some
earlier studies in the terrestrial context \citep{suarez-duffy-1992,
saravanan-1993}.

\item For parameter combinations appropriate to hot exoplanets,
our models yield flows whose hottest regions along the equator
lie to the east of the substellar point.  This property 
results from the eastward group propagation of 
Kelvin waves.  The development of a strong mean flow (equatorial
superrotation) further enhances the offset by its eastward advection
of the temperature pattern.  Together, these effects
can explain the offsets observed on HD 189733b
\citep{knutson-etal-2007b, knutson-etal-2009a}.
\end{itemize}

Despite the ubiquity of superrotation in our models, 
the flow could shift regimes to one with westward
zonal-mean equatorial flow if westward equatorial accelerations caused
by other processes---not considered here---become important.
For example, if baroclinic instabilities occur in midlatitudes, they
could cause Rossby wave radiation at midlatitudes, potentially allowing
absorption of Rossby waves near the equator.  This would contribute
a westward wave-induced acceleration near the equator.  
If this westward acceleration dominates over the eastward equatorial
acceleration caused by
the day-night heating contrast, then the net wave-induced
acceleration at the equator could be westward and equatorial
superrotation would not occur.  
This could occur on hot Jupiters if the planetary rotation rate 
is sufficiently fast and heating rates are sufficiently low.
Note that baroclinic instabilities cannot occur in a one-layer model such
as the shallow-water model, helping to explain why no such transitions
to westward zonal-mean equatorial flow were seen in the shallow-water models
presented here.  In some cases, Hadley cells may also force the 
circulation into a regime of westward equatorial flow, particularly if
the planetary obliquity is non-zero; this is the case in Earth's
troposphere.   On the other hand,
if the planetary rotation is sufficiently slow, additional mechanisms 
for generating equatorial superrotation become possible, even when the 
forcing is axisymmetric \citep{delgenio-zhou-1996, mitchell-vallis-2010}.
Exploring the combinations of orbital semimajor axes,
stellar fluxes, planetary rotation rates, and atmospheric compositions 
under which such regime transitions can occur in 3D models is an 
important goal for future work, as such transitions could have
major implications for visible and infrared light curves.

It is worth mentioning that the presence of a physical surface---and
entropy gradients on that surface---promote the
existence of baroclinic instabilities in midlatitudes
\citep[see for example][Chapter 6]{vallis-2006}, so tidally locked
terrestrial planets may be more prone than hot Jupiters to experience
a regime transition to a flow containing midlatitude eastward jets in
addition to, or instead of, a flow dominated by equatorial superrotation.

In the geophysical-fluid-dynamics (GFD) literature, the generation of 
midlatitude eastward zonal jets is often described theoretically
using simple barotropic considerations analogous to those summarized in 
\S\ref{background}, involving the propagation of waves.  
While such barotropic considerations
seem to work reasonably well for midlatitude jets,
the barotropic framework fundamentally fails to explain the emergence 
of equatorial superrotation
in 3D models of synchronously rotating exoplanets, where the flow
is often steady and symmetric about the equator.  
The theory presented here overcomes this obstacle and 
represents an extension of the barotropic
framework to a thermally forced, stratified, ageostrophic flow with 
finite Rossby deformation radius.

Specifically, the generation of eastward jets is often
interpreted in terms of the meridional propagation of Rossby waves 
away from a source region and their dissipation at other latitudes 
\citep[e.g.,][]{held-2000, vallis-2006}.   As described in
\S\ref{background}, this interpretation seems to flow naturally
from barotropic theory, in which free Rossby waves, even at the 
equator, can propagate not only in longitude but also in latitude.
In contrast, although our work clearly shows how superrotation can emerge 
on tidally locked planets, 
meridional wave propagation plays no obvious role in our theory.
Unlike barotropic Rossby waves, the baroclinic Rossby 
waves in our theory
are equatorially trapped, confined to an equatorial waveguide:
they can propagate in longitude and potentially height but not 
latitude.\footnote{Although the theory presented here is for steady, 
forced conditions, this key
distinction holds even for freely propagating, unforced waves: 
barotropic Rossby waves can generally propagate in latitude---even
at the equator---while baroclinic equatorial Rossby waves tend to 
be confined to an equatorial waveguide.  For a discussion of the
trapping of equatorial waves, see \citet[][pp.~394-400, 429-432]{holton-2004}
or \citet[][pp.~200-208]{andrews-etal-1987}. }
Moreover, under conditions appropriate to typical hot Jupiters, 
these waves exhibit meridional scales typically stretching from the
equator to the pole.  It is not at all clear that the paradigm of waves 
propagating from one latitude to another applies in this context.
Rather, the velocity tilts that allow the meridional momentum
fluxes to generate superrotation
appear to be explained by the differential {\it zonal}---rather
than meridional---propagation of equatorially trapped Kelvin and 
Rossby waves.  Future work may further clarify the issue.

\appendix
\section{A. Non-dimensionalization}
\label{nondim}

The nondimensional solutions to Eqs.~(\ref{u-mom-nd})--(\ref{h-nd})
involve three dimensionless
parameters --- $k$, $\tau_{\rm rad}$, and $\tau_{\rm drag}$.  Here
we relate these dimensionless
parameters to physical properties for
exoplanets.  For synchronously locked exoplanets, we expect
the forcing to correspond to a zonal planetary wavenumber $1$,
implying a dimensional wavenumber of $a^{-1}$, where $a$ is the
planetary radius.  Thus, our dimensionless wavenumber $k$ has
the value $a^{-1} (\sqrt{gH}/\beta)^{1/2}$, that is
\begin{equation}
k = \left({\sqrt{gH}\over 2 \Omega a}\right)^{1/2}
\end{equation}
If we assume for purposes of illustration that the mean layer 
thickness $H$ is a scale height, then $gH$ is just $RT$ where
$R$ is the specific gas constant and $T$ is the mean temperature
of the atmosphere.
For a hydrogen atmosphere where $R=3700\J\kg^{-1}\K^{-1}$,
this gives
\begin{equation}
k = 0.75 \left({T\over 1000\K}\right)^{1/4} \left({P\over
3\,\rm days}\right)^{1/2}\left({R_{\rm J}\over a}\right)^{1/2}
\end{equation}
where $P$ is the rotational period and 
$R_J$ is Jupiter's radius.  For a carbon dioxide atmosphere 
where $R=189\J\kg^{-1}\K^{-1}$,
\begin{equation}
k=1.2\left({T\over 1000\K}\right)^{1/4} \left({P\over
3\,\rm days}\right)^{1/2}\left({R_{\oplus}\over a}\right)^{1/2}
\end{equation}
where $R_{\oplus}$ is Earth's radius.  Thus, relevant values
of $k$ for a hot Jupiter and a hot Earth are similar, in the
range of $\sim$0.5--2.  Hotter atmospheres, longer rotation
(=orbital) periods, and smaller planetary radii would promote
larger values of $k$.

Consider now $\tau_{\rm rad}$ and $\tau_{\rm drag}$.  
Their nondimensional values
are their dimensional values times $(\sqrt{gH}\beta)^{1/2}$.
Considering $\tau_{\rm nondim}$ to be either dimenionless time constant
($\tau_{\rm rad}$ or $\tau_{\rm drag}$) and $\tau_{\rm dim}$ to be
its dimensional counterpart, we have
\begin{equation}
\tau_{\rm nondim} = \tau_{\rm dim} \left({2 \Omega \sqrt{gH}\over a}.
\right)^{1/2}
\end{equation}
Again equating $gH$ with $RT$, for a hot Jupiter with a hydrogen atmosphere 
we obtain
\begin{equation}
\tau_{\rm nondim} = 3.6 \left({\tau_{\rm dim}
\over 10^5\sec}\right)\left({R_J\over a}\right)^{1/2} 
\left(T\over{1000{\K}}\right)^{1/4} \left({3\,{\rm days}\over P}\right)^{1/2}.
\end{equation}
while for a hot Earth with a CO$_2$ atmosphere
\begin{equation}
\tau_{\rm nondim} = 5.7 \left({\tau_{\rm dim}
\over 10^5\sec}\right)\left({R_{\oplus}\over a}\right)^{1/2} 
\left(T\over{1000{\K}}\right)^{1/4} \left({3\,{\rm days}\over P}\right)^{1/2}.
\end{equation}

\section{B. Analytic solutions for general $\tau_{\rm rad}$ and $\tau_{\rm drag}$}
\label{gill-solutions}

\def\R{{\cal P}}
\def\qtwor{\hat q_{2_{\rm real}}}
\def\qtwoi{\hat q_{2_{\rm imag}}}

Here we present solutions to the nondimensional, linearized 
shallow-water equations, (\ref{u-mom-nd})--(\ref{h-nd}), subject to
thermal forcing and drag.  We follow the solution method outlined by 
\citet{gill-1980} and \citet{wu-etal-2001}. For notational brevity, we 
define $\alpha\equiv \tau_{\rm drag}^{-1}$ and $\gamma\equiv\tau_{\rm rad}^{-1}$. 
Defining
\begin{eqnarray}
\label{qdef}
q=\sqrt{\gamma}\eta + \sqrt{\alpha}u \\
\label{rdef}
r=\sqrt{\gamma}\eta - \sqrt{\alpha}u,
\end{eqnarray}
we convert the coupled equations for $u$, $v$, and $\eta$ 
(Eqs.~\ref{u-mom-nd}--\ref{h-nd}) to equivalent
equations for $q$, $r$, and $v$:
\begin{eqnarray}
\label{qrv1}
{\partial q\over \partial x} + \sqrt{\alpha}{\partial v\over\partial y} = \sqrt{\gamma}yv
+ \sqrt{\alpha} S(x,y) - \sqrt{\alpha\gamma} q\\
\label{qrv2}
{\partial r\over \partial x} - \sqrt{\alpha}{\partial v\over\partial y}=\sqrt{\gamma}yv
+ \sqrt{\alpha \gamma} r - \sqrt{\alpha} S(x,y) \\
\label{qrv3}
yq - yr + \sqrt{\alpha\over \gamma}{\partial q\over\partial y} + \sqrt{\alpha\over\gamma}
{\partial r\over\partial y} = -2\alpha^{3/2} v.
\end{eqnarray}
\citet{gill-1980} and \citet{wu-etal-2001} neglected the drag term in
the meridional momentum equation, which is equivalent to dropping the
term on the right side of Eq.~(\ref{qrv3}).  However, we retain the
full form of Eqs.~(\ref{qrv1})--(\ref{qrv3}).
These equations are separable, and we adopt series solutions 
\begin{equation}
\{q, r, v, S\} = \sum_{n=0}^{\infty} \{q_n(x), r_n(x), v_n(x), S_n(x)\}\psi_n(y).
\label{series-solns}
\end{equation}
Recursion relations for the parabolic cylinder functions are 
\begin{eqnarray}
\label{rr1}
{d\psi_n\over dy}={2n\psi_{n-1}\over \R} - {y \psi_n\over \R^2}\\
\label{rr2}
{d\psi_n\over dy}=-{\psi_{n+1}\over \R} + {y\psi_n\over \R^2}.
\end{eqnarray}
Inserting expressions (\ref{series-solns})
into Eqs.~(\ref{qrv1})--(\ref{qrv3}), using
the recursion relations (\ref{rr1})--(\ref{rr2}), and invoking the
orthogonality of the parabolic cylinder functions leads to the system
\begin{eqnarray}
\label{n-qrv1}
{dq_n\over dx} + \sqrt{\alpha\gamma}q_n - (\alpha\gamma)^{1/4} v_{n-1}=\sqrt{\alpha}S_n 
\qquad\qquad n\ge1\\
\label{n-qrv2}
{dr_n\over dx} - \sqrt{\alpha\gamma} r_n - 2(n+1)(\alpha\gamma)^{1/4} 
v_{n+1}= - \sqrt{\alpha}S_n \qquad\qquad n\ge 0\\
\label{n-qrv3}
2(n+1)q_{n+1} - r_{n-1}= -2\alpha^{3/2} \left({\gamma\over\alpha}\right)^{1/4}  v_n.  
\qquad\qquad n\ge 1
\end{eqnarray}
Equations~(\ref{n-qrv1}) and (\ref{n-qrv3}) do not apply for $n=0$, 
and Eqs.~(\ref{qrv1})--(\ref{qrv3}) instead yield for that case
\begin{eqnarray}
\label{q_0}
{dq_0\over dx} + \sqrt{\alpha\gamma}q_0 = \sqrt{\alpha}S_0.\\
\label{v_0}
(\alpha/\gamma)^{1/4}q_1 = -\alpha^{3/2} v_0
\end{eqnarray}

Given a specified longitude and latitude dependence of the forcing
(and hence given $S_n(x)$ for all $n\ge0$), our goal is to solve for $q_n(x)$, 
$r_n(x)$, and $v_n(x)$.  To determine $q_0$, we simply use Eq.~(\ref{q_0}).
For $n=1$, Eq.~(\ref{n-qrv1}) relates $q_1$ and $v_0$ to $S_1$.  To determine 
$v_0$, use Eq.~(\ref{v_0}).  Inserting into
Eq.~(\ref{n-qrv1}), we obtain
\begin{equation}
\label{q_1}
{dq_1\over dx} + \left(\sqrt{\alpha\gamma} + {1\over\alpha}\right)q_1 
= \sqrt{\alpha}S_1.
\end{equation}
Obtaining an equation for $q_n$ for $n\ge2$ requires full use of 
Eqs.~(\ref{n-qrv1})--(\ref{n-qrv3}).  First, rewrite Eq.~(\ref{n-qrv2}) 
as an equation for $dr_{n-2}/dx$ in terms of $r_{n-2}$,
$v_{n-1}$, and $S_{n-2}$.  Next, obtain an equation for $r_{n-2}$ 
from Eq.~(\ref{n-qrv3}), and differentiate this expression to
obtain an equation for $dr_{n-2}/dx$.  Inserting these two expressions
into the equation derived from (\ref{n-qrv2}) yields
\begin{equation}
\label{coupled-rv}
2n{dq_n\over dx} + 2\alpha^{3/2}\left({\gamma\over\alpha}\right)^{1/4}
{dv_{n-1}\over dx} - \sqrt{\alpha\gamma}\left[2nq_n + 2\alpha^{3/2}\left({\gamma
\over\alpha}\right)^{1/4}v_{n-1}\right]-2(n-1)(\alpha\gamma)^{1/4}v_{n-1}=
-\sqrt{\alpha}S_{n-2}
\end{equation}
Equations~(\ref{n-qrv1}) and (\ref{coupled-rv}) form two coupled differential
equations for $q_n$ and $v_{n-1}$ in terms of the known coefficients
$S_i$.  We solve for
$v_{n-1}$ from Eq.~(\ref{n-qrv1}) and insert this into Eq.~(\ref{coupled-rv})
to obtain a single ordinary differential equation for $q_n$ in terms
of $S_n$, $dS_n/dx$, and $S_{n-2}$:
\begin{eqnarray}
\label{q_n}
\alpha{d^2q_n\over dx^2} + {dq_n\over dx} - [(2n-1)\alpha^{1/2}\gamma^{1/2} +
\alpha^2\gamma]q_n = 
-[\alpha^2\gamma^{1/2} + (n-1)\alpha^{1/2}]S_n
+\alpha^{3/2} {dS_n\over dx} - {\alpha^{1/2}\over 2} S_{n-2}
\end{eqnarray}
for $n\ge2$.  With Eqs.~(\ref{q_0}), (\ref{q_1}), and (\ref{q_n}), 
all possible $q_n$ are determined from the specified forcing
terms.  To determine $v_n$ from the $q_n$, we use Eq.~(\ref{v_0})
for $n=0$, while for $n\ge1$, we use Eq.~(\ref{n-qrv1}):
\begin{equation}
\label{v_n}
(\alpha\gamma)^{1/4}v_n = {dq_{n+1}\over dx} + \sqrt{\alpha\gamma}q_{n+1} - 
\sqrt{\alpha}S_{n+1}.
\end{equation}
To determine $r_n$ from the $q_n$ and $v_n$, we use Eq.~(\ref{n-qrv3}):
\begin{equation}
\label{r_n}
r_n = 2(n+2)q_{n+2} + 2\alpha^{3/2}\left({\gamma\over\alpha}\right)^{1/4}v_{n+1}.
\end{equation}
valid for all $n\ge0$.

We now specify the forcing and solve for the response.
For simplicity, consider a sinusoidal variation of all the variables
in longitude:
\begin{equation}
\{q_n(x), r_n(x), v_n(x), S_n(x)\} = \{\hat q_n, \hat r_n, \hat v_n,
\hat S_n\}e^{ikx}
\end{equation}
where $\hat q_n$, $\hat r_n$, $\hat v_n$, and $\hat S_n$ are 
complex constants and $k$ is the dimensionless zonal wavenumber
associated with the day-night heating contrast.  We take the
forcing to be symmetric about the equator (appropriate for 
a planet with zero obliquity) and, to keep the mathematics tractable,
assume that the forcing is represented solely by the term $S_0(x)$,
corresponding to pattern of heating and cooling that is 
a Gaussian, centered about the equator,
with a half-width (in latitude) of the equatorial
Rossby radius of deformation modified by frictional and radiative
effects:
\begin{equation}
S(x,y)=\hat S_0\psi_0(y)e^{ikx}.
\end{equation}
While the full solution would require consideration
of $S_n$ for all $n\ge0$, the first term, $S_0$, will be the dominant
term for cases where the deformation radius is similar to a planetary
radius, as is the case on typical hot Jupiters. 
Consideration of this term alone will therefore suffice to illustrate the 
qualitative features relevant for pumping the equatorial jet on hot Jupiters.

With these assumptions, Eq.~(\ref{q_0}) implies
\begin{equation}
\hat q_0 = {\sqrt{\alpha}(\sqrt{\alpha\gamma} - ik)\over 
\alpha\gamma + k^2}\hat S_0.
\end{equation}
Given that $S_1=0$, Eq.~(\ref{q_1}) implies that $q_1=0$.
Similarly, Eq.~(\ref{q_n}) implies that 
\begin{equation}
\label{q_2}
\hat q_2 = {\alpha^{3/2} k^2 + 3\alpha \gamma^{1/2} + \alpha^{5/2}\gamma + ik\alpha^{1/2}
\over 2[(\alpha k^2 + 3\alpha^{1/2}\gamma^{1/2} + \alpha^2\gamma)^2 + k^2]}
\hat S_0 
\end{equation}
All $\hat q_n$, with $n\ge3$, equal zero.  From Eq.~(\ref{v_0}),
$v_0=0$, whereas
\begin{equation}
\hat v_1 = {(ik + \sqrt{\alpha\gamma})\over (\alpha\gamma)^{1/4}}\hat q_2
\end{equation}
which can be expressed in terms of the real and imaginary
components of $\hat q_2=\qtwor + i\qtwoi$, as
\begin{equation}
\hat v_1 = [(\alpha\gamma)^{1/4}\qtwor - {k\over(\alpha\gamma)^{1/4}}\qtwoi]
+ i[(\alpha\gamma)^{1/4}\qtwoi + {k\over(\alpha\gamma)^{1/4}}\qtwor].
\end{equation}
All $v_n$, with $n\ge2$, equal zero.  From Eq.~(\ref{r_n}),
\begin{equation}
\hat r_0 = (4 + 2\alpha^{3/2}\sqrt{\gamma} + 2\alpha i k)\hat q_2
\end{equation}
or equivalently
\begin{equation}
\hat r_0 = [(4 + 2\alpha^{3/2}\sqrt{\gamma})\qtwor - 2\alpha k\qtwoi]
+ i[(4 + 2\alpha^{3/2} \sqrt{\gamma})\qtwoi + 2\alpha k\qtwor]
\end{equation}
All $r_n$, with $n\ge1$ equal zero.  This completes the solutions
for $q$, $r$, and $v$.  The solutions for $u$ and $\eta$ are then
determined from (see Eqs.~\ref{qdef}--\ref{rdef})
\begin{eqnarray}
\label{h}
\eta={q+r\over 2\sqrt{\gamma}}\\
\label{u}
u={q-r\over 2\sqrt{\alpha}}.
\end{eqnarray}

It is interesting to consider limits of these solutions
as $\gamma$ and $\alpha$ become infinitely large (implying 
that the radiative or drag time constants go to zero).  
The solution presented above implies that, in the limit  $\alpha\to\infty$ 
at constant $\gamma$, or in the limit $\gamma\to\infty$ at 
constant $\alpha$, the quantities $\hat q_0 \to \hat S_o/\sqrt{\gamma}$, 
$\hat q_2\to \hat S_0/(2\alpha^{3/2}\gamma)$, and 
$\hat r_0\to \hat S_0/\sqrt{\gamma}$.  Noting that
$\eta=(2\sqrt{\gamma})^{-1}
[\hat q_0\psi_0(y) + \hat q_2\psi_2(y) + \hat r_0\psi_0(y)]e^{ikx}$,
these limits imply that $\eta\to \hat S_0 \gamma^{-1} e^{ikx}$,
which, by definition, is simply the radiative-equilibrium
height field.  Thus, when either time constant goes to
zero, the height field becomes the radiative-equilibrium 
height field---even if the other time constant is non-zero.

\section{C. Analytic solutions in the absence of drag}
\label{gill-nodrag}

The steady, linearized, nondimensional shallow
water equations on an equatorial $\beta$ plane
(Eqs.~\ref{u-mom-nd}--\ref{h-nd}) have a particularly
simple analytic solution in the case when frictional drag is set to zero.    
In this case, the nondimensional equation governing the meridional 
velocity (Eq.~\ref{v-equation}) becomes
\begin{equation}
{y^2 v\over \tau_{\rm rad}} - {\partial v\over\partial x}= 
y{\partial S\over\partial x}.
\label{v-equation-no-drag}
\end{equation}
This equation differs vastly from Eq.~(\ref{v-equation}) because
it no longer contains any $y$ derivatives.  As before, we seek separable
solutions that are sinusoids in $x$.   We specify the
forcing amplitude $S(x,y)=\tilde S(y)e^{ikx}$ (where $\tilde S(y)$ is
a specified function that describes the latitude dependence of
the radiative heating/cooling), and search for solutions
$u=\tilde u(y)e^{ikx}$, $v=\tilde v(y) e^{ikx}$, and
$\eta=\tilde \eta(y)e^{ikx}$, $\tilde u(y)$, $\tilde v(y)$, and 
$\tilde \eta(y)$ are complex functions of latitude that we seek to determine.
Inserting these expressions into Eq.~(\ref{v-equation-no-drag}), we
obtain
\begin{equation}
\left({y\over \tau_{\rm rad}}- {ik\over y}\right)\tilde v = ik\tilde S
\label{v-structure}
\end{equation}
from which we have the solution
\begin{equation}
\tilde v = {\tilde S(y)} k \tau_{\rm rad}{i y^3 - k\tau_{\rm rad}y
\over y^4 + k^2\tau_{\rm rad}^2}.
\label{v-soln-no-drag}
\end{equation}
The height field is determined by the fact that $\partial \eta/\partial x
= yv$ (Eq.~\ref{u-mom-nd}), which yields
\begin{equation}
\tilde \eta = \tilde S(y)\tau_{\rm rad}{y^4 + iy^2 k \tau_{\rm rad}\over
y^4 + k^2\tau_{\rm rad}^2}.
\label{h-soln-no-drag}
\end{equation}
We can then determine $u$ using $yu=-\partial\eta/\partial y$ 
(Eq.~\ref{v-mom-nd}), which yields
\begin{equation}
\tilde u=\tilde S(y)\tau_{\rm rad}\left[4{y^6+ iy^4k\tau_{\rm rad}\over
(y^4 + k^2\tau_{\rm rad})^2} - {4y^2 + 2ik\tau_{\rm rad}\over
y^4 + k^2\tau_{\rm rad}^2}\right] - {\partial \tilde S(y)\over\partial y}
\tau_{\rm rad}\left[{y^3 + iyk\tau_{\rm rad}\over y^4 + k^2\tau_{\rm rad}^2}
\right]
\label{u-soln-no-drag}
\end{equation}
Note that $\tilde S(y)$ can be any function; there is no need to
decompose the solution into a summation over normal modes (e.g.,
parabolic cylinder functions), as in the case of finite $\tau_{\rm drag}$.

The solutions take on a particularly simple form in the 
limit $\tau_{\rm rad}\to\infty$:
\begin{eqnarray}
\tilde v=-\tilde S(y)\\
\tilde \eta={i\tilde S(y) y^2\over k}\\
\tilde u =4\tilde S i\left[{y^4\over k^3} - {2\over k}\right]
- {\partial\tilde S\over\partial y}{iy\over k}
\end{eqnarray}
It can be seen that, in this limit, the amplitudes of $v$ are
in phase with $S$.  The peaks in $\eta$ and $u$ are shifted
by $90^{\circ}$ in longitude relative to $S(x,y)$ and are
zero at $x=0$.  The solution in this case is mirror symmetric about the 
$y$ axis, unlike the case with finite time constants.

\begin{figure*}
\vskip 10pt
\includegraphics[scale=0.7, angle=0]{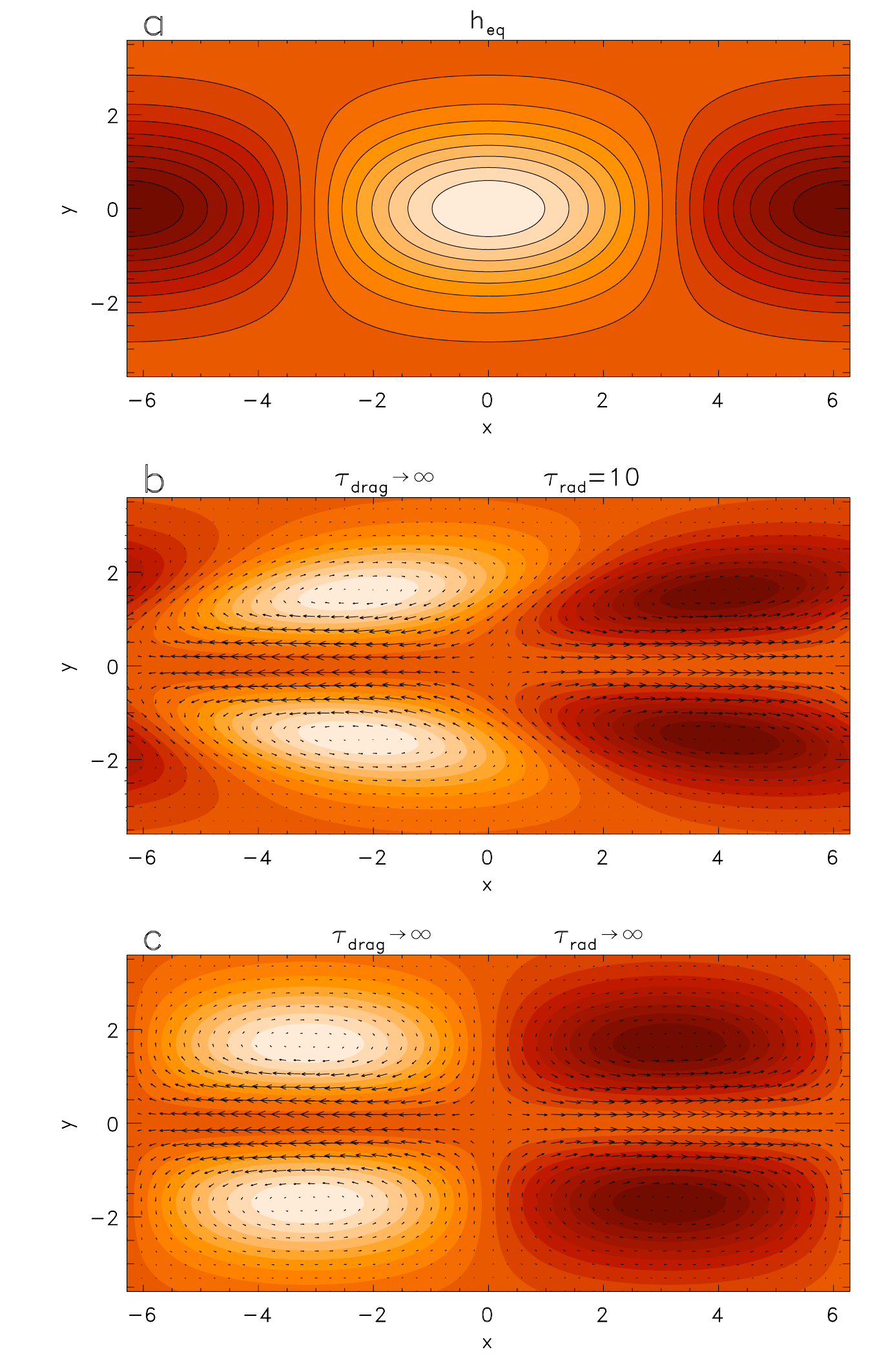}
\caption{Analytic solutions of the linearized shallow-water
equations (\ref{u-mom-nd})--(\ref{h-nd}) in the limit of
zero frictional drag.  ({\it a}) shows the radiative-equilibrium
height field.  ({\it b}) and ({\it c}) show the solutions 
(height field in colorscale and velocity as arrows) for dimensionless 
$\tau_{\rm rad}$=10 and $\infty$, respectively.
}
\label{analyt-nodrag}
\end{figure*}

Figure~\ref{analyt-nodrag} illustrates the solutions for the
case $\tilde S(y)=\hat Se^{-l^2y^2}$ with $l=0.6$ and $\hat S$ being
a real constant.  As in 
Figs.~\ref{matsuno}--\ref{cases}, the solutions exhibit
midlatitude cyclones and anticyclones, with equatorial flow
that is zonally aligned and diverges from a longitude near the
substellar point.  Because drag is zero, the layer is
flat at the equator (see Eq.~\ref{h-soln-no-drag}, which shows that
$\eta=0$ for $y=0$).  
Interestingly, when $\tau_{\rm rad}$ is finite, the westward
phase shift of the Rossby waves is large at low latitudes
and approaches zero at high latitudes (Fig.~\ref{analyt-nodrag}b).  
This leads to phase
tilts that are southwest-northeast in the northern hemisphere
and northwest-southeast in the southern hemisphere---opposite
to the cases shown in Figs.~\ref{matsuno}--\ref{cases}.  In
the limit $\tau_{\rm rad}\to\infty$, these phase tilts disappear
and the solution strongly resembles that shown in the bottom
right corner of Fig.~\ref{cases}.

Analysis of the continuity equation explains  
these behaviors.  In addition to the mass source/sink
caused by the forcing $S$, two processes affect the local layer
thickness---radiative damping and horizontal convergence/divergence.  
Consider how their relative strengths depend on latitude.  
In the absence of drag, the linearized force balance is geostrophic,
i.e., $yv = \partial \eta/\partial x$ and $yu=-\partial\eta/\partial y$
(see Eqs.~\ref{u-mom-nd}--\ref{v-mom-nd}).  This means that,
for a given velocity amplitude, the thickness gradients---and hence
lateral thickness variations themselves---will
be larger farther from the equator.  As a result, for a given
velocity amplitude, the radiative
damping (which is $-\eta/\tau_{\rm rad}$) is stronger farther
from the equator.  On the other hand, in geostrophic balance,
the wind flows parallel to contours of constant height, which
severely limits the horizontal convergence/divergence; convergence
can only come about due to variations of Coriolis parameter $f$
with latitude, which are stronger near the equator.  In
geostrophic balance, the horizontal divergence is $-\beta v/f$,
which is just $-v/y$ for the equatorial beta plane considered here.
Thus, for a given velocity amplitude, the amplitude of
horizontal convergence
is large near the equator but small at high latitudes.  Given
these latitude dependences, we thus expect that the thickness changes
caused by the forcing ($S$) will predominantly be balanced by
radiative relaxation at high latitude but horizontal convergence/divergence
at low latitude.

Equations~(\ref{v-equation-no-drag}--\ref{v-structure}), which represent
the zonal derivative of the continuity equation, describe exactly this
balance.  The three terms correspond to local changes
in layer thickness due to the forcing $S$ (right side),
mass source/sinks due to radiative damping (first term on left side),
and changes in the layer thickness due to horizontal convergence
or divergence of the fluid flow (second term on left side).
For a given amplitude of $v$, 
the radiative damping increases with latitude (scaling with $y$),
while the effect of horizontal convergence decreases with latitude
(scaling with $y^{-1}$).  The transition occurs
at $y\sim\sqrt{k\tau_{\rm rad}}$ (or
$y\sim \sqrt{gH k \tau_{\rm rad}/\beta}$ using dimensional quantities):
radiative relaxation balances the forcing poleward of this latitude,
whereas thickness changes due to horizontal convergence balance the
forcing equatorward of this latitude.

When radiative relaxation balances the forcing ($y\gtrsim 
\sqrt{k\tau_{\rm rad}}$), the
height field is in phase (in longitude)
with the radiative equilibrium height field.  When convergence/divergence
balances the forcing ($y\lesssim \sqrt{k\tau_{\rm rad}}$),
the convergence $\partial u/\partial x + \partial v/\partial y$,
and therefore $v$ itself, are phase with the radiative-equilibrium height 
field.  Since $\partial\eta/\partial x = yv$, this implies that
the height field is phase shifted westward by $90^{\circ}$ relative
to the radiative equilibrium height field.  In the transition
zone ($y\sim\sqrt{k\tau_{\rm rad}}$), these arguments imply that
the phase lines of thickness---and
therefore the velocities themselves---tilt from southwest-to-northeast
in the northern hemisphere and northwest-to-southeast in the southern
hemisphere.  This explains the phase tilts seen in Fig.~\ref{analyt-nodrag}b.
In the limit $\tau_{\rm rad}\to\infty$, the whole domain lies in the 
region where convergence balances forcing, explaining the lack of phase 
tilts of $\tilde \eta$ in Fig.~\ref{analyt-nodrag}c.

\section{D. Analytic solution in limit $\tau_{\rm rad}\to0$}
\label{zero-taurad}

There also exist simple analytic solutions to the dimensionless,
linearized shallow-water equations (\ref{u-mom-nd})--(\ref{h-nd}),
for general $\tau_{\rm drag}$,
in the limit $\tau_{\rm rad}\to0$.  This limit is particularly
relevant for the hottest of tidally locked exoplanets, which,
due to high temperature and/or significant visible-wavelength
opacity, are expected to have short radiative time constants.
Moreover, as a simplification of the full system
it provides insights into the dynamical mechanisms.

In Appendix~\ref{gill-solutions}, we showed that, in the
limit $\tau_{\rm rad}\to0$, the height field converges
toward the radiative-equilibrium height field.  Thus, in
the momentum equations, we can replace the height with
radiative-equilibrium height.  The continuity equation
involves the {\it difference} between $\eta$ and $\eta_{\rm eq}$,
however, and so we must retain $\eta$ in that equation.
This leads to the system:
\begin{eqnarray}
\label{u-zero-taurad}
{\partial\eta_{\rm eq}\over\partial x}-yv=-{u\over\tau_{\rm drag}}\\
\label{v-zero-taurad}
{\partial\eta_{\rm eq}\over\partial y}+yu=-{v\over\tau_{\rm drag}}\\
\label{h-zero-taurad}
\left({\partial u\over\partial x}+ {\partial v\over\partial y}
\right)=  {\eta_{\rm eq}-\eta\over\tau_{\rm rad}}
\end{eqnarray}
Equations~(\ref{u-zero-taurad})--(\ref{v-zero-taurad}) constitute
an algebraic system for $u$ and $v$ that can directly be solved to
yield
\begin{eqnarray}
\label{tau-rad-u}
u=-{{\partial\eta_{\rm eq}\over\partial x}+ y\tau_{\rm drag}{\partial
\eta_{\rm eq}\over\partial y}\over {1\over\tau_{\rm drag}} + 
y^2\tau_{\rm drag}}\\
\label{tau-rad-v}
v={-{\partial\eta_{\rm eq}\over\partial y}+ y\tau_{\rm drag}{\partial
\eta_{\rm eq}\over\partial x}\over {1\over\tau_{\rm drag}} + 
y^2\tau_{\rm drag}}.
\end{eqnarray}
Given $u$ and $v$, Eq.~(\ref{h-zero-taurad}) can then be solved
to determine the (tiny) mismatch between $\eta$ and $\eta_{\rm eq}$.

Consider the limits of this solution for extreme values of
the drag time constant.  As $\tau_{\rm drag}\to0$, 
Eqs.~(\ref{u-zero-taurad})--(\ref{v-zero-taurad}) imply that
$u=-\tau_{\rm drag}^{-1}\partial\eta_{\rm eq}/\partial x$ and
$v=-\tau_{\rm drag}^{-1}\partial\eta_{\rm eq}/\partial y$.  In this
strong-drag limit, the winds simply flow down the pressure gradient.
On the other hand, in the limit of large $\tau_{\rm drag}$, then
away from the equator, we obtain
$u=-y^{-1}\partial\eta_{\rm eq}/\partial y$ and $v=y^{-1}\partial
\eta_{\rm eq}\partial x$---implying geostrophic balance.  Note
that the approximate solutions (\ref{tau-rad-u})--(\ref{tau-rad-v}) 
are singular at the equator in the limit $\tau_{\rm drag}\to\infty$;
in this limit, the divergence becomes infinite at the equator,
and the assumption that $\eta=\eta_{\rm eq}$ used to derive
 (\ref{tau-rad-u})--(\ref{tau-rad-v}) no longer holds.

\begin{figure*}
\vskip 10pt
\includegraphics[scale=0.85, angle=0]{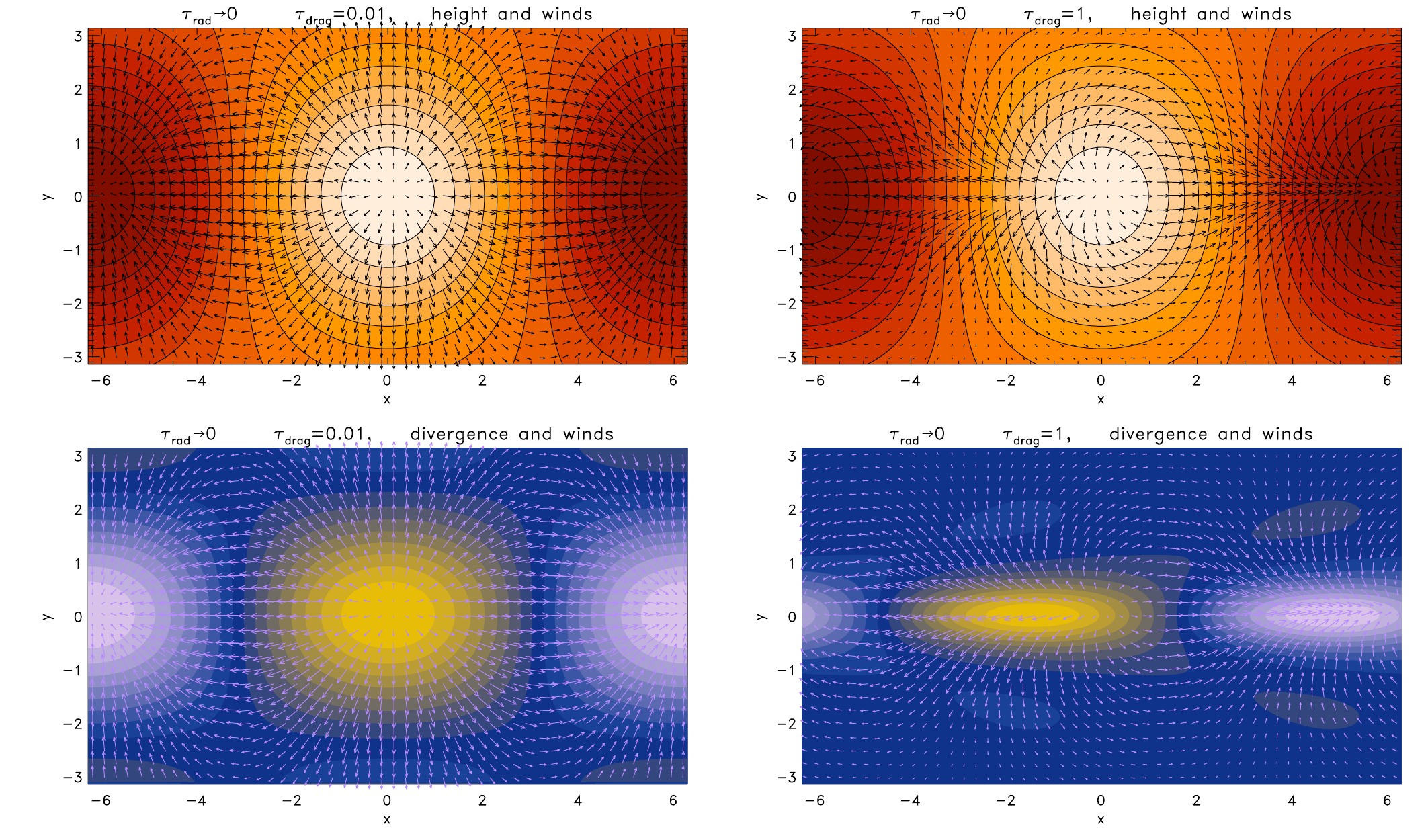}
\caption{Analytic solution of the linearized, dimensionless
shallow-water equations on an equatorial beta plane
(\ref{u-mom-nd})--(\ref{h-nd}) in the limit $\tau_{\rm rad}\to0$.
Left column shows solution for $\tau_{\rm drag}=0.01$, corresponding
to the drag-dominated limit; the right column shows solution
for $\tau_{\rm drag}=1$, implying that Coriolis and drag forces
are comparable over much of the plotted region.  In both
cases, top panel shows winds and height field ($\equiv$radiative-equilibrium
height field) and bottom panel shows divergence.  The 
radiative-equilibrium heightfield is taken as 
$\eta_{\rm eq}=\eta_{\rm eq_0}e^{-l^2y^2}e^{ikx}$ with $l=0.316$
and $k=0.5$.  For $\eta_{\rm eq_0}=0.8$, the maximum dimensionless
wind speed is 0.0045 on the left and 0.45 on the right.  The
dimensionless divergence ranges between $-0.0036$ to $0.0036$ on
the left and between $-0.54$ and $0.54$ on the right.
}
\label{analyt-zero-taurad}
\end{figure*}

Figure~\ref{analyt-zero-taurad} displays this solution for
a drag-dominated case ($\tau_{\rm drag}=0.01$, left column) 
and case where drag is comparable to Coriolis forces over much of 
the domain ($\tau_{\rm drag}=1$, right column).  The top row
shows the winds with the assumed height field, and the bottom
row shows the winds and the horizontal divergence.  As expected, when
drag is strong (left column), the air flows directly from 
the substellar point toward 
the antistellar point and is almost parallel to the pressure gradients.
When drag is less dominant, however (right column), the solution forms a 
Matsuno-Gill-type wind pattern which exhibits velocities that
tilt northwest-southeast in the northern hemisphere
and southwest-northeast in the southern hemisphere.  As discussed
in \S\ref{sw}, this pattern of velocity tilts would generate equatorial 
superrotation.

The mechanism for generating these velocity tilts is, simply,
the three-way force balance between pressure-gradient, Coriolis,
and drag forces.  Because drag points in the opposite direction 
of the velocities, and Coriolis forces are perpendicular to the velocities,
this three-way balance {\it must} give a velocity direction that 
is rotated clockwise of $-\nabla \eta$ in the northern hemisphere
and counterclockwise of $-\nabla\eta$ in the southern hemisphere.
A visual inspection of Fig.~\ref{analyt-zero-taurad} makes clear
that, given the expected pattern of $\nabla\eta_{\rm eq}$, this 
rotation forces the flow pattern to adopt velocity tilts that 
are northwest-southeast (southwest-northeast) in the
northern (southern) hemisphere.



\acknowledgements
This research was supported by NASA Origins grant NNX08AF27G and 
PATM grant NNX10AB91G to APS. 

\bibliographystyle{apj}
\bibliography{showman-bib}



\end{document}